\documentclass[12pt]{article}

\usepackage{spaccent}
\usepackage{fullpage,multicol}
\usepackage{hyperref}
\usepackage{graphicx}
\usepackage{amssymb}
\usepackage{amsmath}
\usepackage{latexsym}
\usepackage{color}
\usepackage[english]{babel}
\usepackage{comment}
\usepackage{dirtytalk}
\usepackage{dblfloatfix}
\usepackage{ntheorem}
\usepackage{siunitx}
\newcommand{\medidas}[0]{\unit[unit-font-command = \mathit]{\second}}

\newcommand{\medidahz}[0]{\unit[unit-font-command = \mathit]{\hertz}}
\newcommand{\medidabits}[0]{\unit[unit-font-command = \mathit]{bits}}

\newcommand{\medidagigabytes}[0]{\unit[unit-font-command = \mathit]{Gb}}

\theoremseparator{:}

\usepackage{tocloft}
\usepackage{enumitem}
\usepackage[dvipsnames]{xcolor}
\usepackage{framed}
\usepackage{listings}
\usepackage{makecell}
\usepackage{multirow}
\usepackage{pifont}
\newcommand{\ignore}[1]{}

\newcommand{\blue}[1]{\textcolor{blue}{#1}}

\usepackage{lipsum}
\newcommand\blfootnote[2]{%
  \begingroup
  \renewcommand\thefootnote{#2}\footnote{#1}%
  \addtocounter{footnote}{-1}%
  \endgroup
}
\let\nvec\vec
\def\vec#1{\nvec{\vphantom t\smash{#1}}}
\newcommand{\encirculado}[1]{\raisebox{.5pt}{\textcircled{\raisebox{-.10pt} {{\footnotesize $#1$}}}}}
\newcolumntype{?}{!{\vrule width 2pt}}
\newlength{\Oldarrayrulewidth}
\newcommand{\Cline}[2]{%
  \noalign{\global\setlength{\Oldarrayrulewidth}{\arrayrulewidth}}%
  \noalign{\global\setlength{\arrayrulewidth}{#1}}\cline{#2}%
  \noalign{\global\setlength{\arrayrulewidth}{\Oldarrayrulewidth}}}
\newcolumntype{M}[1]{>{\centering\arraybackslash}m{#1}}

\newcommand{\yesj}[0]{\textcolor{OliveGreen}{\footnotesize\ding{52}}}
\newcommand{\noj}[0]{}

\begin{document}
\specialaccent
\thispagestyle{empty}

% \begin{multicols}{2}
% \includegraphics[width=5cm]{Cosas del informe/LogoUAI.JPG}

% \begin{center}\bf \footnotesize
% Facultad de Ingeniería y Ciencias\\
% Doctorado en Data Science\\
% \end{center}
% \end{multicols}

\begin{center}
\vspace{4mm}
\textbf{\large Testing chatbots on the creation of encoders for audio conditioned image generation}

\vspace{2mm}
\textbf{Jorge E. León}\textsuperscript{1}\blfootnote{E-mail: jorgleon@alumnos.uai.cl}{*} and \textbf{Miguel Carrasco}\textsuperscript{2}

\vspace{2mm}
\footnotesize{\textsuperscript{1} Adolfo Ibánez University (UAI), Santiago, Chile}

\footnotesize{\textsuperscript{2} Diego Portales University (UDP), Santiago, Chile}
\end{center}

\begin{abstract}

On one hand, recent advances in chatbots has led to a rising popularity in using these models for coding tasks.
On the other hand, modern generative image models primarily rely on text encoders to translate semantic concepts into visual representations, even when there is clear evidence that audio can be employed as input as well.
Given the previous, in this work, we explore whether state-of-the-art conversational agents can design effective audio encoders to replace the CLIP text encoder from Stable Diffusion 1.5, enabling image synthesis directly from sound.
We prompted five publicly available chatbots (namely, ChatGPT o3-mini, Claude 3.7 Sonnet, DeepSeek-R1, Gemini 2.5 Pro Preview 03-25, and Grok 3) to propose neural architectures to work as these audio encoders, with a set of well-explained shared conditions.
Each valid suggested encoder was trained on over two million context related audio–image–text observations, and evaluated on held-out validation and test sets using various metrics, together with a qualitative analysis of their generated images.
Although almost all chatbots generated valid model designs, none achieved satisfactory results, indicating that their audio embeddings failed to align reliably with those of the original text encoder.
Among the proposals, the Gemini audio encoder showed the best quantitative metrics, while the Grok audio encoder produced more coherent images (particularly, when paired with the text encoder).
Our findings reveal a shared architectural bias across chatbots and underscore the remaining coding gap that needs to be bridged in future versions of these models.
We also created a public demo so everyone could study and try out these audio encoders.
Finally, we propose research questions that should be tackled in the future, and encourage other researchers to perform more focused and highly specialized tasks like this one, so the respective chatbots cannot make use of well-known solutions and their creativity/reasoning is fully put to the test.

\end{abstract}

% \tableofcontents
% \newpage

\section{Introduction}\label{sec:Introduction}

\begin{figure*}[t]
\centering
\includegraphics[page=13,width=16.4cm,trim={0 30mm 0 30mm},clip]{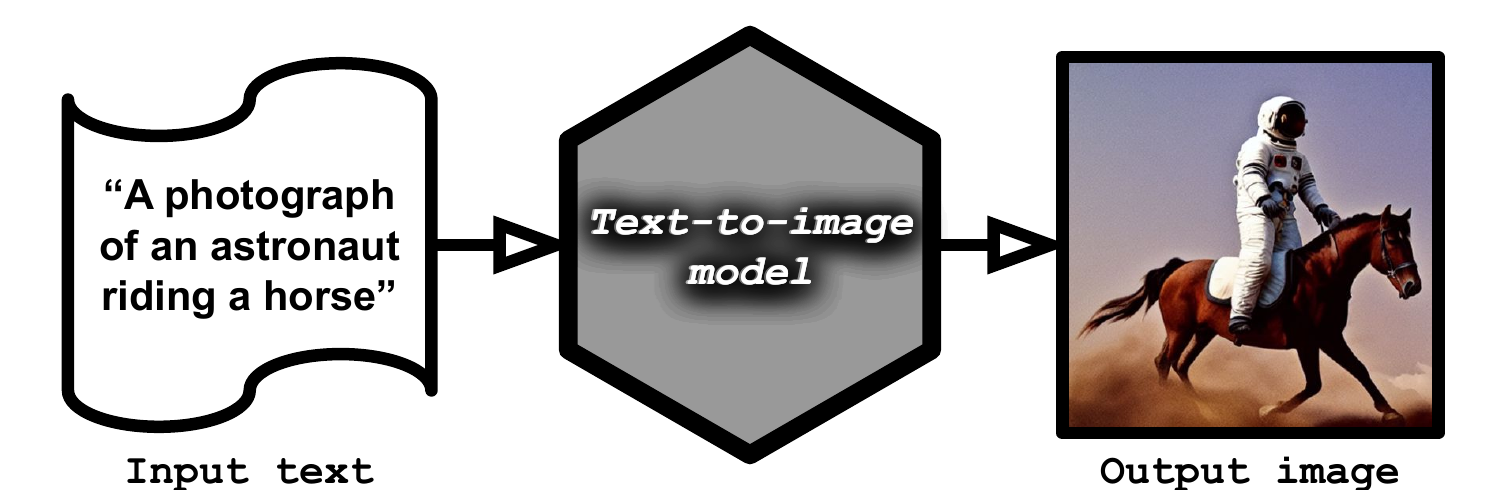}
\caption{Text-to-image generation example, created with Stable Diffusion 1.5. Text-to-image is a technique that generates images from textual descriptions, allowing users to create visual content based on their written prompts. Some popular models that perform this task are Stable Diffusion \cite{High-Resolution_Image_Synthesis, SDXL, Stable_diffusion_3, Stable_Diffusion}, DALL·E \cite{Zero-Shot_Text-to-Image, Improving_Image_Generation_with_Better_Captions, DALLE_3}, Imagen \cite{Photorealistic_Text-to-Image_Diffusion_Models, Imagen3} and FLUX \cite{Flux}.}
\label{fig:texto-a-imagen}
\end{figure*}

In the latest years, there has been an unprecedented development in the world of machine learning \cite{From_Classical_Machine_Learning_to_Deep_Neural_Networks}.
Several models have begun to excel in creative activities (previously considered exclusive to human minds by many) \cite{Text-to-image_Diffusion_Models, Creativity_and_Machine_Learning}, and even using non-specialized hardware \cite{A_Survey_of_On-Device_Machine_Learning}.
In this scenario, models have emerged that can generate text associated with an image \cite{CLIP, BLIP, BLIP-2}; just as others have appeared that, based on texts/prompts, are capable of generating images that can fairly faithfully represent said texts \cite{Text-to-image_Diffusion_Models, High-Resolution_Image_Synthesis, SDXL, Imagen3, Flux}.
An example of this can be seen in Figure \ref{fig:texto-a-imagen}.

From this last task, usually referred to as text-to-image, several others emerge, such as: inpainting \cite{Image_inpainting}, outpainting \cite{Outpainting_Images_and_Videos}, or image-to-image \cite{Image-to-Image_Translation, Comparison_and_Analysis_of_Image-to-Image}.
All of these make use of texts and images as input, meaning that they are clear examples of multimodal techniques for image generation.

While there are numerous image synthesis works that give images and texts as input, there are not many that include audio in the equation (whether with or without additional input texts or images involved).
Moreover, it has even been mentioned that, relative to other image datasets, audio-visual datasets are few and far between \cite{Audio-to-Image_Cross-Modal}.
As an added point, working with audio is not as intuitive as doing so with text \cite{MusicLM, Intuitive_Multilingual}.
Due to this, as alluded to in \cite{Make-an-audio}, audio-related generative models in general lag behind in research; fact that can be corroborated while exploring fields such as image generation conditioned by audio \cite{Codi2, Any-to-Any_Generation, BindDiffusion}, in contrast with image generation conditioned by text and image \cite{Image-to-Image_Translation, Text-to-image_Diffusion_Models, High-Resolution_Image_Synthesis, SDXL, Imagen3, Flux}.

Our literature review provided clear evidence on the existence of relationships between audio and text that represent the same situation, as well as between audio and image, that should be further exploited by research and modern models (for a small summary on generative tasks that involve said modality combinations, consult Table \ref{tab:Research_summary}).
This could have an impact on: multimodal data analysis, correction of low-quality/low-resolution recordings, video generation for various purposes (virtual assistants, music videos, video transitions, etc.), democratization of artificial intelligence, augmented reality that incorporates the environmental audio of the user, transfer learning with multimodal models, among others \cite{Remote_Sensing_Image_Generation_From_Audio, Audio_deepfakes, I_Hear_Your_True_Colors, Deep_Audio-visual_Learning, AudioGen, Jukebox, A_Survey_on_Audio_Synthesis, A_survey_of_multimodal_deep_generative_models, Multimodal_Learning_With_Transformers, Multimodal_Image_Synthesis_and_Editing}.

\begin{table}[t!]
    \begin{tabular}{m{0.17\textwidth}m{0.21\textwidth}m{0.49\textwidth}m{0.0\textwidth}}
        \hline
        \centering \textbf{Task} & \centering \textbf{Description} & \centering \textbf{Nuances} &\\ \hline
        Image-to-audio & Based on an image, an audio is generated that conveys the same semantic information as the input image. & Advances have been made in the generation of audios that mimic the possible soundscape for a given image \cite{I_Hear_Your_True_Colors, Any-to-Any_Generation}. In a similar fashion, audios can also be generated from videos, which are nothing more than an ordered collection of images \cite{A_Survey_on_Audio_Synthesis, Deep_Audio-visual_Learning}. &\\ \hline
        Text-to-audio & Based on a text, an audio is generated that conveys the same semantic information as the input text. & Some models are able to resemble a human voice reading the text given as input (subtask usually referred to as text-to-speech \cite{Audio_deepfakes, A_Survey_on_Audio_Synthesis, Vall_e, Towards_audio_language_modeling}). Moreover, some even make music \cite{Mustango} and generate the lyrics based on text input \cite{Jukebox}, or generate sounds that accommodate to a given description \cite{Fugatto, AudioGen, Any-to-Any_Generation, AudioLDM}. &\\ \hline
        Audio-to-image & Based on an audio, an image is generated that conveys the same semantic information as the input audio. & Voice recordings can be used to condition the modification of human faces so their mouths adapt to the corresponding sounds (i.e. lip sync \cite{Audio_deepfakes, Multimodal_Image_Synthesis_and_Editing}), and even the whole face can be created from scratch with the aforementioned recordings \cite{A_Survey_on_Audio_Synthesis}. In addition, some models are capable of representing scenarios where a specific audio is produced \cite{Any-to-Any_Generation, Deep_Audio-visual_Learning}. &\\ \hline
        Audio-to-text & Based on an audio, a text is generated that conveys the same semantic information as the input audio. & The most popular subtask here probably is speech transcription (or recognition) \cite{Audio_deepfakes, Multimodal_Learning_With_Transformers, transcripter_generation, whisper}. However, models that remarkably generate text description (or captions) from audios in general have begun to arise in recent years \cite{Any-to-Any_Generation, AudioSetCaps, BLAP, BLAT}. &\\ \hline
    \end{tabular}
    \caption{A summary on the most common generative audio-text and audio-image tasks.}
    \label{tab:Research_summary}
\end{table}

Despite the above, we can still come up with ways to adapt the use of existing models to work with different data modalities than the ones that were originally intended for.
For instance, given the mentioned advancements in image-to-image models that are conditioned on textual inputs, it could be worth considering a new approach for scenarios where the objective is to perform image-to-image generations using audio instead of text.
A logical strategy for this goal could be to transcribe the audios into the corresponding textual representations/descriptions, which could then be utilized within existing text-image models.
This method should leverage the strengths of well established text-image models, potentially validating the addition of audio.
Although, it is crucial to acknowledge that, in addition to the fact that fields like audio-to-text conversion are still evolving and have not received as much attention as their visual counterparts \cite{Audio_Describing_Sound, Audio_Text_Models_Do_Not_Yet, Deep_Audio-visual_Learning, transcripter_generation, Cacophony}, such approach presents several challenges that should be kept in mind.
Let us review the main ones:

\begin{enumerate}[label=\Alph*]
    \item Word limit in current models: currently, the problem of increasing the token window (i.e., words and characters) of text-to-image and audio-to-text models is open. For example, Stable Diffusion (an open-source neural network model that generates images based on text and/or image \cite{High-Resolution_Image_Synthesis}) has a context window of 75 tokens \cite{SD_Akashic_Records}.
    \item Compatibility between text-image and audio-text models: even if a capacity of hundreds of thousands of tokens is reached to describe any audio (as can be seen analogously in certain current text generation models \cite{Llama, The_Llama_3_Herd_of_Models, Mamba, Mistral_Models, Claude}), the syntax of the text obtained with such an audio-text model must match that used by the respective text-image model with which it is to be combined, in order to maximize communication between the two \cite{AudioGen, High-Resolution_Image_Synthesis, SD_Akashic_Records, The_Dawn_of_LMMs, gemini_1_5, Audio_Text_Models_Do_Not_Yet, Benchmarking_Cognitive_Biases}.
    \item Noise incorporation\footnote{See \cite{What_is_noise} for a brief classical exploration of the definition of the term.}: in addition to the above, it has repeatedly been shown that transforming one modality to another is prone to incorporating noise or failing (to some extent) due to the noise that the data contains beforehand \cite{transcripter_generation, Audiobox, Nlip, Bicro, From_Association_to_Generation, Noise_Aware_Learning}. As a result, the more transformations we make, the more noise we risk adding in the process.
    \item Incorporation of biases: finally, it is pertinent to highlight that, influenced both by the data and their training architectures and configurations, models tend to prioritize and specialize in certain types of audio and have their own preferences for describing them \cite{Are_Models_Biased_on_Text, Word_Level_Explanations, Cacophony, Large_pre_trained_language_models, On_Some_Biases_Encountered, Pre_trained_Speech_Processing_Models, Look_Listen_and_Answer, Dont_Just_Assume}. For example, typical cases of this can be seen in the underestimation/distortion of the order of events \cite{Audio_Text_Models_Do_Not_Yet, Benchmarking_Cognitive_Biases} or in the omission of details considered irrelevant \cite{Mirrorgan, Benchmarking_Cognitive_Biases}.
\end{enumerate}

It is because of these reasons that even if in some cases audios could/can be converted to texts for image generation, this is a significantly more problematic approach than just using the original audios instead.
For this reason, in this research we claim that, when working with a given set of modalities, it is convenient to perform the least number of data modality conversions possible.
Furthermore, we believe that more audio-to-image research is needed to better address the respective tasks, instead of just trying to get by with what is already available.

Nevertheless, text-to-image is not the only field with great advancements, but text generation as well.
This is particularly noticeable with the surge of multiple publicly available chatbots, which are commonly put to the test in different coding tasks \cite{Do_Large_Language_Models_Pay_Similar_Attention_Like_Human_Programmers, A_Survey_on_Large_Language_Models_for_Code_Generation, Large_Language_Models_for_Code_Generation}.
However, a constant concern that looms over these models is running out of tasks to truly explore their limitations, in order to find points of improvement \cite{Line_Goes_Up, Web-Bench, Inadequacies_of_Large_Language_Model_Benchmarks}.

The previous translates into the clear need for more specialized coding tasks for chatbots, with coherent methods to assess the quality of the obtained results.

In light of the above, we have formalized and conducted an experiment for chatbots to generate audio encoders that can replace the text encoder of Stable Diffusion 1.5, and test the properly trained models under various metrics (both quantitative and qualitative).
This paper delves into all of that, assuming that the reader only possess general knowledge regarding the inner workings of artificial neural networks.

In summary, in this paper we address the need for more research in audio-to-image, as well as for more chatbot tests on coding tasks (which are in constant danger of running out of methods to search for their flaws).
Keeping in mind the importance of fairness in our experiments, we employed a shared set of well defined conditions across all of them.
It is worth mentioning that we were also careful of merely using data which is free of copyright conflicts.
Finally, we discuss about our results, together with some possible improvements and lines of research that could follow this work.
\section{Preliminaries}\label{sec:Preliminaries}

Large language models (LLMs), a type of deep neural networks \cite{Grok_Gemini_ChatGPT_and_DeepSeek}, have made notorious breakthroughs on conversational artificial intelligence by using a dual-phase strategy: first undergoing extensive pre-training on vast human-curated datasets, and then being fine-tuned with targeted human guidance.
This methodology has empowered them to produce varied and lifelike text.
The most prominent use case for LLMs is in the form of chatbots \cite{A_Systematic_Review_and_Comprehensive_Analysis_of_Pioneering_AI_Chatbot_Models, A_contemporary_review_on_chatbots, A_comprehensive_review_of_large_language_models}.

One task with high interest in the community for these chatbots is code generation, which is a common point of reference regarding the quality of the models \cite{Do_Large_Language_Models_Pay_Similar_Attention_Like_Human_Programmers, A_Survey_on_Large_Language_Models_for_Code_Generation, Large_Language_Models_for_Code_Generation}.
However, there have been several tests and benchmarks that have become pointless as chatbots keep rapidly improving and reaching consistently perfect scores on them.
This means that there is always room for more sophisticated and specialized code generation tests for these models \cite{Line_Goes_Up, Web-Bench, Inadequacies_of_Large_Language_Model_Benchmarks}.

In parallel to the previous, during the last decade, image generation has experienced enormous growth, driven by significant advances in fields such as artificial intelligence, machine learning and computer vision \cite{RenAIssance, The_Pile}.
This progress has led to the creation of increasingly realistic and stylized images \cite{Image_Generation}.
While, thanks to advances in the quality of computer-generated images (with recent examples like Stable Diffusion XL \cite{SDXL} or 3 \cite{Stable_diffusion_3}, DALL·E 3 \cite{DALLE_3, Improving_Image_Generation_with_Better_Captions}, Imagen 3 \cite{Imagen3} or FLUX \cite{Flux}), the level of these images has reached a degree that makes it difficult to differentiate them from human-generated images; there is still much work to be done in terms of improving quality consistency, reducing biases, lowering computational costs, and facilitating user control over the generations (i.e. generating what the user actually expects/wants) \cite{Text-to-image_Diffusion_Models}.

\begin{figure*}[t]
\vspace*{-6mm}
\centering
\includegraphics[page=1,width=16.4cm]{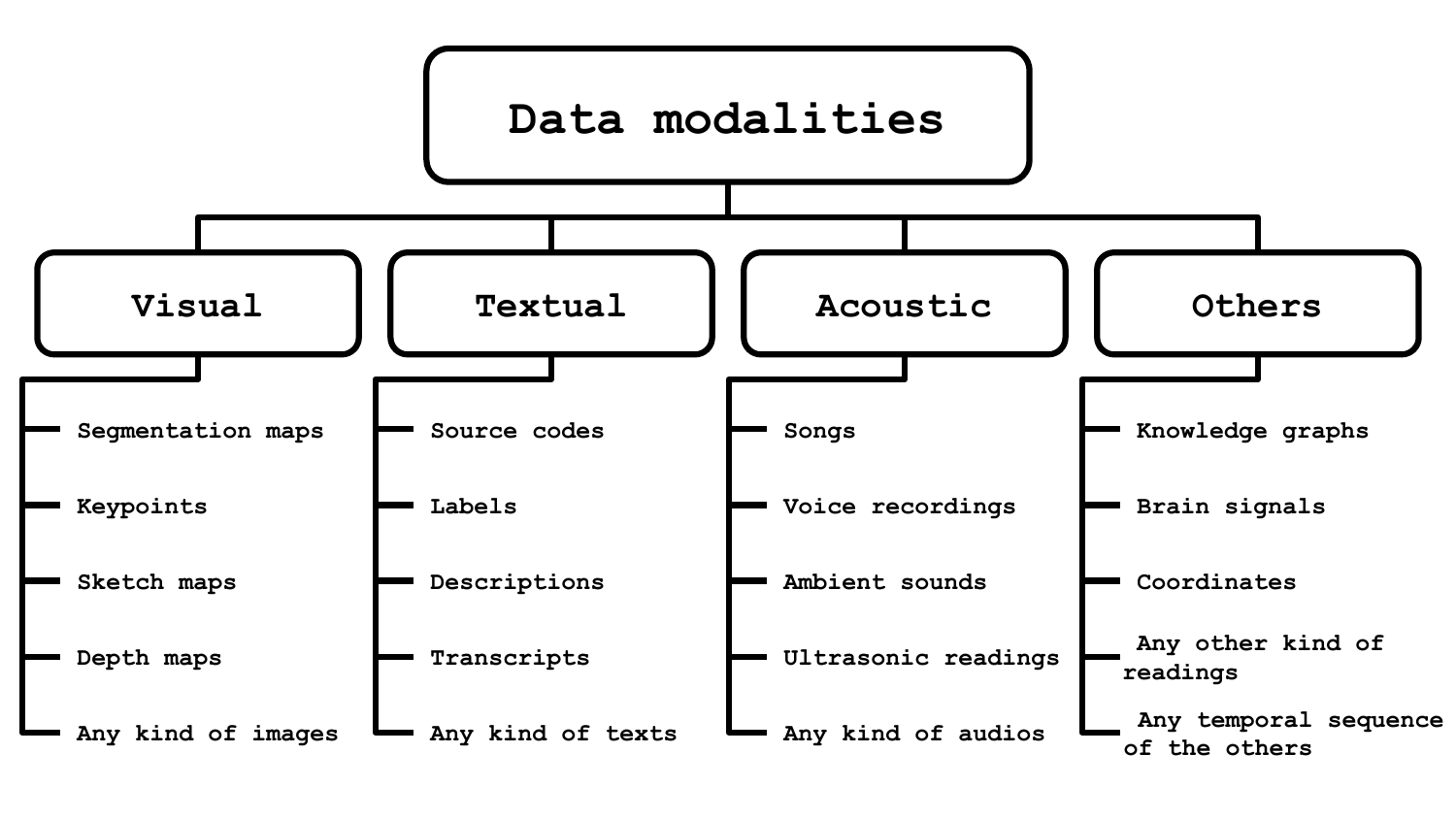}
\vspace*{-10mm}
\caption{Types of data modalities.}
\label{fig:modalidades}
\end{figure*}

To address this last challenge, one of the strategies that has been adopted is to increase the number of data modalities that the models receive (i.e. the types of data that are taken as input; e.g. text, image, audio, etc.) \cite{A_survey_of_multimodal_deep_generative_models, Any-to-Any_Generation, Multimodal_Learning_With_Transformers, Multimodal_Image_Synthesis_and_Editing, gemini_1_5}.
It is pertinent to comment that this increase in the number of modalities not only allows for greater control on the respective tasks, but also opens a way to perform new ones (for example, a detailed analysis can be seen in \cite{The_Dawn_of_LMMs}; where the capabilities of GPT-4V, a colossal multimodal model of text and images, are particularly studied).
In order to better illustrate the concept of data modalities, and inspired by the classification of data types explained in \cite{Multimodal_Image_Synthesis_and_Editing}, in Figure \ref{fig:modalidades} we present a conceptual map of the types of data modalities that can be used, along with examples for each.\footnote{For the sake of brevity, in our conceptual map we are just including the most popular examples.}

An example of the use of multiple data modalities tends to be seen in image-to-image generation, where an image is taken as a reference to generate a new image, since the input image is usually accompanied by a text or a label to better condition/guide the final result \cite{Image-to-Image_Translation}.

In contrast, audio conditioned image-to-image generation has not been explored as much as text conditioned image-to-image generation.
The latter may be because working with audio is not as intuitive as working with text \cite{MusicLM, Intuitive_Multilingual}, or, relative to other image datasets, audio-visual datasets are few and far between \cite{Audio-to-Image_Cross-Modal}.
Nevertheless, that does not invalidate the potential benefit that could be obtained by using audio in certain scenarios.
For example, this could have an impact on: multimodal data analysis \cite{Multimodal_Image_Synthesis_and_Editing, Remote_Sensing_Image_Generation_From_Audio}, correction of low-quality/low-resolution recordings \cite{A_survey_of_multimodal_deep_generative_models, Multimodal_Learning_With_Transformers}, video generation for various purposes (virtual assistants, music videos, video transitions, etc.) \cite{Deep_Audio-visual_Learning, A_Survey_on_Audio_Synthesis}, democratization of artificial intelligence \cite{Audio_deepfakes}, augmented reality that incorporates the environmental audio of the user, transfer learning with multimodal models \cite{I_Hear_Your_True_Colors}, among others.

The literature also presents significant advances in exploiting audio-text or audio-image relationships.
This is corroborated by multiple audio-image works. Some examples are audio-based image generations \cite{Audio-to-Image_Cross-Modal, Remote_Sensing_Image_Generation_From_Audio, Audio-to-Visual_Cross-Modal}, sound source localizations in audiovisual recordings (which not only identify which sector of the image is emitting sounds but also which sounds), audio-image pairings (which detect the most relevant audio for a particular image or vice versa) \cite{Deep_Audio-visual_Learning}, or audio-based image generations \cite{I_Hear_Your_True_Colors, A_Survey_on_Audio_Synthesis}.
Similarly, audio-text cases can be seen, such as audio generation based on text \cite{AudioGen, Jukebox} or text generation based on audio \cite{Audio_Describing_Sound, Audio_Text_Models_Do_Not_Yet, Deep_Audio-visual_Learning, transcripter_generation, whisper}.
There are some examples of image generation based on audio and text \cite{AudioToken, TimbreCLIP}, and there are even cases of image-to-image generation assisted by audio, but for specific cases such as face changes (which replace a person's features with another's while maintaining consistency with the original voice recording) or lip synchronizations (where, for an image of a person, a video is generated while simulating mouth movement according to a voice recording) \cite{Audio_deepfakes, A_Survey_on_Audio_Synthesis}, which could be labeled more as a case of inpainting than image-to-image.
Finally, advances in other similar areas can also be highlighted (such as text-to-video, appreciable with models like Sora \cite{sora_analisis, sora_reporte}, Veo \cite{Veo}, Gen-3 \cite{Gen3} and Movie Gen \cite{Movie_Gen}), and more information on some of these developments can be found at \cite{A_Survey_on_Generative_Diffusion_Models, A_survey_of_multimodal_deep_generative_models}.

Currently, image-to-image generation conditioned by audio is a little explored area of high interest in the community.
To the best of our knowledge, one of the best models to date for this task is the recent CoDi model \cite{Any-to-Any_Generation}.
This is a model that can take any combination of audio, image, text, and video inputs, and create material of any of those types (a task they called any-to-any).
Additionally, a new version (CoDi-2) has also been published, which is more flexible and adapted to conversations \cite{Codi2}.
Another similar option is NExT-GPT, which also allows for a conversational creative process, and it works as well with audio, image, text, and video inputs \cite{NExT_GPT}.
Despite their promising results for future iterations, they have not yet reached a quality that could be considered ideal.
Probably, the best open-source model for this task is BindDiffusion \cite{BindDiffusion}.
This model is both based on the image generation model Stable Diffusion \cite{High-Resolution_Image_Synthesis}, and on the multimodal encoder ImageBind, which incorporates six modalities, including, predictably, audio and image \cite{Imagebind}. 
Notwithstanding, its apparently higher quality than CoDi or NExT-GPT, it also has room for improvement, and it is not evident that it is always advisable to include the largest possible number of data modalities in these models (as seems to have been attempted in all of these cases).

In light of the above, we have come to the conclusion that the design of encoders for image-to-image generation conditioned by audio would be an excellent subject to further test the coding capabilities of modern chatbots.
To the best of our knowledge, only a few researches have specifically attempted this, and none with optimal results \cite{AudioCLIP, Wav2CLIP, TimbreCLIP}.
Meaning that there is no well-known solution and progress could also be made in the field through this assessment.
Due to this, in this research we propose to formalize and perform such test in some of the most modern and popular chatbots available.
\section{Methodology}\label{sec:Methodology}

\begin{figure*}[t]
% \vspace*{-6mm}
\centering
\includegraphics[page=14,width=16.4cm]{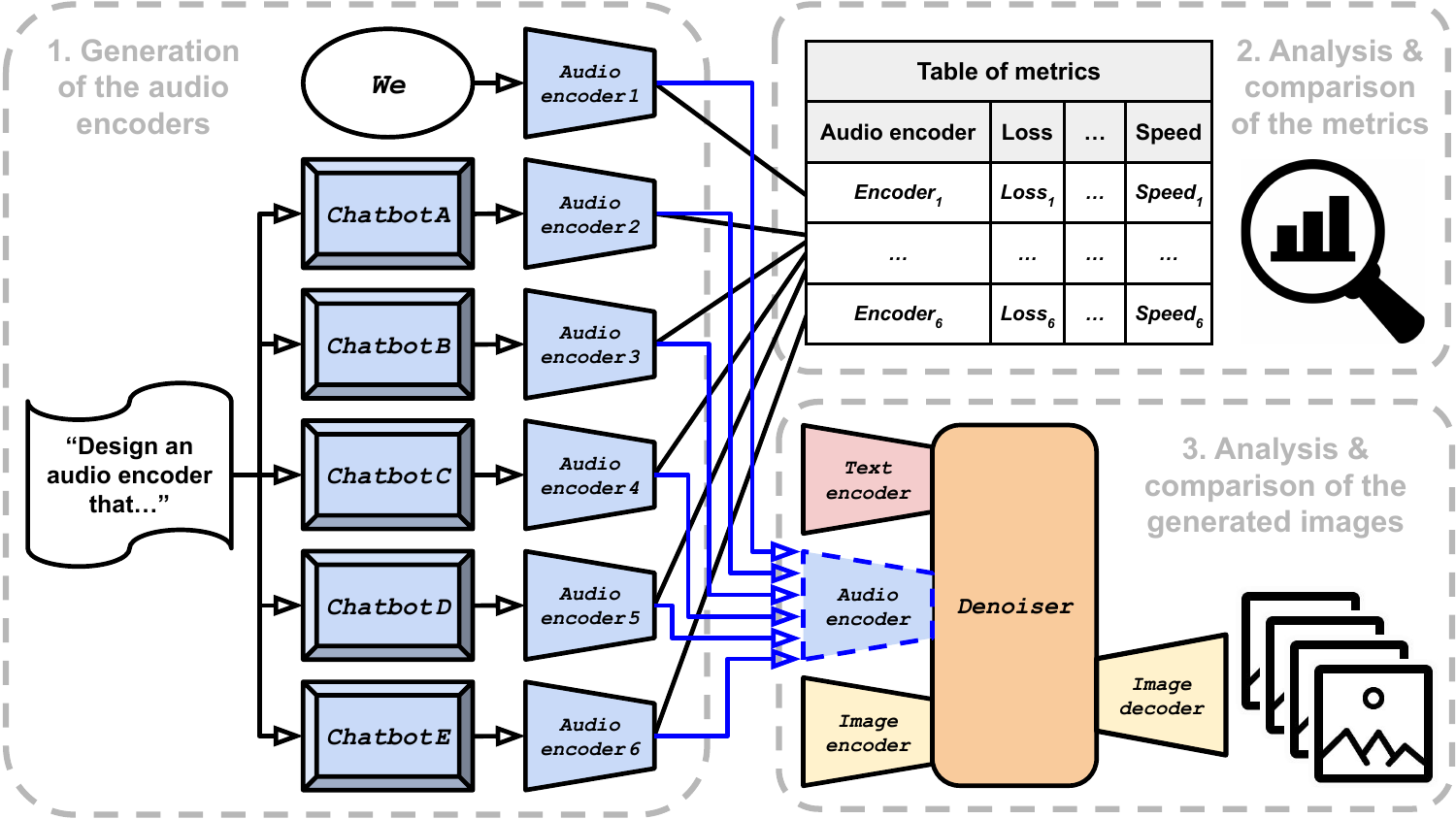}
\vspace*{-6mm}
\caption{Summary of our methodology. \textbf{1. Generation of the audio encoders:} This initial phase involves the design of all the audio encoders to be tested and compared (some chatbots may be unable to come up with a suitable architecture, and thus they would be ruled out of the following tests). \textbf{2. Analysis \& comparison of the metrics:} In this step, we measure and compare several metrics based on the encodings obtained from each audio encoder (we dive into more detail about them in Subsubsection \ref{subsubsec:Evaluating_the_Audio_Encoders}). \textbf{3. Analysis \& comparison of the generated images:} The final phase consists of generating multiple images, in various ways and repurposing Stable Diffusion 1.5 to do so, to analyze and compare them (once again, let us postpone the details until Subsubsection \ref{subsubsec:Evaluating_the_Audio_Encoders}).}
\label{fig:Summary_of_methodology}
\end{figure*}

For this research, we decided to test the coding capabilities of five different chatbots by asking them to design an audio encoder for Stable Diffusion 1.5 (a fairly known open-source text-to-image model \cite{Stable_Diffusion_v1-5_Model_Card}), and compare them on common audio-to-image (including texts and images as inputs in some generations), as well as on different metrics.
A summary of our methodology itself can be seen in Figure \ref{fig:Summary_of_methodology}, and the aforementioned models are: ChatGPT o3-mini \cite{OpenAI_o3-mini}, Claude 3.7 Sonnet \cite{Claude_3.7}, DeepSeek-R1 \cite{DeepSeek-R1}, Gemini 2.5 Pro Preview 03-25 \cite{Gemini_2.5}, and Grok 3 \cite{Grok_3_Beta}.
However, before diving into the details, we better contextualize our experiments by explaining how Stable Diffusion 1.5 works (in Subsection \ref{subsec:How_Stable_Diffusion_1.5_Works}), followed by the formal description of our experiments (in Subsection \ref{subsec:Our_Experiments}), and finally we speak a bit more about the chosen chatbots, together with the data and hardware used for training and testing the architectures (in Subsection \ref{subsec:About_our_Chosen_Chatbots_Data_and_Hardware}).

\subsection{How Stable Diffusion 1.5 Works}\label{subsec:How_Stable_Diffusion_1.5_Works}

\begin{figure*}[t]
% \vspace*{-6mm}
\centering
\includegraphics[page=2,width=16.4cm]{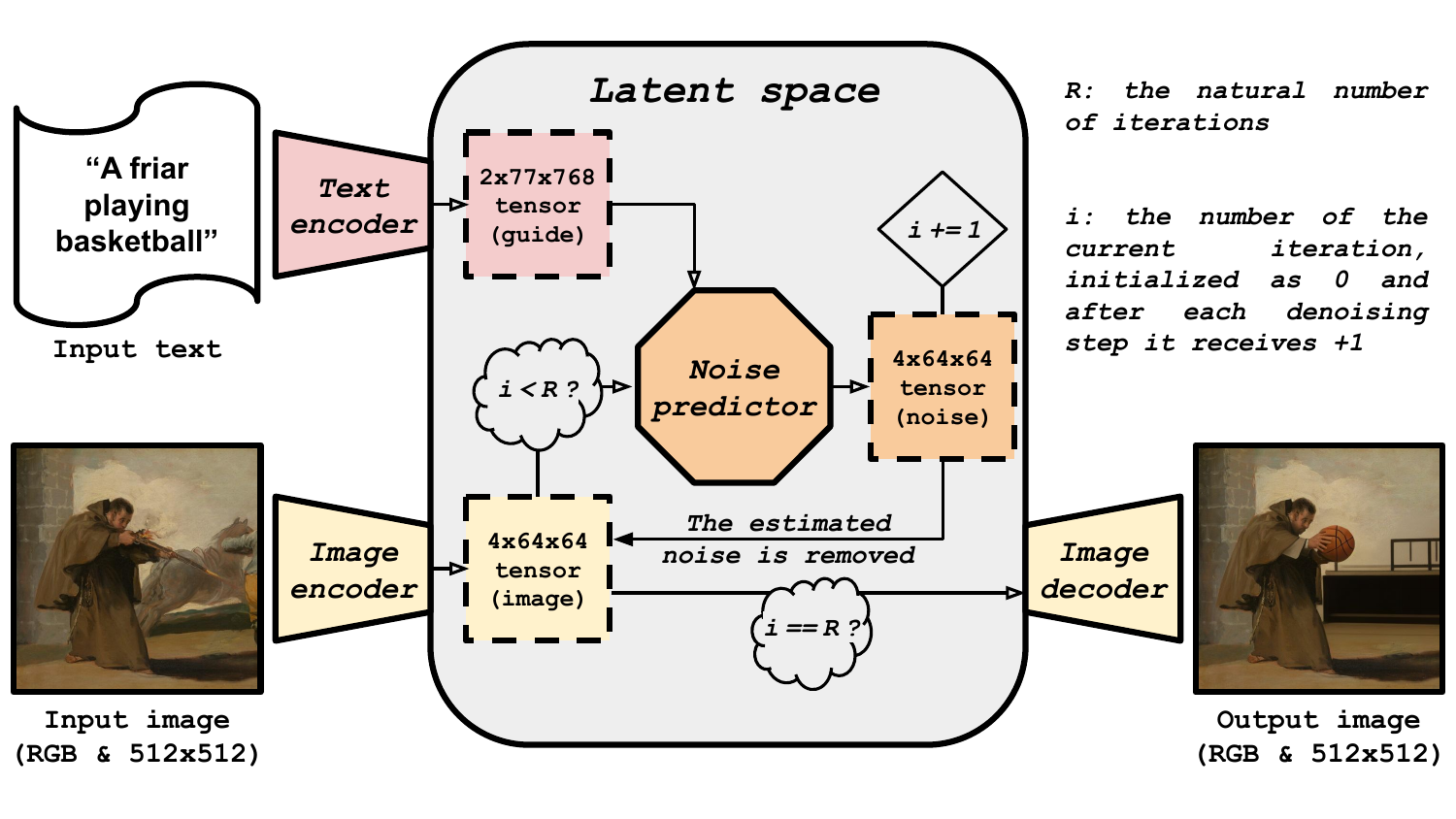}
\vspace*{-8mm}
\caption{Inner workings of Stable Diffusion 1.5. As we can see, both an input text and input image were considered (i.e. image-to-image conditioned by text). Nevertheless, it is still possible to skip the input image (i.e. text-to-image), replacing the initial image encoding by a tensor of random values, according to a normal distribution with $\mu = 0$ and $\sigma = 1$. For the shown example, we decided to use an image of $3\!\times\!512\!\times\!512$. However, other $3\!\times\!a\!\times\!b$ dimensions can be used without issue and the latent dimensions would be equal to $4\!\times\!\left\lfloor a/8 \right\rfloor\!\times\!\left\lfloor b/8 \right\rfloor$, but there would be some distortion in the decoding phase if $a\bmod{}8\neq{}0$ or $b\bmod{}8\neq{}0$. Regarding the text, there is a maximum of 75 tokens (any token after that one is discarded).}
\label{fig:SD_inner_working}
\end{figure*}

Let us see Figure \ref{fig:SD_inner_working} to explore the general inner workings of Stable Diffusion 1.5.
This model is composed of three submodels \cite{High-Resolution_Image_Synthesis, Stable_Diffusion_v1-5_Model_Card}, which we explain below.

First of all, we have the CLIP model \cite{Learning_From_Noisy_Correspondence}.
Specifically, it is CLIP ViT-L/14; from which a tokenizer is used to give each token an id and generate an attention mask to only consider tokens across the length of the original text, and a Transformer \cite{Attention_is_All_you_Need} to encode the tokenized text.
As a whole, we refer to it as the text encoder of Stable Diffusion 1.5 and it actually just produces $77\!\times\!768$-matrices.
The technicality of having a $2\!\times\!77\!\times\!768$-tensor per text comes from the need to consider an empty text for reference during the noise predictions, meaning that one $77\!\times\!768$-matrix is derived from our actual text (i.e. the conditional embedding) and the other $77\!\times\!768$-matrix is the product of inputting an empty text (i.e. the unconditional embedding).

Secondly, we have the variational autoencoder (VAE) \cite{Auto-Encoding_Variational_Bayes}.
It is also formed by two parts, which are the image encoder and the image decoder.
As our diagram illustrates, the first one compresses $3\!\times\!a\!\times\!b$-images into $4\!\times\!\left\lfloor a/8 \right\rfloor\!\times\!\left\lfloor b/8 \right\rfloor$-embeddings, while the latter performs the reverse process.
This is useful, as it allows to work with a smaller representation of the input and output images, with very low information loss (lowering the computational cost).

Thirdly, we have the last neural network component, which is the denoising U-Net \cite{U-Net}.
This U-Net is a sort of mixture of ResNet \cite{Deep_Residual_Learning_for_Image} and Transformer blocks, and, as one could expect, it correspond to what we call the noise predictor in our figure.
This final submodel has the role of estimating the corresponding noise in the current image embedding, working with both text encodings to produce an average noise prediction to subtract from the image embedding.
Here we control preponderance of our input text (usually also called prompt) with a \textit{guidance scale} parameter (the bigger the value, the closer the resulting image should depict our prompt).

It only remains to clarify that this noise reduction is repeated $R$ times, being $R \in \mathbb{N}$ of our choice; and in each iteration the resulting values are scaled by a scheduler, so the changes become less severe as the process is repeated (therefore, hopefully converging into a coherent image).

Complementary, we can point out that this same architecture is repeated in the other versions of Stable Diffusion 1.X, changing just how much training was carried out on each one, added to some other minor tweaks in the training settings \cite{Stable_Diffusion_v1-5_Model_Card, High-Resolution_Image_Synthesis}.
More meaningful changes can be seen in derived works, such as Stable Diffusion XL \cite{SDXL} or 3 \cite{Stable_diffusion_3}, although that is not of our concern here.

\subsection{Our Experiments}\label{subsec:Our_Experiments}

\begin{figure*}[t]
% \vspace*{-6mm}
\centering
\includegraphics[page=3,width=16.4cm]{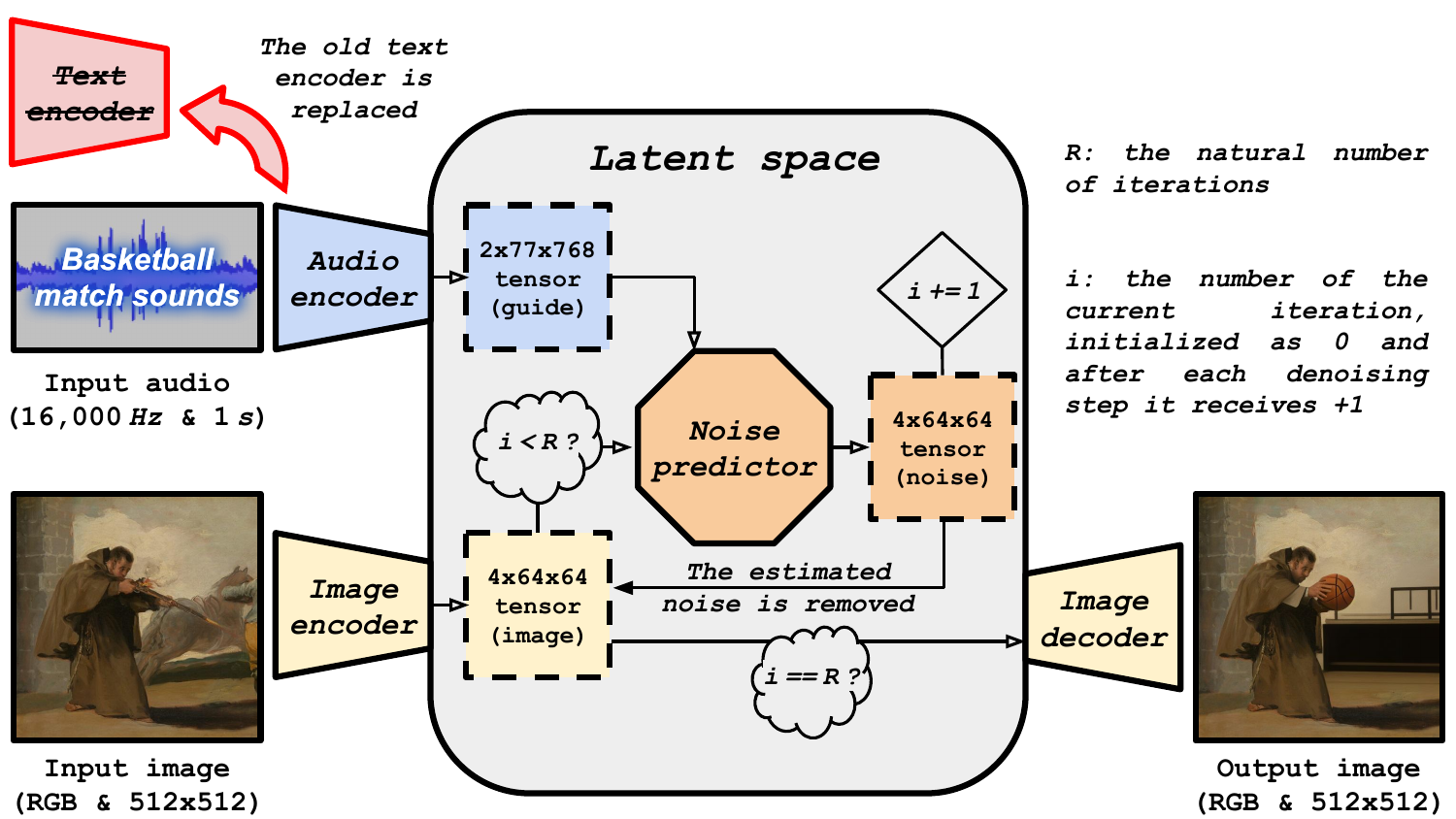}
% \vspace*{-6mm}
\caption{The main concept behind our experiments. The only difference with Figure \ref{fig:SD_inner_working} is that we are replacing the original text encoder with an original audio encoder. As seen later, this exchange was performed with different audio encoders, and the original text encoder can also be used in addition to any of the audio ones.}
\label{fig:Encoder_replacement}
\end{figure*}

The main concept of our experiments can be seen in Figure \ref{fig:Encoder_replacement}.
As shown in \cite{AudioCLIP, Wav2CLIP, TimbreCLIP}, it is possible to imitate to some capacity the embeddings performed by the CLIP text encoder with a new audio one.
The exact match between the two (i.e. audio and text) is virtually impossible, as there are always some small nuances that cannot be fully replicated.
Nevertheless, it has already been demonstrated that a high level of likeness is possible and there is hope for a lot more improvement in the area.

For the sake of order, this subsection is divided in two parts: the first one describes how we obtain the audio encoders from the chatbots (Subsubsection \ref{subsubsec:Obtaining_the_Audio_Encoders}), and the second dives into the tests we performed to study their capabilities and quality (Subsubsection \ref{subsubsec:Evaluating_the_Audio_Encoders}).

\subsubsection{Obtaining the Audio Encoders}\label{subsubsec:Obtaining_the_Audio_Encoders}

\begin{figure*}[t]
% \vspace*{-6mm}
\centering
\includegraphics[page=4,width=16.4cm]{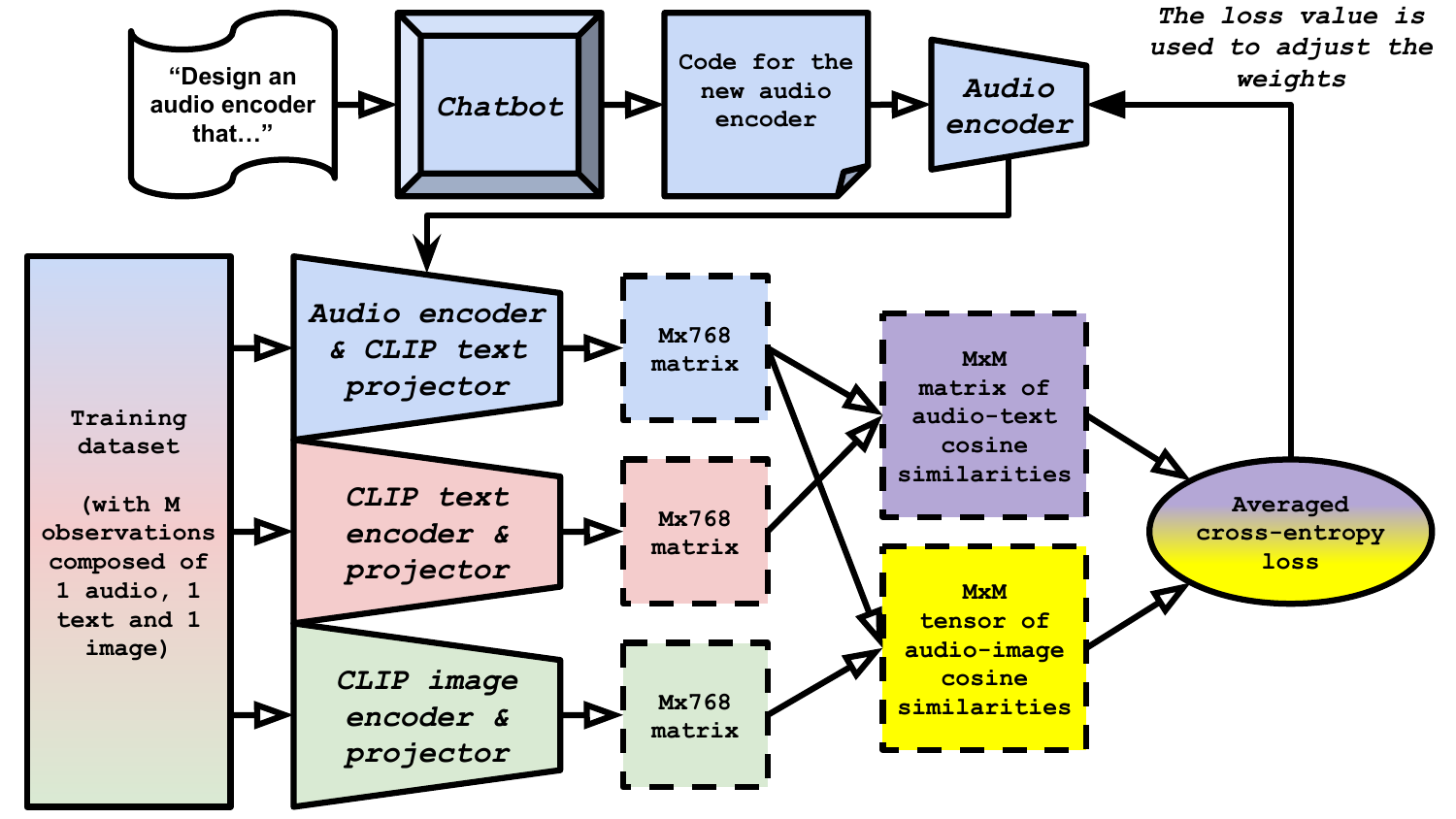}
% \vspace*{-6mm}
\caption{Workflow for a chatbot to create an audio encoder. This is repeated on each chatbot under evaluation.}
\label{fig:Encoder_training}
\end{figure*}

Due to the previous and inspired by \cite{AudioCLIP}, we decided to design a workflow, visible in Figure \ref{fig:Encoder_training}, for chatbots to create audio encoders.
In it, we give each chatbot some shared set of instructions, and if they are able to produce an audio encoder coherent with what was asked (mainly complying with the input and output shapes), said encoder is trained to mimic the embeddings of the image encoder and text encoder of CLIP.
Please notice that the CLIP image encoder employed here is different from the VAE encoder used for generation.
Every accepted audio encoder undergoes a training with 32 epochs, mini-batches of $\mbox{1,151}$ observations and a learning rate of $0.001$.
The dataset used for training contains observations composed by an audio (with length of 1 \medidas{}, sample rate of $\mbox{16,000}$ \medidahz{}, 16 \medidabits{} of depth and monophonic channel), an image (with dimensions of 512x512 \& RGB), and a text (in english and with a maximum length of 16 words), the three of them associated by context (more details in Subsection \ref{subsec:About_our_Chosen_Chatbots_Data_and_Hardware}).

The following is the prompt given to each chatbot, which we decided to re-enter at a maximum of three times and until a valid encoder, that complies with the requirements we asked for in the prompt, is generated (otherwise, it was ruled out from the following tests):

\begin{framed}
    Hey, mate, I have an interesting Python task for you.
    
    I want you to replace the text encoder from Stable Diffusion 1.5 (which is basically the one from CLIP ViT Large Patch14) for one that works with audio instead. Particularly, the input audios are 1 \medidas{}, with sample rate of $\mbox{16,000}$ \medidahz{}, 16 \medidabits{} of depth and monophonic channel. The output of each encoding should be a $77\!\times\!768$ matrix. Keep also in mind that your model must be created with PyTorch.
    
    I already have the dataset ready (its samples are trios of images, texts and audios) and the training figured out. I will be using a symmetric cross entropy loss over their cosine similarity scores, comparing to both the text and image encodings, which means that I will be working with the outputs of the new encoder as logits (just like CLIP did for its training, and with a learning rate of 0.001 and 32 epochs). In summary, you just need to effectively take $\mbox{16,000}$ dimensional vectors and convert them to $77\!\times\!768$ matrices, in a way that maximizes the chance of obtaining similar encodings between the original text and image encoders with the new audio encoder (consider the best current techniques for this).
    
    As a final point, I have written a bit of the code for you, so just fill the respective spaces I have reserved for you and feel free to add as many lines as you deem necessary, but do not add code anywhere else (for example, the inputs of the methods are untouchable), except for importing more libraries if you need them. You only have one chance, so take your time and think thoroughly. Good luck.

    \small
    \begin{lstlisting}
import torch
import torch.nn as nn
class NewAudioEncoder(nn.Module):
  def \_\_init\_\_(self):
    super(NewAudioEncoder, self).\_\_init\_\_() 
    ### 
    ### (YOUR CODE GOES HERE) 
    ###
  def forward(self, x):
    x = x.view((-1,1,16000))/32767
    ### 
    ### (YOUR CODE GOES HERE) 
    ###
    assert (x.shape[1:] == (77, 768)), f"Expected shape (-1, 77, 
    768), but got {x.shape}."
    return x
    \end{lstlisting}
\end{framed}

Now, let us formalize the loss function used.
Given two $M\!\times\!768$-matrices (each one also interpretable as $M$ ordered vectors of projections with length $768$), $A$ and $B$, we first compute their matrix product $P := AB^{\top}$.
Keep in mind that the dot product of two vectors $\vec{a}$ and $\vec{b}$, with a shared origin and an angle $\theta$ formed between them, can be calculated in the following way:
\begin{equation*}
    \vec{a} \cdot \vec{b} = |\vec{a}||\vec{b}| cos(\theta)
\end{equation*}

Given that every vector in $A$ and $B$ is normalized from their projections, we are left with:
\begin{equation*}
    \vec{a} \cdot \vec{b} = cos(\theta)
\end{equation*}

This is relevant to us, as the cosine of the angle formed by two vectors can be seen as a measurement of similarity between them.
To name the main cases, $\vec{a}$ and $\vec{b}$ are aligned if $cos(\theta) \approx 1$, $\vec{a}$ and $\vec{b}$ are orthogonal if $cos(\theta) \approx 0$, and $\vec{a}$ and $\vec{b}$ are opposites if $cos(\theta) \approx -1$ (hence, $cos(\theta)$ is also know as the cosine similarity between $\vec{a}$ and $\vec{b}$).
From the previous, we can derive that our matrix $P$ is a matrix of cosine similarities for all the encodings of our $M$ observations.

Having understood the above and making the assumption that we only want to ensure that embeddings, from $A$ and $B$, that come from the same observations are close to each other, while distant from others, we can resort to the cross-entropy of $P$ and an identity matrix $I_{M}$ to minimize our loss.
We define the cross-entropy function in the following fashion:
\begin{equation}
    cross\textnormal{-}entropy(A, B) := -\frac{1}{M}\sum_{j=1}^{M}\sum_{k=1}^{N} ln \left ( \frac{e^{a_{j,k}}}{\sum_{l=1}^{M} e^{a_{j,l}}} \right ) b_{j,k},
    \label{eq:cross-entropy}
\end{equation}

\noindent where $N$ is the number of columns in our matrices $A$ and $B$, and $a_{j, k}$ and $b_{j, k}$ represent the elements in row $j$ and column $k$ of $A$ and $B$, respectively.
However, we must also take into account the distance between rows, so we also calculate the cross-entropy of $P^{\top}$ and an identity matrix $I_{M}$.
This means that the total cross-entropy of cosine similarities (\textit{TCEOCS}) between $A$ and $B$ is $cross\textnormal{-}entropy(P, I_{M}) + cross\textnormal{-}entropy(P^{\top}, I_{M})$.
It is worth mentioning that this loss function is quite close to the original one from CLIP \cite{Learning_Transferable_Visual_Models_From_NL}, although some factors are disregarded in our case.

At this point, we need to remember that we are actually working with three $M\!\times\!768$-matrices (for audio, image and text, separately), so let us add a $C$ matrix to the equation.
Consider $Q := AC^{\top}$, which translates to the \textit{TCEOCS} between $A$ and $C$ being $cross\textnormal{-}entropy(Q, I_{M}) + cross\textnormal{-}entropy(Q^{\top}, I_{M})$.

Lastly, we average these values to obtain the loss function seen below:
\begin{equation}
    \begin{matrix}
        \scriptsize
        loss(A, B, C) := \frac{cross\textnormal{-}entropy(P, I_{M}) + cross\textnormal{-}entropy(P^{\top}, I_{M}) + cross\textnormal{-}entropy(Q, I_{M}) + cross\textnormal{-}entropy(Q^{\top}, I_{M})}{6}
    \end{matrix},
    \label{eq:loss}
\end{equation}

\noindent being $\frac{1}{6}$ a scale factor replicated from \cite{AudioCLIP}.

As an additional reference point, we estimated interesting to design our own audio encoder (avoiding its refinement through trial and error, in order to keep the conditions fair) and compare it to what the chatbots come up with.
And thus we did so.

\subsubsection{Evaluating the Audio Encoders}\label{subsubsec:Evaluating_the_Audio_Encoders}

For our experiments, we split our data into three subsets: training, validation and test.
Here we are just concerned with the latter two, as the validation subset was used to measure the evolution of our encoders during training, and the test subset was destined to more carefully explore their results on more metrics after training.
It is worth mentioning that from our $\mbox{2,240,231}$ observations, we reserved $\mbox{3,774}$ for validation and $\mbox{23,524}$ for test.
This relatively low quantity for the validation subset is justified by the fact that we mainly wanted to maximize the performance of the encoders, enabling them to learn from the widest collection of observations possible, but we also needed a representative $\sim 1\%$ of random samples for the final test.

In our two main validations (the one before training and the one when it was completed), we did not merely registered the \textit{loss}, as well as the \textit{TCEOCS}s, but also three additional metrics based on the resulting projections.
To understand the first one, let us swiftly explain the mean squared error (\textit{MSE}).
Keep in mind that our motivation lies in the fact that the \textit{MSE} is a popular metric to evaluate how well the prediction of a model aligns with real outcomes ($\textit{MSE} \in [0,\infty)$, with lower values meaning a closer match).
For a variable to be predicted, $x$, with $M$ samples, this metric is commonly defined by the next formula:
\begin{equation}
    \textit{MSE}_{x} := \frac{1}{M}\sum_{j=1}^{M}(\hat{x}_j - x_j)^2,
    \label{eq:MSE1}
\end{equation}

\noindent where $\hat{x}_j$ is the prediction made for sample $j$, and $x_j$ is the corresponding ground truth.
Nonetheless, we must consider that our output is not a single value per observation, but the vector with length $768$ that we mentioned previously (see Subsubsection \ref{subsubsec:Obtaining_the_Audio_Encoders}).
This presents a small inconvenient, as analyzing $768$ \textit{MSE}s individually is rather impractical.
To amend this, we employed a slightly different version of the \textit{MSE} than (\ref{eq:MSE1}), intended for a vector of variables $\vec{x} = \langle x_1, \ldots, x_{N}\rangle$, and it is the following:
\begin{equation}
    \textit{MSE}_{\vec{x}} := \frac{1}{MN}\sum_{j=1}^{M}\sum_{k=1}^{N}(\hat{x}_{j,k} - x_{j,k})^2,
    \label{eq:MSEN}
\end{equation}

\noindent where $\hat{x}_{j,k}$ is the prediction of variable $x_k$ made for sample $j$, and $x_{j,k}$ is the corresponding ground truth.
Obviously, in this case $N = 768$.

Please also note that $\textit{MSE}_{\vec{x}} = \frac{1}{N}\sum_{k=1}^{N}\textit{MSE}_{x_k}$, so for convenience and generalization purposes, we will simply refer to this metric as $\mu(\textit{MSE})$.

Now, tackling the aforementioned second metric, it is imperative to talk about the coefficient of determination ($R^2$).
This metric has multiple definitions, depending on the field of math where it appears.
Regardless, for our situation we can interpret it as a metric that shows how much better the predictions of a model align with the ground truth values, when compared with just consistently returning the average of said values as a prediction instead ($R^2 \in (-\infty, 1]$, with higher values meaning a closer match).
For a variable to be predicted, $x$, with $M$ samples, $R^2$ is defined in this way:
\begin{equation}
    R_{x}^2 := 1 - \frac{\sum_{j=1}^{M}(x_j - \hat{x}_j)^2}{\sum_{k=1}^{M}(x_k - \mathbb{E}(x))^2},
    \label{eq:R2}
\end{equation}

\noindent where $\hat{x_j}$ is the prediction made by the model for sample $j$, and $\mathbb{E}(x)$ is the expected value of the variable $x$, which is equal to $\frac{1}{M}\sum_{l=1}^{M}x_l$.
Once again, we are faced with the impracticality of properly checking $768$ values individually.
Analogous to (\ref{eq:MSEN}), we chose to work with the average $R^2$ across all variables (let us call it $\mu(R^2)$), but also with their standard deviation (let us call it $\sigma(R^2)$).

So, in summary, in each validation we calculate the \textit{loss}, together with the \textit{TCEOCS}s, the $\mu(\textit{MSE})$s, the $\mu(R^2)$s and the $\sigma(R^2)$s of the respective audio encoder compared with the projections of the CLIP text encoder and the projections of the CLIP image encoder, separately.

Moving on to speak of the test phase, we computed the same metrics of the validations on the projections obtained based on the test subset, but also some additional ones.

One of these complementary metrics is the average time for the audio encoders to generate their outputs ($\tau$).
Specifically, due to hardware constraints, we passed our $\mbox{23,524}$ samples in batches of $\mbox{1,000}$, with a last one of $524$.
Besides, to ensure a good representativeness of their speeds, we repeated this measurement 100 times and averaged the results.

And, once again, we measured the $\mu(\textit{MSE})$, the $\mu(R^2)$ and the $\sigma(R^2)$, with the difference that this time we worked with the raw audio embeddings and compared them with just the text embeddings (without additional projections, as they share the same tensor dimensions).
As we previously stated, both types of encoders output matrices of shape $77\!\times\!768$, so we can recycle the concept of averaging and getting standard deviations from the metrics of the different variables (the only difference being that here we consider $77*768 = \mbox{59,136}$ variables, instead of just $768$).

So, to summarize the last paragraphs, in the test we calculate \textit{loss}, the average time to encode a batch of observations $\tau$, the \textit{TCEOCS}s, and the $\mu(\textit{MSE})$s, the $\mu(R^2)$s and the $\sigma(R^2)$s of the audio encoder compared with the projections of the CLIP text encoder, the projections of the CLIP image encoder and the raw outputs of the CLIP text encoder, separately.

\begin{figure*}[t]
% \vspace*{-6mm}
\centering
\includegraphics[page=6,width=16.4cm]{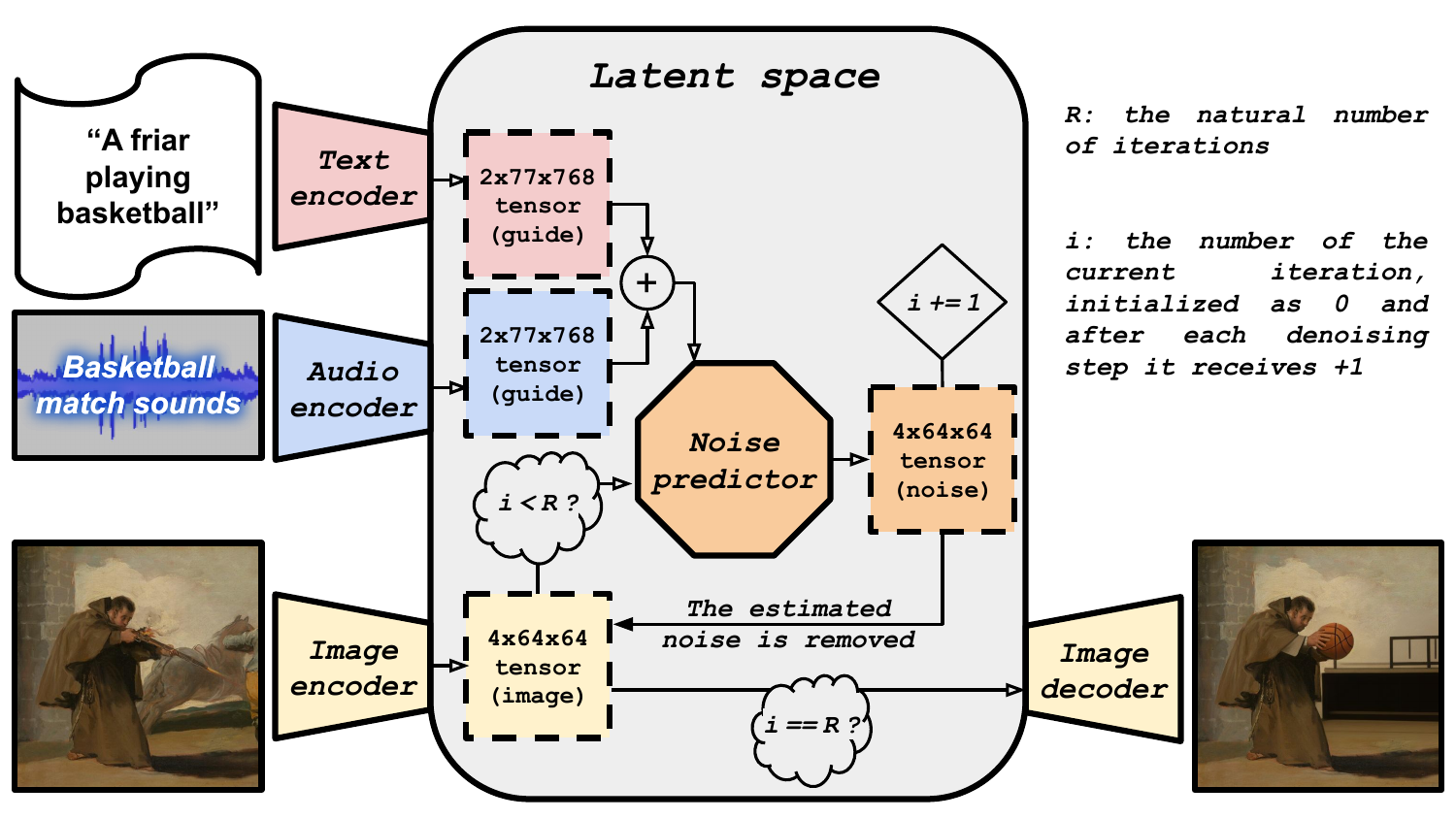}
% \vspace*{-6mm}
\caption{Example of generation based on multiple encoders for guidance. Consider that more than two encoder could be used and even with the same type of input data repeated (e.g. we could assemble a combination of the CLIP text encoder and three new audio encoders). The \encirculado{+} gate represents some kind of weighted sum of the respective embeddings. For our experiments, consider that \encirculado{+} outputs a matrix with the same input shape ($2\!\times\!77\!\times\!768$) with the corresponding arithmetic means of the inputs.}
\label{fig:Multiple_encoders}
\end{figure*}

Ultimately, the only aspect that remains for us to analyze are their architectures and, more importantly, the quality of the images they generate.
Note that the generated images can be synthesized considering multiple encoders.
For reference, take a look at Figure \ref{fig:Multiple_encoders}, where we evidence how these multiple encoder can collaborate effectively.
Essentially, the workflow is the same, but we combine multiple guidance embeddings into one (under the assumption that their latent spaces are similar enough).
As the audio encoders are originally intended to replace the text encoder, both share the same output dimensions and, thus, these can be directly averaged.
Namely, we decided to employ arithmetic means when averaging multiple of these guidance embeddings.

\begin{figure*}[t]
% \vspace*{-6mm}
\centering
\includegraphics[page=9,width=16.4cm]{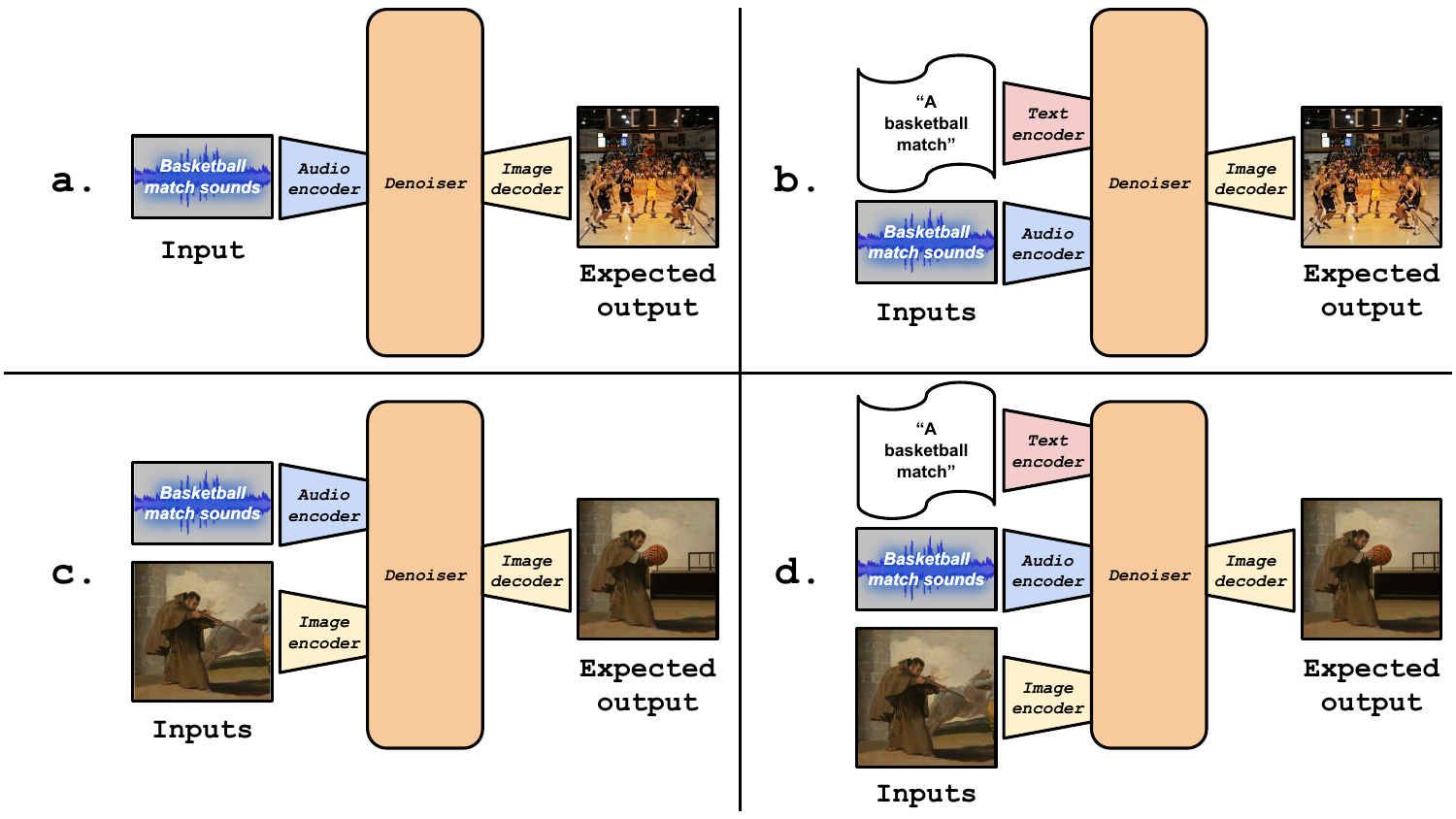}
% \vspace*{-6mm}
\caption{Methods of generation we intend to use for each audio encoder. First we will generate images using just audio fragments (a.), then we will accompany the input audios with brief textual descriptions of the respective audios (b.), followed by a generation considering just the input audios and unrelated reference images (c.), and, finally, we will use the three aforementioned types of data together (d.).}
\label{fig:All_generations}
\end{figure*}

More precisely, we opted to do four types of generations per audio encoder (see Figure \ref{fig:All_generations}).
These are with just audio as input, with audio and text as input, with audio and image as input, and with audio, image and text as input.
For this purpose, we collected 10 different images and 10 audios of various situations, and manually wrote brief descriptions for each audio.
As each pair of audio and text convey similar information, we intended to assess if the encoders were able to reinforce each other constructively (meaning that they share a similar latent space) or not.
We repeated our generations with each one of the 110 possible combinations of inputs ($10\textnormal{ images} * 10\textnormal{ audios} + 10\textnormal{ audios without images}$).
However, to have some sort of benchmarks, we also replicated the experiments without audio encodings, considering just the text encodings as guidance embeddings (serving as good case examples), and using only random values from a normal distribution $\mathcal{N}(\mu=0,\sigma^{2}=1)$ instead (serving as bad case examples).\footnote{We found that $\mathcal{N}(\mu=0,\sigma^{2}=1)$ somewhat resembles the distribution of real embeddings obtained with the text encoder, but random embeddings based on said distribution rarely generate coherent images.}
To reach fairer conclusions, we also duplicated each generation once (resulting in two different output images per case), and, to explore more possibilities, we also replicated the experiments averaging the encodings of all encoders obtained from the chatbots and even with the one designed by us.

It is pertinent to remark that for generations that did not involve image-to-image we used a guidance scale of $7.5$ and $100$ denoising steps.
For the others, we employed a guidance scale of $10$, a strength of $0.7$ for the input image and $200$ denoising steps.

\subsection{About our Chosen Chatbots, Data and Hardware}\label{subsec:About_our_Chosen_Chatbots_Data_and_Hardware}

As we previously announced, we decided to test five models.\footnote{We here declare that we are not paid by any of the companies behind these chatbots and we do not posses any particular affinity with any of them, so our judgment is unbiased and purely scientific.}
We list them below with some basic information:

\begin{itemize}
    \item ChatGPT o3-mini  \cite{OpenAI_o3-mini}: It was created by OpenAI and contains a reasoning option. We worked with it on April 7, 2025.
    
    \item Claude 3.7 Sonnet \cite{Claude_3.7}: It was created by Anthropic and contains hybrid reasoning. We worked with it on April 12, 2025.
    
    \item DeepSeek-R1 \cite{DeepSeek-R1}: It was created by DeepSeek and contains reasoning by default. We worked with it on April 12, 2025.
    
    \item Gemini 2.5 Pro Preview 03-25 \cite{Gemini_2.5}: It was created by Google DeepMind and contains reasoning by default. We worked with it on April 9, 2025.
    
    \item Grok 3 \cite{Grok_3_Beta}: It was created by xAI and contains a reasoning option. We worked with it on April 6, 2025.
\end{itemize}

We activated the reasoning option whenever possible, but, to avoid copies of other works, we kept disabled the options to search on the internet.
Keep in mind that we included the dates in which we asked the chatbots to generate the audio encoders because soon they could be subject to updates that drastically change their capabilities on this task (as has been the case for others in the past).

We also required to push the boundaries on what has been accomplished in other similar research, in order to avoid any of the chatbots just repeating the architectures they memorized during their trainings.
Due to this, we designed our own dataset, with $\mbox{2,240,231}$ audio-image-text observations from videos (for more details on the process and the dataset itself, please consult \cite{Obtaining_acoustic_visual_and_textual_data}).
Despite a common length for audios in these cases is around 5 \medidas{} \cite{AudioCLIP, Wav2CLIP, MusicLM} or more \cite{whisper, Any-to-Any_Generation}, we chose 1 \medidas{} instead to challenge the chatbots, while also facilitating the convergence during training.
Likewise, we omitted usual preprocessing steps like generating a spectrogram \cite{TimbreCLIP}, leaving such choices to the chatbots themselves.
It is crucial to also point out that we discovered a small amount of noise in our data, particularly in the generated texts \cite{Obtaining_acoustic_visual_and_textual_data}.
Because of this, we expected that the audio encoders would align more easily with the CLIP image encoder, than with the CLIP text encoder.

For the audios and images of the final generation tests, we carefully searched for material without copyright issues and adapted them to coincide with the properties that the encoders required.
Namely, the images are from \cite{Pexels, Rawpixel, Picryl} and the audios are from \cite{Pixabay, Freesound}.

Lastly, regarding our hardware, we had access to a NVIDIA H100 NVL (with 94 \medidagigabytes{}), which allowed us to conduct everything about the training, validation and test that we described in Subsection \ref{subsec:Our_Experiments}.
\section{Results and Analysis}\label{sec:Results_and_Analysis}

Only four of the five chatbots we selected were able to come up with an encoder that meet our conditions.
These were ChatGPT o3-mini, DeepSeek-R1, Gemini 2.5 Pro Preview 03-25 and Grok 3.
From this point forward, for the sake of brevity, we refer to these chatbot as ChatGPT, DeepSeek, Gemini and Grok, respectively. Similarly, when we use the Ours label, we will be referring to the encoder created by us.

Unfortunately, the resulting architectures of the encoders ended up being too complex to be reasonably graphed in this article in any useful way.
Due to this, we have resorted to create the Table \ref{tab:Generated_encoders}, which summarizes the layers present in each model, the number of trainable parameters, the number of branchings (i.e. the number of times an output feeds into multiple layers in the model), and the number of retries to generate a suitable encoder.
At first glance, we can already notice a few differences between the encoders (mostly regarding the sizes they opted for).
Nevertheless, Grok and Deepseek seem to have designed surprisingly similar architectures, and it was unexpected that most chatbots were able to come up with an appropriate encoder on their first try (only Gemini needing one retry).
In any case, Transformer encoders with GELU activations appear as one of the most common practices \cite{Attention_is_All_you_Need, Learning_Transferable_Visual_Models_From_NL}, idea that we actually overlooked in our own design.
Based solely on the number of trainable parameters and branchings, we can form the following complexity order (the first being the model that we could say is more complex, and, therefore, with more flexibility and risk of overfitting): 1. DeepSeek, 2. Grok, 3. Gemini, 4. ChatGPT and 5. Ours.\footnote{For more details, feel free to see their full architecture in this \href{https://github.com/Jorvan758/A-SD-Alt/blob/main/LLM\%20AVT\%20Stable\%20Diffusion\%20Demo.ipynb}{\blue{\underline{Jupyter Notebook demo}}}, where you can also try the whole models and generate images like we did.}

\begin{table}[t]
\vspace*{-2mm}
\begin{center}
    \renewcommand{\arraystretch}{1.05}
    \begin{tabular}{|wc{11.0em}?wc{4.3em}|wc{4.3em}|wc{4.3em}|wc{4.3em}|wc{4.3em}|}
        \cline{2-6}
        \multicolumn{1}{c|}{} & \textbf{Ours} & \textbf{ChatGPT} & \textbf{DeepSeek} & \textbf{Gemini} & \textbf{Grok}
        \\
        \cline{1-1}
        \Cline{2pt}{2-6}
        \multicolumn{1}{|c?}{\textit{AdaptiveAvgPool1d}}
        & 1 & 0 & 0 & 0 & 0 \\
        \hline
        \textit{AdaptiveMaxPool1d} & 2 & 0 & 0 & 0 & 0 \\
        \hline
        \textit{AmplitudeToDB} & 0 & 0 & 0 & 1 & 0 \\
        \hline
        \textit{Conv1d} & 14 & 1 & 1 & 0 & 0 \\
        \hline
        \textit{Dropout} & 1 & 13 & 36 & 24 & 36 \\
        \hline
        \textit{GELU} & 0 & 1 & 1 & 1 & 1 \\
        \hline
        \textit{LayerNorm} & 7 & 8 & 24 & 17 & 24 \\
        \hline
        \textit{Linear} & 3 & 8 & 24 & 17 & 25 \\
        \hline
        \textit{MelScale} & 0 & 0 & 0 & 1 & 0 \\
        \hline
        \textit{MelSpectrogram} & 0 & 0 & 0 & 1 & 0 \\
        \hline
        \textit{ModuleList} & 0 & 1 & 1 & 1 & 1 \\
        \hline
        \textit{MultiheadAttention} & 0 & 4 & 12 & 8 & 12 \\
        \hline
        \textit{NonDynamicQuantiLinear} & 0 & 4 & 12 & 8 & 12 \\
        \hline
        \textit{SiLU} & 1 & 0 & 0 & 0 & 0 \\
        \hline
        \textit{Spectrogram} & 0 & 0 & 0 & 1 & 0 \\
        \hline
        \textit{TransformerEncoder} & 0 & 1 & 1 & 1 & 1 \\
        \hline
        \textit{TransformerEncoderLayer} & 0 & 4 & 12 & 8 & 12 \\
        \hline
        Trainable parameters & $\mbox{2,043,692}$ & $\mbox{22,275,584}$ & $\mbox{85,274,112}$ & $\mbox{56,825,856}$ & $\mbox{85,213,440}$
        \\
        \hline
        Branchings & 8 & 8 & 24 & 17 & 24
        \\
        \hline
        Retries & 0 & 0 & 0 & 1 & 0
        \\
        \hline
    \end{tabular}
\end{center}
\vspace*{-6mm}
\caption{Relevant information from the designed audio encoders. The italic texts denote the names of layers from the PyTorch and Torchaudio libraries, and we also included the names and quantities of the nested layers. Trainable parameters considers all weights and biases that are adjustable during training, we call branching to the cases where the output of a layer feeds into multiple layers, and the retries are the account of times we had to re-enter the generation prompt so the chatbot can try to come up an acceptable audio encoder.}
\label{tab:Generated_encoders}
\end{table}

Moving on to the training, we have prepared Table \ref{tab:Validations} to display the validation metrics registered before and after training.
From these values, we can first of all note how similar is the performance between all the encoders in general.
The DeepSeek encoder seems to have initialized with a particularly favorable configuration, given the fact that it had the best initial performance in all metrics, with the exception of the $\sigma(R^2)_{i}$.
However, that advantange seems to quickly disappear, as the ChatGPT encoder takes the best score on half of the final text-related metrics, while ours outperformed everyone in the final image-related ones and even in the final loss values.
As we had foreseen, most encoders appear to more easily align with the CLIP image encoder, but the DeepSeek and Grok encoders challenged our expectations regarding this, with the DeepSeek one even straying away of its original \textit{TCEOCS} with the CLIP image encoder.
The reason to this phenomenon is unclear, but we suspect that this is likely a sign of a lack of enough training time, specially when considering that these two are the biggest models.
Nevertheless, let us remind ourselves that these values are only intended to partake in a short exploration of the changes during the training, and we should not jump to further conclusions on the real performance of the encoders based solely on the small number of observations we destined to these validations.

\begin{table}[t!]
\vspace*{-2mm}
\begin{center}
    \small
    \renewcommand{\arraystretch}{1.05}
    \begin{tabular}{|wc{5.5em}|wc{4.5em}?wc{4.5em}|wc{4.5em}|wc{4.5em}|wc{4.5em}|wc{4.5em}|}
        \cline{3-7}
        \multicolumn{1}{c}{} & \multicolumn{1}{c|}{} & \textbf{Ours} & \textbf{ChatGPT} & \textbf{DeepSeek} & \textbf{Gemini} & \textbf{Grok}
        \\
        \cline{1-2}
        \Cline{2pt}{3-7}
        \multirow{2}{*}{\textit{loss}} & Before & 5.49114 & 5.49157 & \textbf{5.49101} & 5.49136 & 5.49106 \\
        \cline{2-7}
         & After & \textbf{5.45905} & 5.48628 & 5.49103 & 5.47614 & 5.49097 \\
        \Cline{2pt}{1-7}
        \multirow{2}{*}{$\textit{TCEOCS}_{t}$} & Before & 16.47362 & 16.47350 & \textbf{16.47290} & 16.47338 & 16.47296 \\
        \cline{2-7}
         & After & 16.47289 & 16.47288 & \textbf{16.47286} & 16.47287 & \textbf{16.47286} \\
        \Cline{2pt}{1-7}
        \multirow{2}{*}{$\mu(\textit{MSE})_{t}$} & Before & 0.00269 & 0.00276 & \textbf{0.00231} & 0.00250 & 0.00282 \\
        \cline{2-7}
         & After & 0.00261 & \textbf{0.00260} & 0.00266 & \textbf{0.00260} & 0.00359 \\
        \Cline{2pt}{1-7}
        \multirow{2}{*}{$\mu(R^2)_{t}$} & Before & -1.01E10 & -1.04E10 & \textbf{-8.71E09} & -9.42E09 & -1.06E10 \\
        \cline{2-7}
         & After & -9.84E09 & \textbf{-9.79E09} & -1.00E10 & -9.80E09 & -1.35E10 \\
        \Cline{2pt}{1-7}
        \multirow{2}{*}{$\sigma(R^2)_{t}$} & Before & 3.57E10 & 4.64E10 & \textbf{1.99E10} & 2.68E10 & 4.39E10 \\
        \cline{2-7}
         & After & 3.32E10 & 3.37E10 & \textbf{2.34E10} & 3.56E10 & 1.23E11 \\
        \Cline{2pt}{1-7}
        \multirow{2}{*}{$\textit{TCEOCS}_{i}$} & Before & 16.47324 & 16.47595 & \textbf{16.47315} & 16.47479 & 16.47340 \\
        \cline{2-7}
         & After & \textbf{16.28143} & 16.44480 & 16.47334 & 16.38396 & 16.47298 \\
        \Cline{2pt}{1-7}
        \multirow{2}{*}{$\mu(\textit{MSE})_{i}$} & Before & 0.00256 & 0.00267 & \textbf{0.00254} & 0.00256 & 0.00255 \\
        \cline{2-7}
         & After & \textbf{0.00232} & 0.00252 & 0.00257 & 0.00248 & 0.00275 \\
        \Cline{2pt}{1-7}
        \multirow{2}{*}{$\mu(R^2)_{i}$} & Before & -3.31930 & -3.69429 & \textbf{-3.07476} & -3.20882 & -3.44299 \\
        \cline{2-7}
         & After & \textbf{-1.76477} & -2.10325 & -3.10393 & -1.92407 & -7.35781 \\
        \Cline{2pt}{1-7}
        \multirow{2}{*}{$\sigma(R^2)_{i}$} & Before & 6.70230 & 13.19494 & 5.68107 & \textbf{5.04533} & 10.68663 \\
        \cline{2-7}
         & After & \textbf{4.27730} & 4.30765 & 5.23513 & 4.69817 & 100.88818 \\
        \hline
    \end{tabular}
\end{center}
\vspace*{-6mm}
\caption{Validation metrics before and after training. Subindex $t$ denotes values measured between the projections of the corresponding audio encoder and the ones from the CLIP text encoder, while subindex $i$ is for the values that use the projections of the CLIP image encoder instead of the latter. The best case in each row is marked in bold.}
\label{tab:Validations}
\end{table}

\begin{table}[t!]
\vspace*{-0mm}
\begin{center}
    \small
    \renewcommand{\arraystretch}{1.05}
    \begin{tabular}{|wc{6.0em}?wc{5.3em}|wc{5.3em}|wc{5.3em}|wc{5.3em}|wc{5.3em}|}
        \cline{2-6}
        \multicolumn{1}{c|}{} & \textbf{Ours} & \textbf{ChatGPT} & \textbf{DeepSeek} & \textbf{Gemini} & \textbf{Grok}
        \\
        \cline{1-2}
        \Cline{2pt}{2-6}
        \textit{loss} & \textbf{6.67858} & 6.70695 & 6.71061 & 6.69679 & 6.71056 \\
        \cline{1-6}
        $\tau$ (in \medidas{}) & \textbf{0.38958} & 0.62417 & 0.55958 & 1.29292 & 0.65667 \\
        \Cline{1pt}{1-6}
        $\textit{TCEOCS}_{t}$ & 20.13163 & 20.13162 & 20.13163 & \textbf{20.1316} & 20.13163 \\
        \cline{1-6}
        $\mu(\textit{MSE})_{t}$ & 0.00261 & \textbf{0.00260} & 0.00266 & \textbf{0.00260} & 0.00359 \\
        \cline{1-6}
        $\mu(R^2)_{t}$ & -6.02E10 & \textbf{-5.99E10} & -6.19E10 & -6.00E10 & -8.22E10 \\
        \cline{1-6}
        $\sigma(R^2)_{t}$ & 2.08E11 & 2.03E11 & \textbf{1.44E11} & 2.23E11 & 7.41E11 \\
        \cline{1-6}
        $\textit{TCEOCS}_{i}$ & \textbf{19.93983} & 20.11005 & 20.13205 & 20.04915 & 20.13172 \\
        \cline{1-6}
        $\mu(\textit{MSE})_{i}$ & \textbf{0.00231} & 0.00257 & 0.00257 & 0.00241 & 0.00276 \\
        \cline{1-6}
        $\mu(R^2)_{i}$ & \textbf{-1.78525} & -2.21997 & -3.11264 & -1.87543 & -6.91678 \\
        \cline{1-6}
        $\sigma(R^2)_{i}$ & \textbf{4.38500} & 4.39950 & 5.24137 & 4.74028 & 91.50352 \\
        \cline{1-6}
        $\mu(\textit{MSE})_{rt}$ & 6.35E04 & 3.15E00 & 2.17E00 & \textbf{2.11E00} & 2.37E00 \\
        \cline{1-6}
        $\mu(R^2)_{rt}$ & -1.84E16 & -5.71E11 & -3.27E11 & \textbf{-3.17E11} & -3.36E11 \\
        \cline{1-6}
        $\sigma(R^2)_{rt}$ & \textit{invalid} & 1.45E13 & \textbf{4.86E12} & 8.44E12 & 5.05E12 \\
        \hline
    \end{tabular}
\end{center}
\vspace*{-6mm}
\caption{Test metrics. Same subindexes as Table \ref{tab:Tests}, with the addition of $rt$, which stands for the values measured between the raw outputs of the corresponding audio encoder and the ones from the CLIP text encoder. The best case in each row is marked in bold, and \textit{invalid} means that the value was too close to $-\infty$ or $+\infty$ to be registered.}
\label{tab:Tests}
\end{table}

In Table \ref{tab:Tests}, we can actually corroborate that the top encoders of most of the previous metrics remain unchanged (except for $\textit{TCEOCS}_{t}$), but we now have some clearer insight on what to expect in the generations.
On the new metrics that measure the similarity of the raw outputs of each encoder with the ones from the CLIP text encoder ($\mu(\textit{MSE})_{rt}$, $\mu(R^2)_{rt}$ and $\sigma(R^2)_{rt}$) our encoder performs strikingly poorly; which is rather unfortunate, as a close resemblance to these raw outputs is what in practice should translate into clear images.
The most probable explanation to these low scores may be that our model is missing some sort of limiter or normalizer to the outputs it yields; while being able to do without it in the other metrics, as the projection includes a normalization step.
Even so, at least the relatively small size of our model translated into a faster batch-processing time, although not by such a large margin.
For reference, the CLIP text encoder has a $\tau$ of 0,65792 \medidas{}, meaning that almost all encoders are faster than that one, excluding the Gemini one (probably due to the preprocessing layers it included with Mel spectrograms).
Taking into account that a $R^2$ is usually considered slightly positive when it takes a value $\geq 0.4$ \cite{The_Use_of_Partial_Least_Squares_Path_Modeling, Structural_Equation_Modeling_in_Information_Systems, The_Partial_Least_Squares_Approach_to_Structural} and that all the $\mu(R^2)$s we obtained are negative, we can confidently say that none of the trained audio encoders is a good replacement for the original text encoder.
Otherwise, based merely on the other metrics and ignoring our own encoder, Gemini seems to have done the best job overall.

\begin{figure}[t!]
% \vspace*{-9mm}
\begin{center}
    \setlength{\tabcolsep}{0pt}
    \renewcommand{\arraystretch}{0.0}
    \begin{tabular}{|M{6.5em}|}
        \hline
        \vspace*{2.0mm} \textbf{Input I} \\ $\color{white}a$ \\
        \hline
        \includegraphics[width=\linewidth]{\detokenize{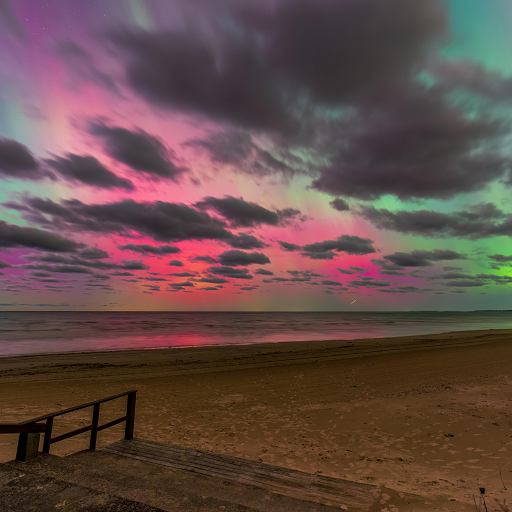}} \\
        \hline
    \end{tabular}
    \hspace{8mm}
    \begin{tabular}{|M{6.5em}|M{6.5em}|}
        \hline
        \vspace*{2.0mm} \textbf{R to I} & \vspace*{2.0mm} \textbf{R\&I to I} \\
        $\color{white}a$ &  \\
        \hline
        \includegraphics[width=\linewidth]{\detokenize{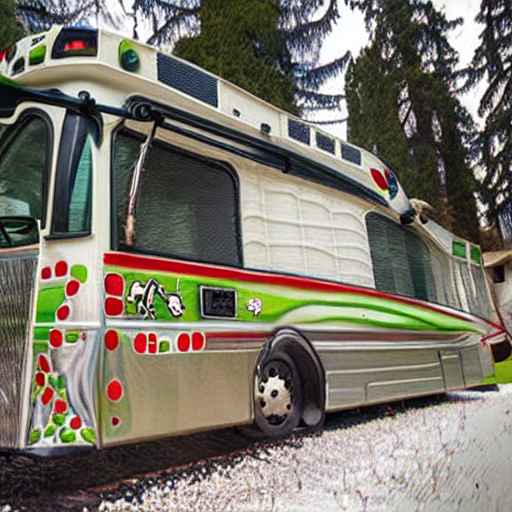}} & \includegraphics[width=\linewidth]{\detokenize{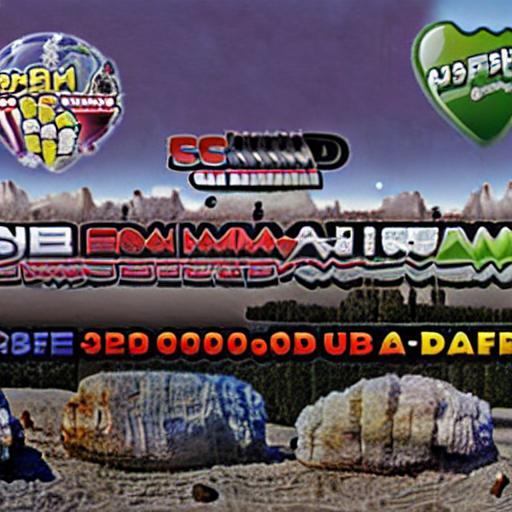}} \\
        \hline
    \end{tabular}
    \hspace{8mm}
    \begin{tabular}{|M{6.5em}|M{6.5em}|}
        \hline
        \vspace*{2.0mm} \textbf{T to I} & \vspace*{2.0mm} \textbf{T\&I to I} \\
        $\color{white}a$ &  \\
        \hline
        \includegraphics[width=\linewidth]{\detokenize{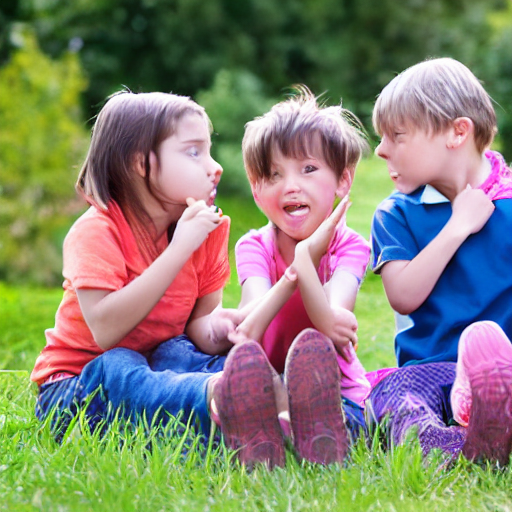}} & \includegraphics[width=\linewidth]{\detokenize{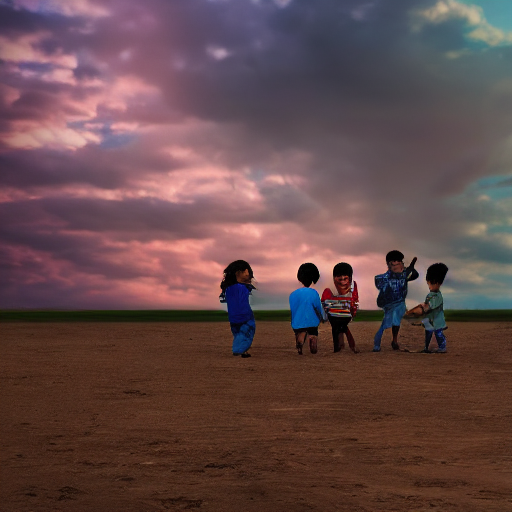}} \\
        \hline
    \end{tabular}
\end{center}
\caption{Benchmark sample of generations for the input text of ``Children talking and playing" (used only in the synthesized images of the right box) and input image given in the left box (used only in the synthesized images on the right side of the respective boxes). \textbf{I} stands for image, \textbf{R} for random, and \textbf{T} for text. Samples generated with \textbf{T} serve as good case examples, while the ones generated with \textbf{R} serve as a bad case examples.}
\label{fig:Benchmarks_1}
\end{figure}

\begin{figure}[t!]
% \vspace*{-9mm}
\begin{center}
    \setlength{\tabcolsep}{0pt}
    \renewcommand{\arraystretch}{0.0}
    \begin{tabular}{|M{6.5em}|}
        \hline
        \vspace*{2.0mm} \textbf{Input I} \\ $\color{white}a$ \\
        \hline
        \includegraphics[width=\linewidth]{\detokenize{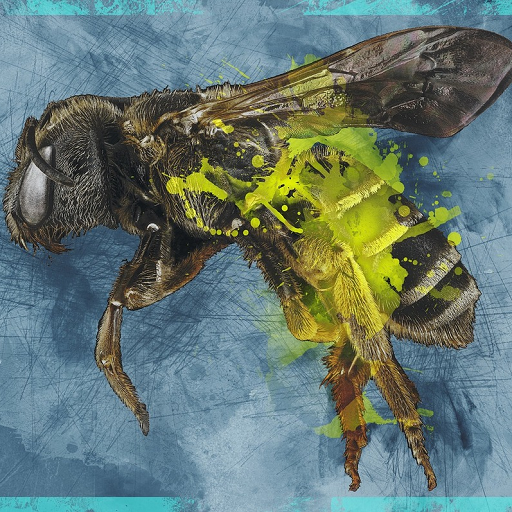}} \\
        \hline
    \end{tabular}
    \hspace{8mm}
    \begin{tabular}{|M{6.5em}|M{6.5em}|}
        \hline
        \vspace*{2.0mm} \textbf{R to I} & \vspace*{2.0mm} \textbf{R\&I to I} \\
        $\color{white}a$ &  \\
        \hline
        \includegraphics[width=\linewidth]{\detokenize{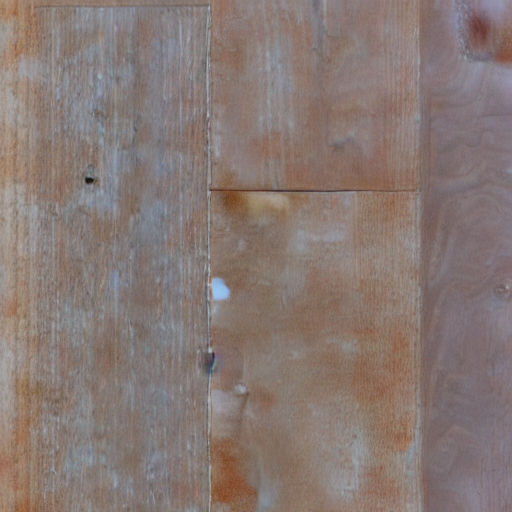}} & \includegraphics[width=\linewidth]{\detokenize{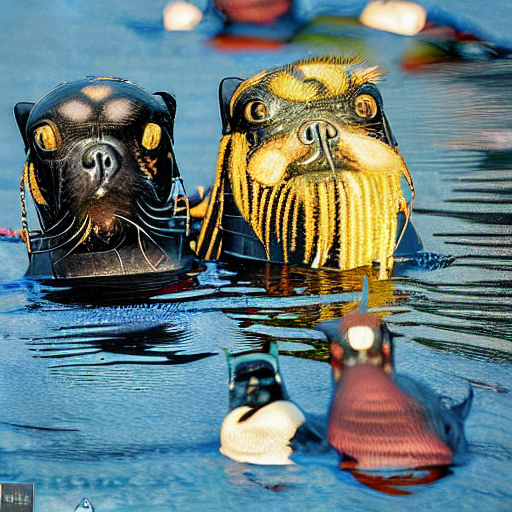}} \\
        \hline
    \end{tabular}
    \hspace{8mm}
    \begin{tabular}{|M{6.5em}|M{6.5em}|}
        \hline
        \vspace*{2.0mm} \textbf{T to I} & \vspace*{2.0mm} \textbf{T\&I to I} \\
        $\color{white}a$ &  \\
        \hline
        \includegraphics[width=\linewidth]{\detokenize{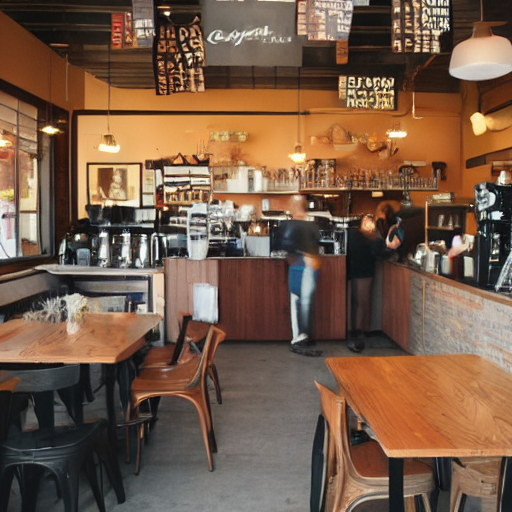}} & \includegraphics[width=\linewidth]{\detokenize{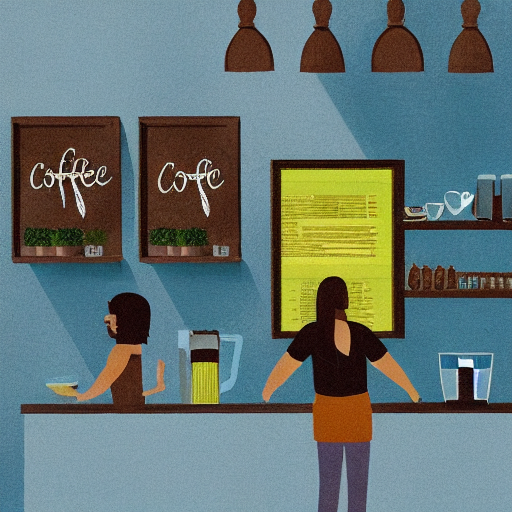}} \\
        \hline
    \end{tabular}
\end{center}
\caption{Benchmark sample of generations for the input text of ``The interior of a coffee shop" (used only in the synthesized images of the right box) and input image given in the left box (used only in the synthesized images on the right side of the respective boxes). \textbf{I} stands for image, \textbf{R} for random, and \textbf{T} for text. Samples generated with \textbf{T} serve as good case examples, while the ones generated with \textbf{R} serve as a bad case examples.}
\label{fig:Benchmarks_2}
\end{figure}

\begin{figure}[t!]
\vspace*{-14mm}
\begin{center}
    \setlength{\tabcolsep}{0pt}
    \renewcommand{\arraystretch}{0.0}
    \begin{tabular}{|M{6.0em}|M{6.5em}|M{6.5em}|M{6.5em}|M{6.5em}|}
        \cline{2-5}
        \multicolumn{1}{c|}{$\color{white}\begin{matrix} a \\ b \\ c \end{matrix}$} & \textbf{A to I} & \textbf{A\&T to I} & \textbf{A\&I to I} & \textbf{A\&T\&I to I}
        \\
        % \cline{1-2}
        % \Cline{2pt}{2-5}
        \hline
        \vspace*{-2.0mm}\rotatebox[origin=c]{40}{\textbf{Ours}} & \includegraphics[width=\linewidth]{\detokenize{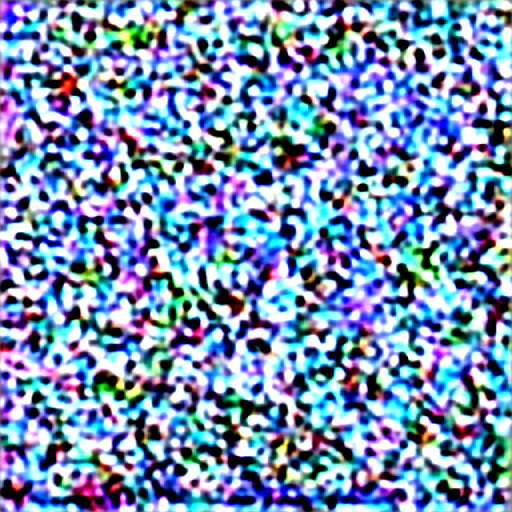}} & \includegraphics[width=\linewidth]{\detokenize{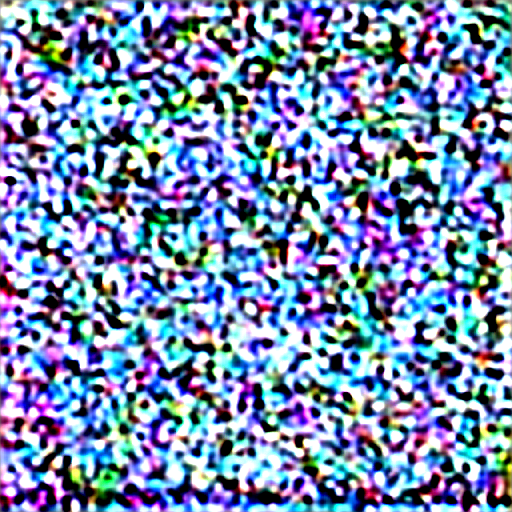}} & \includegraphics[width=\linewidth]{\detokenize{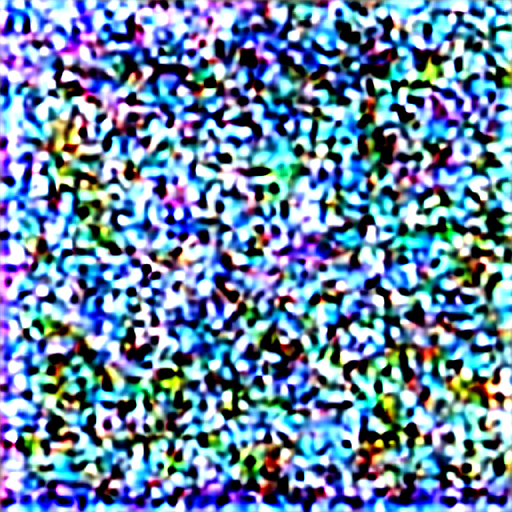}} & \includegraphics[width=\linewidth]{\detokenize{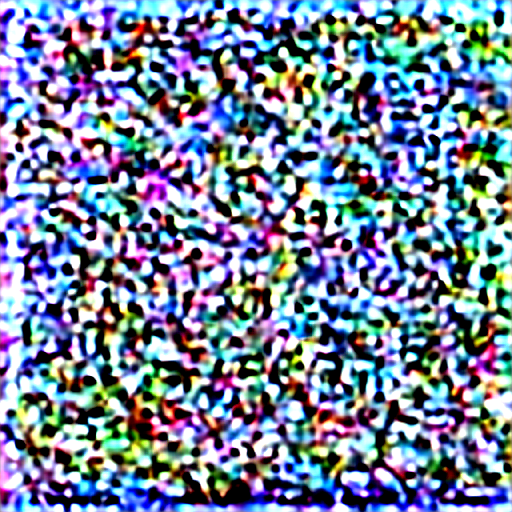}} \\
        \hline
        \vspace*{-5.05mm}\rotatebox[origin=c]{40}{\textbf{ChatGPT}} & \includegraphics[width=\linewidth]{\detokenize{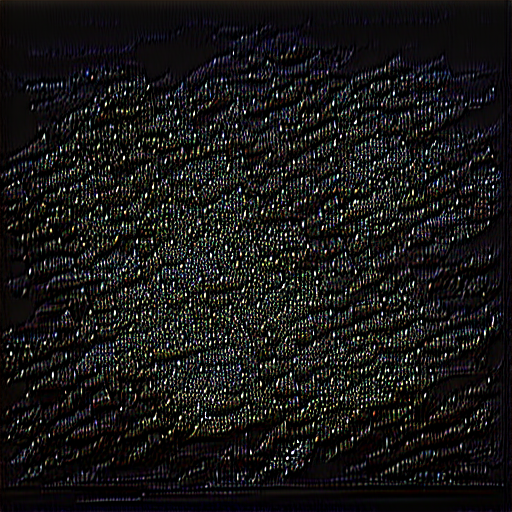}} & \includegraphics[width=\linewidth]{\detokenize{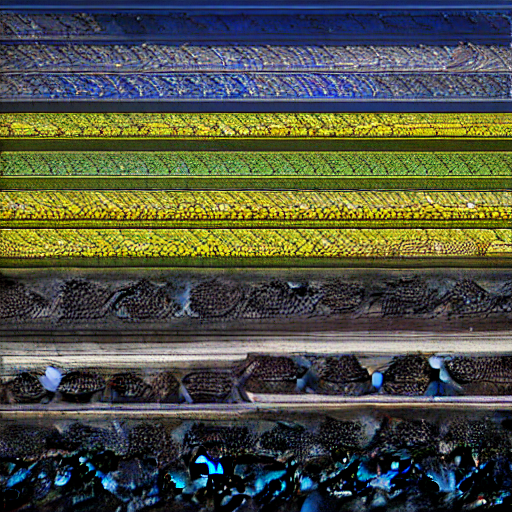}} & \includegraphics[width=\linewidth]{\detokenize{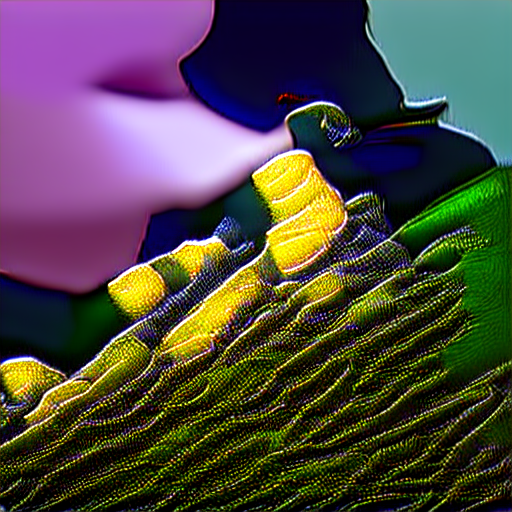}} & \includegraphics[width=\linewidth]{\detokenize{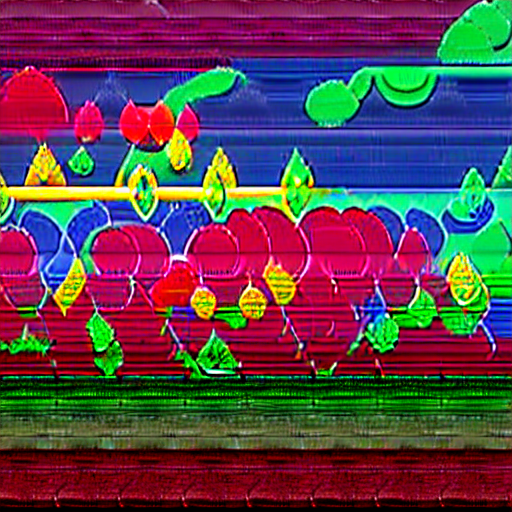}} \\
        \hline
        \vspace*{-5.05mm}\rotatebox[origin=c]{40}{\textbf{DeepSeek}} & \includegraphics[width=\linewidth]{\detokenize{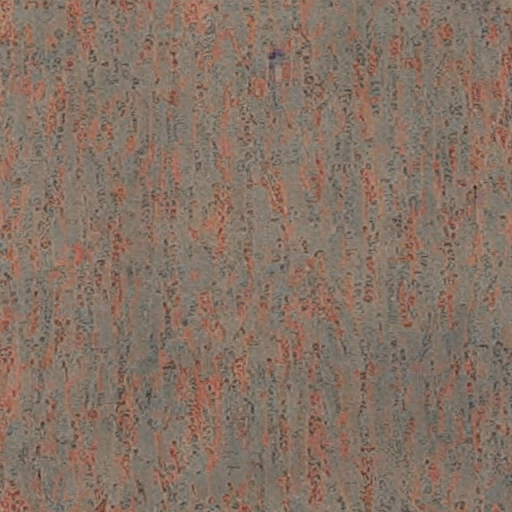}} & \includegraphics[width=\linewidth]{\detokenize{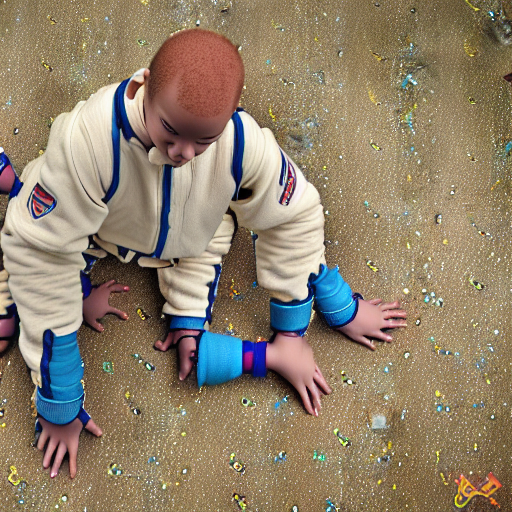}} & \includegraphics[width=\linewidth]{\detokenize{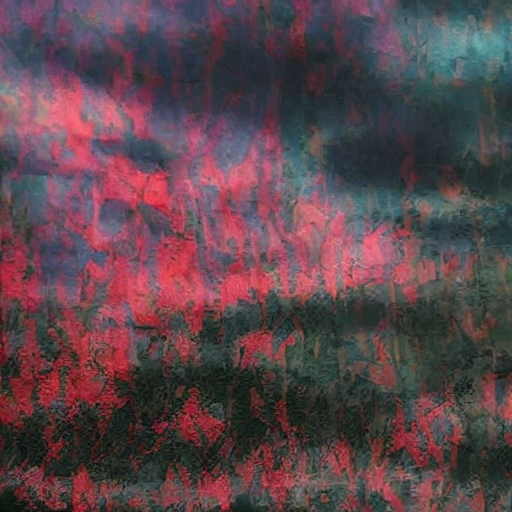}} & \includegraphics[width=\linewidth]{\detokenize{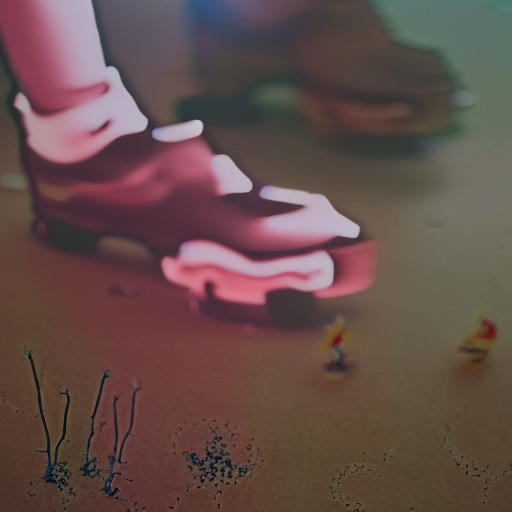}} \\
        \hline
        \vspace*{-3.05mm}\rotatebox[origin=c]{40}{\textbf{Gemini}} & \includegraphics[width=\linewidth]{\detokenize{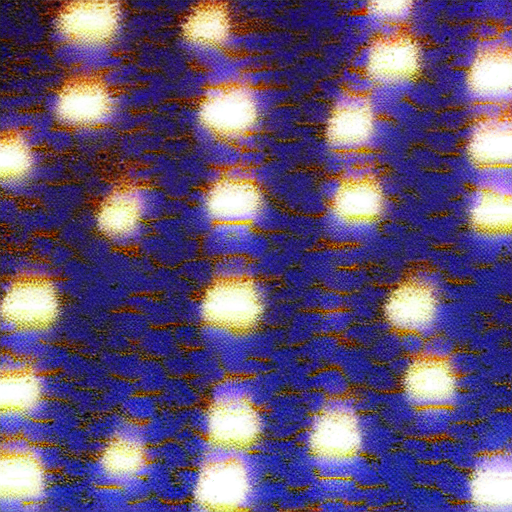}} & \includegraphics[width=\linewidth]{\detokenize{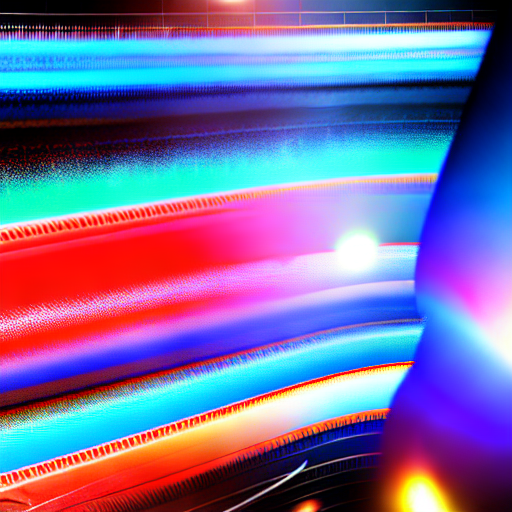}} & \includegraphics[width=\linewidth]{\detokenize{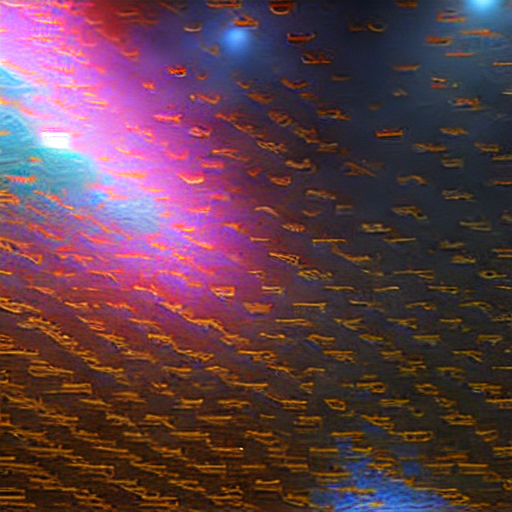}} & \includegraphics[width=\linewidth]{\detokenize{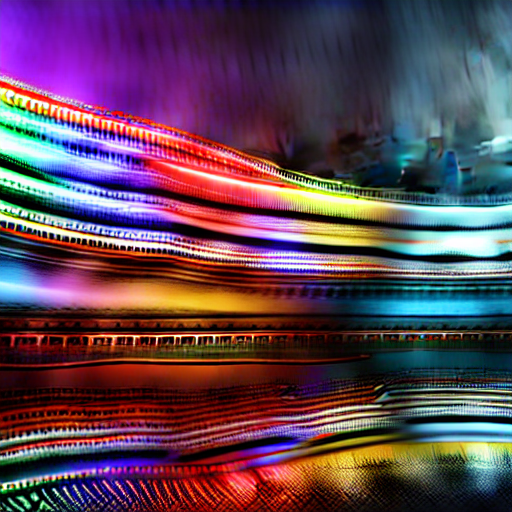}} \\
        \hline
        \vspace*{-2.0mm}\rotatebox[origin=c]{40}{\textbf{Grok}} & \includegraphics[width=\linewidth]{\detokenize{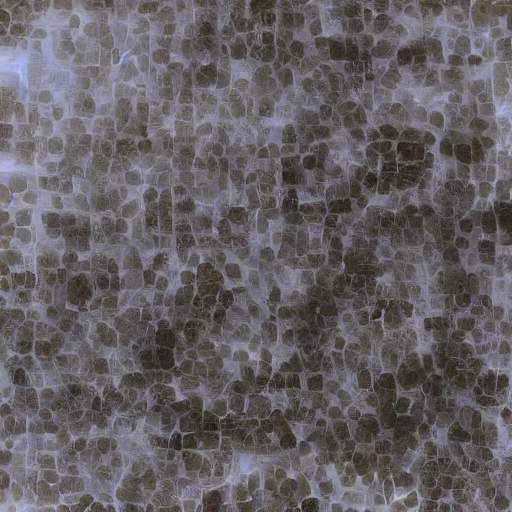}} & \includegraphics[width=\linewidth]{\detokenize{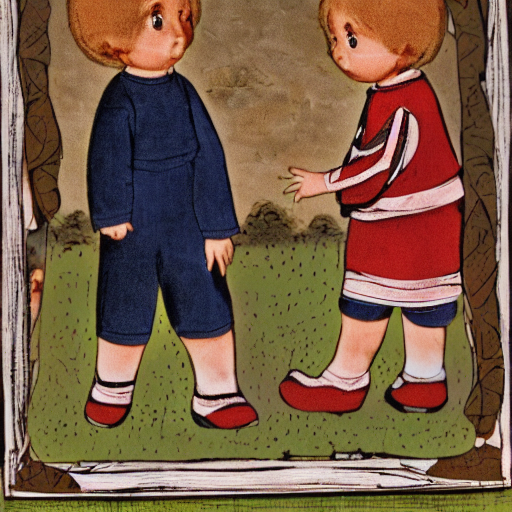}} & \includegraphics[width=\linewidth]{\detokenize{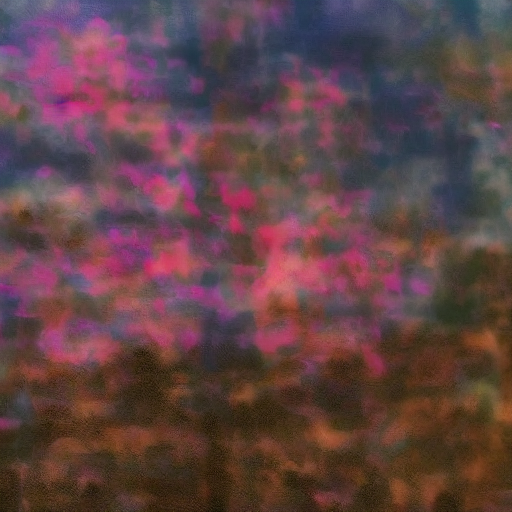}} & \includegraphics[width=\linewidth]{\detokenize{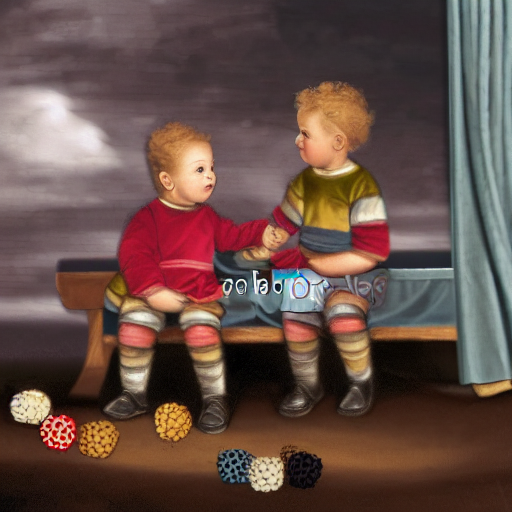}} \\
        \hline
        \vspace*{-6.05mm}\hspace*{-1.5mm}\rotatebox[origin=c]{40}{\textbf{\shortstack{Average\\Chatbots}}} & \includegraphics[width=\linewidth]{\detokenize{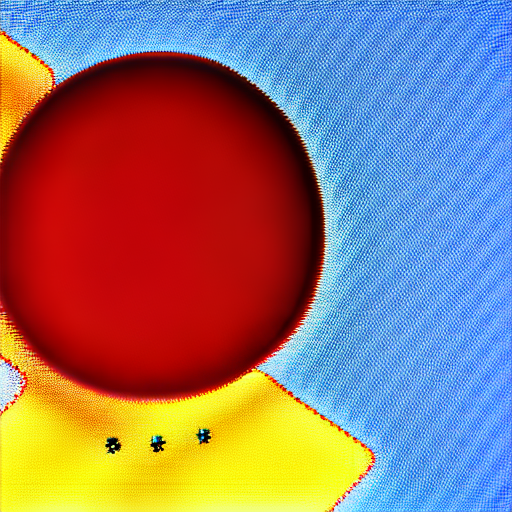}} & \includegraphics[width=\linewidth]{\detokenize{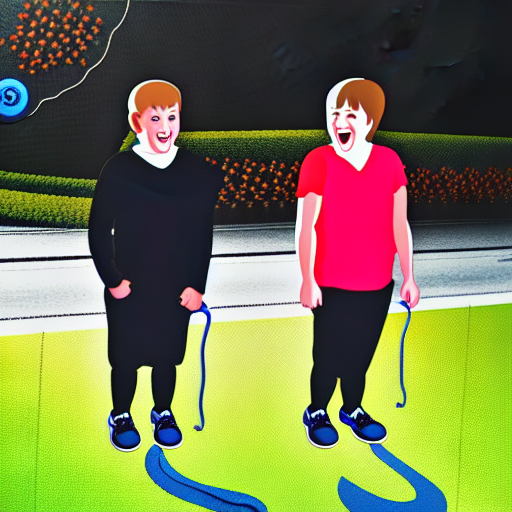}} & \includegraphics[width=\linewidth]{\detokenize{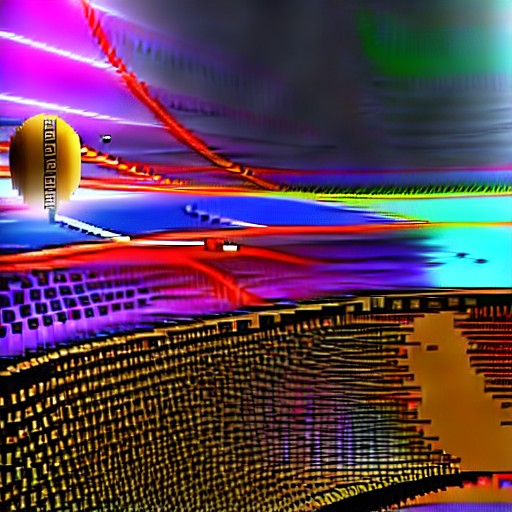}} & \includegraphics[width=\linewidth]{\detokenize{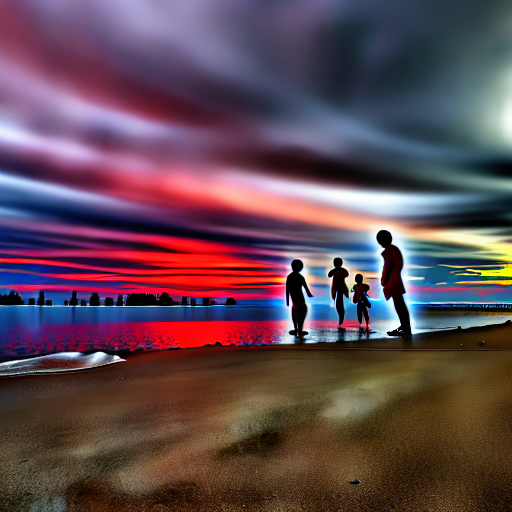}} \\
        \hline
        \vspace*{-6.05mm}\hspace*{-0.5mm}\rotatebox[origin=c]{40}{\textbf{\shortstack{Average\\Everyone}}} & \includegraphics[width=\linewidth]{\detokenize{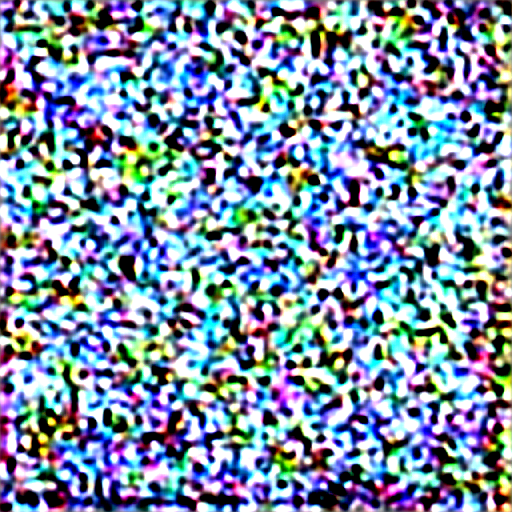}} & \includegraphics[width=\linewidth]{\detokenize{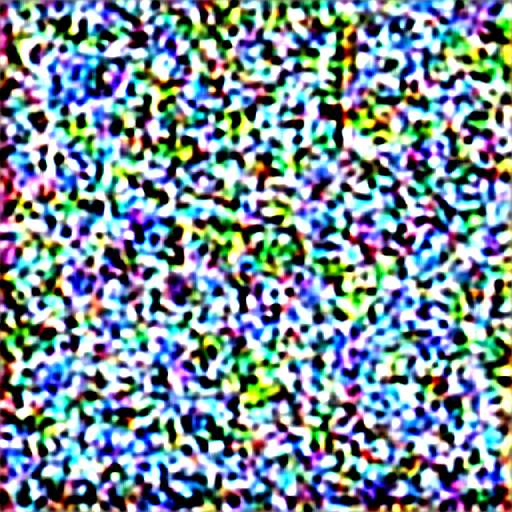}} & \includegraphics[width=\linewidth]{\detokenize{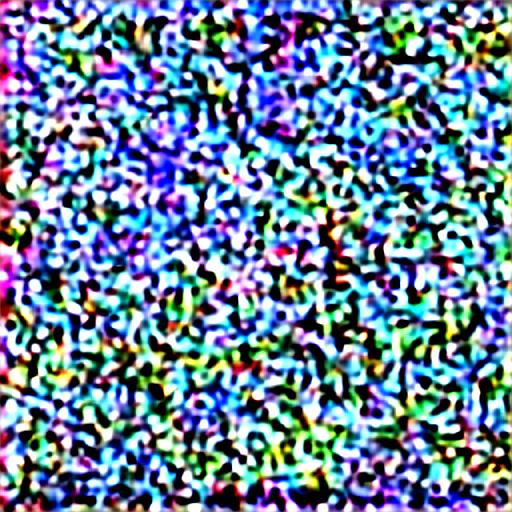}} & \includegraphics[width=\linewidth]{\detokenize{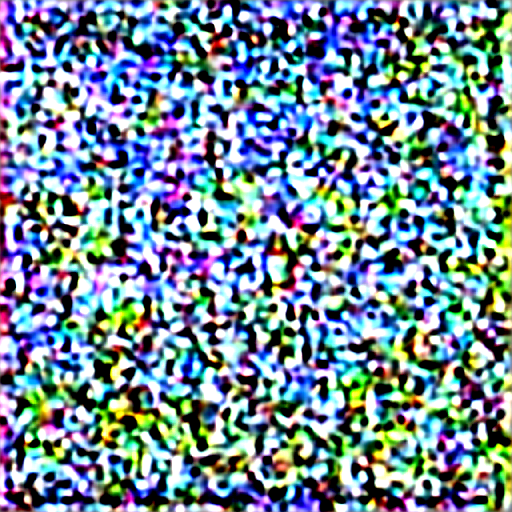}} \\
        \hline
    \end{tabular}
\end{center}
\vspace*{-4mm}
\caption{Sample of generations for the input image shown in Figure \ref{fig:Benchmarks_1}, the text of ``Children talking and playing", and the semantically related audio to the latter (which you can listen \href{https://jorvan758.github.io/A-SD-Alt/Referencias/children\%20talking\%20and\%20playing.wav}{\blue{\underline{here}}}). \textbf{A} stands for audio, \textbf{I} for image, and \textbf{T} for text. The generations from the \textbf{Average Chatbots} row are based on the average embeddings of all the audio encoders that come from chatbots; while \textbf{Average Everyone} does the same, but also adding ours.}
\label{fig:Generation_1}
\end{figure}

\begin{table}[t!]
\vspace*{-4mm}
\begin{center}
    \renewcommand{\arraystretch}{1.05}
    \begin{tabular}{|wc{4.5em}|wc{5.5em}?wc{0.923em}|wc{0.923em}|wc{0.923em}|wc{0.923em}|wc{0.923em}|wc{0.923em}|wc{0.923em}|wc{0.923em}|wc{0.923em}|wc{0.923em}|wc{0.923em}|wc{0.923em}|wc{0.923em}|}
        \cline{3-15}
        \multicolumn{1}{c}{} & \multicolumn{1}{c|}{} & \rotatebox[origin=c]{80}{ \ A bus \ } & \rotatebox[origin=c]{80}{ \ A curtain \ } & \rotatebox[origin=c]{80}{ \ A fence \ } & \rotatebox[origin=c]{80}{ \ A frame \ } & \rotatebox[origin=c]{80}{ \ Balls \ } & \rotatebox[origin=c]{80}{ \ Clouds \ } & \rotatebox[origin=c]{80}{ \ Dots \ } & \rotatebox[origin=c]{80}{ \ Grass \ } & \rotatebox[origin=c]{80}{ \ Kids \ } & \rotatebox[origin=c]{80}{ \ Limbs \ } & \rotatebox[origin=c]{80}{ \ Sand \ } & \rotatebox[origin=c]{80}{ \ Stripes \ } & \rotatebox[origin=c]{80}{ \ Water \ } \\
        \cline{1-2}
        \Cline{2pt}{3-15}
        \multirow{5}{*}{\rotatebox[origin=c]{40}{\textbf{Benchmark}}} & \textbf{Input I} & \noj & \noj & \yesj & \noj & \noj & \yesj & \noj & \noj & \noj & \noj & \yesj & \noj & \yesj \\
        \cline{2-15}
         & \textbf{R to I} & \yesj & \noj & \noj & \noj & \noj & \noj & \noj & \noj & \noj & \noj & \noj & \noj & \noj \\
        \cline{2-15}
         & \textbf{R\&I to I} & \noj & \noj & \noj & \noj & \noj & \noj & \noj & \noj & \noj & \noj & \noj & \noj & \noj \\
        \cline{2-15}
         & \textbf{T to I} & \noj & \noj & \noj & \noj & \noj & \noj & \noj & \yesj & \yesj & \noj & \noj & \noj & \noj \\
        \cline{2-15}
         & \textbf{T\&I to I} & \noj & \noj & \noj & \noj & \noj & \yesj & \noj & \noj & \yesj & \noj & \yesj & \noj & \noj \\
        \Cline{2pt}{1-15}
        \multirow{4}{*}{\rotatebox[origin=c]{40}{\textbf{Ours}}} & \textbf{A to I} & \noj & \noj & \noj & \noj & \noj & \noj & \yesj & \noj & \noj & \noj & \noj & \noj & \noj \\
        \cline{2-15}
         & \textbf{A\&T to I} & \noj & \noj & \noj & \noj & \noj & \noj & \yesj & \noj & \noj & \noj & \noj & \noj & \noj \\
        \cline{2-15}
         & \textbf{A\&I to I} & \noj & \noj & \noj & \noj & \noj & \noj & \yesj & \noj & \noj & \noj & \noj & \noj & \noj \\
        \cline{2-15}
         & \textbf{A\&T\&I to I} & \noj & \noj & \noj & \noj & \noj & \noj & \yesj & \noj & \noj & \noj & \noj & \noj & \noj \\
        \Cline{2pt}{1-15}
        \multirow{4}{*}{\rotatebox[origin=c]{40}{\textbf{ChatGPT}}} & \textbf{A to I} & \noj & \noj & \noj & \noj & \noj & \noj & \yesj & \noj & \noj & \noj & \noj & \noj & \noj \\
        \cline{2-15}
         & \textbf{A\&T to I} & \noj & \noj & \noj & \noj & \noj & \noj & \yesj & \noj & \noj & \noj & \noj & \yesj & \noj \\
        \cline{2-15}
         & \textbf{A\&I to I} & \noj & \noj & \noj & \noj & \noj & \noj & \noj & \noj & \noj & \noj & \noj & \noj & \noj \\
        \cline{2-15}
         & \textbf{A\&T\&I to I} & \noj & \noj & \noj & \noj & \noj & \noj & \noj & \noj & \noj & \noj & \noj & \yesj & \noj \\
        \Cline{2pt}{1-15}
        \multirow{4}{*}{\rotatebox[origin=c]{40}{\textbf{DeepSeek}}} & \textbf{A to I} & \noj & \noj & \noj & \noj & \noj & \noj & \noj & \noj & \noj & \noj & \noj & \noj & \noj \\
        \cline{2-15}
         & \textbf{A\&T to I} & \noj & \noj & \noj & \noj & \noj & \noj & \noj & \noj & \noj & \yesj & \noj & \noj & \noj \\
        \cline{2-15}
         & \textbf{A\&I to I} & \noj & \noj & \noj & \noj & \noj & \yesj & \noj & \noj & \noj & \noj & \noj & \noj & \noj \\
        \cline{2-15}
         & \textbf{A\&T\&I to I} & \noj & \noj & \noj & \noj & \noj & \noj & \noj & \noj & \noj & \yesj & \noj & \noj & \noj \\
        \Cline{2pt}{1-15}
        \multirow{4}{*}{\rotatebox[origin=c]{40}{\textbf{Gemini}}} & \textbf{A to I} & \noj & \noj & \noj & \noj & \noj & \noj & \yesj & \noj & \noj & \noj & \noj & \noj & \noj \\
        \cline{2-15}
         & \textbf{A\&T to I} & \noj & \noj & \noj & \noj & \noj & \noj & \noj & \noj & \noj & \noj & \noj & \yesj & \noj \\
        \cline{2-15}
         & \textbf{A\&I to I} & \noj & \noj & \noj & \noj & \noj & \noj & \noj & \noj & \noj & \noj & \noj & \noj & \noj \\
        \cline{2-15}
         & \textbf{A\&T\&I to I} & \noj & \noj & \noj & \noj & \noj & \noj & \noj & \noj & \noj & \noj & \noj & \yesj & \noj \\
        \Cline{2pt}{1-15}
        \multirow{4}{*}{\rotatebox[origin=c]{40}{\textbf{Grok}}} & \textbf{A to I} & \noj & \noj & \noj & \noj & \noj & \noj & \yesj & \noj & \noj & \noj & \noj & \noj & \noj \\
        \cline{2-15}
         & \textbf{A\&T to I} & \noj & \noj & \noj & \yesj & \noj & \noj & \noj & \yesj & \yesj & \noj & \noj & \noj & \noj \\
        \cline{2-15}
         & \textbf{A\&I to I} & \noj & \noj & \noj & \noj & \noj & \noj & \noj & \noj & \noj & \noj & \noj & \noj & \noj \\
        \cline{2-15}
         & \textbf{A\&T\&I to I} & \noj & \yesj & \noj & \noj & \yesj & \yesj & \noj & \noj & \yesj & \noj & \yesj & \noj & \noj \\
        \Cline{2pt}{1-15}
        \multirow{4}{*}{\rotatebox[origin=c]{40}{\textbf{\shortstack{Average\\Chatbots}}}} & \textbf{A to I} & \noj & \noj & \noj & \noj & \noj & \noj & \noj & \noj & \noj & \noj & \noj & \yesj & \noj \\
        \cline{2-15}
         & \textbf{A\&T to I} & \noj & \noj & \noj & \noj & \noj & \noj & \noj & \yesj & \yesj & \noj & \noj & \noj & \noj \\
        \cline{2-15}
         & \textbf{A\&I to I} & \noj & \noj & \noj & \noj & \noj & \noj & \yesj & \noj & \noj & \noj & \noj & \yesj & \noj \\
        \cline{2-15}
         & \textbf{A\&T\&I to I} & \noj & \noj & \noj & \noj & \noj & \yesj & \noj & \noj & \yesj & \noj & \yesj & \noj & \yesj \\
        \Cline{2pt}{1-15}
        \multirow{4}{*}{\rotatebox[origin=c]{40}{\textbf{\shortstack{Average\\Everyone}}}} & \textbf{A to I} & \noj & \noj & \noj & \noj & \noj & \noj & \yesj & \noj & \noj & \noj & \noj & \noj & \noj \\
        \cline{2-15}
         & \textbf{A\&T to I} & \noj & \noj & \noj & \noj & \noj & \noj & \yesj & \noj & \noj & \noj & \noj & \noj & \noj \\
        \cline{2-15}
         & \textbf{A\&I to I} & \noj & \noj & \noj & \noj & \noj & \noj & \yesj & \noj & \noj & \noj & \noj & \noj & \noj \\
        \cline{2-15}
         & \textbf{A\&T\&I to I} & \noj & \noj & \noj & \noj & \noj & \noj & \yesj & \noj & \noj & \noj & \noj & \noj & \noj \\
        \hline
    \end{tabular}
\end{center}
\vspace*{-4mm}
\caption{Breakdown of the presence of elements in the images from Figures \ref{fig:Benchmarks_1} and \ref{fig:Generation_1}. \yesj{} means the element is at least somewhat visible in the respective image; otherwise, it is not.}
\label{tab:Generation_elements_1}
\end{table}

\begin{figure}[t!]
\vspace*{-14mm}
\begin{center}
    \setlength{\tabcolsep}{0pt}
    \renewcommand{\arraystretch}{0.0}
    \begin{tabular}{|M{6.0em}|M{6.5em}|M{6.5em}|M{6.5em}|M{6.5em}|}
        \cline{2-5}
        \multicolumn{1}{c|}{$\color{white}\begin{matrix} a \\ b \\ c \end{matrix}$} & \textbf{A to I} & \textbf{A\&T to I} & \textbf{A\&I to I} & \textbf{A\&T\&I to I}
        \\
        % \cline{1-2}
        % \Cline{2pt}{2-5}
        \hline
        \vspace*{-2.0mm}\rotatebox[origin=c]{40}{\textbf{Ours}} & \includegraphics[width=\linewidth]{\detokenize{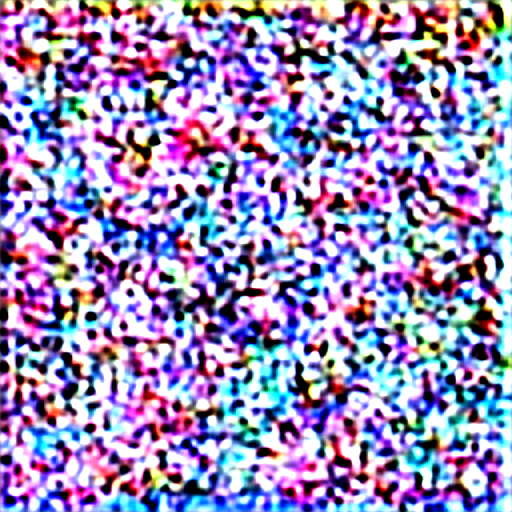}} & \includegraphics[width=\linewidth]{\detokenize{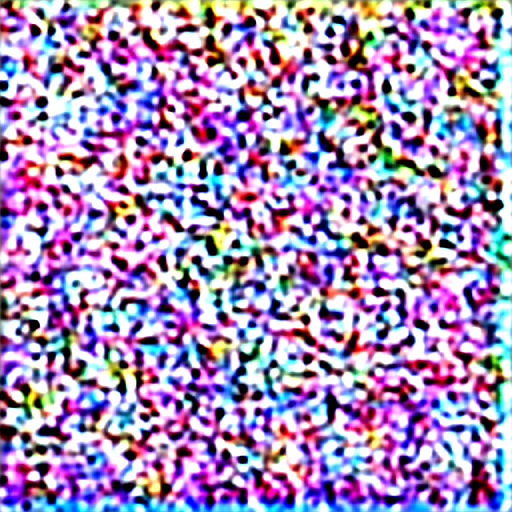}} & \includegraphics[width=\linewidth]{\detokenize{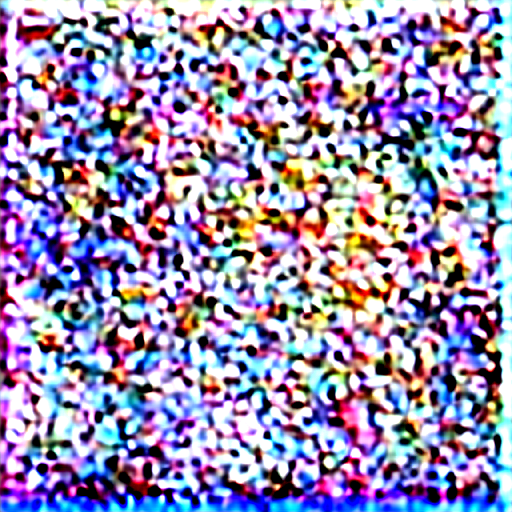}} & \includegraphics[width=\linewidth]{\detokenize{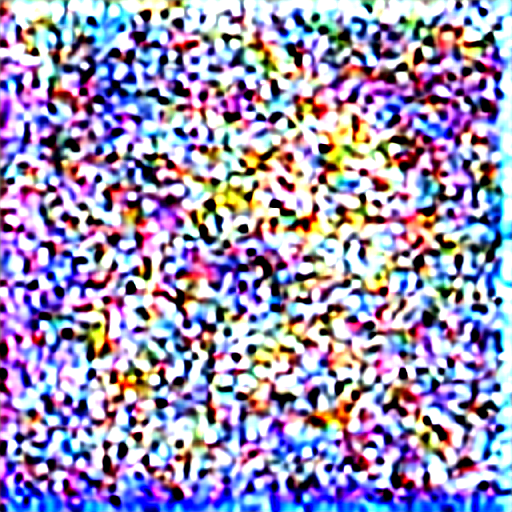}} \\
        \hline
        \vspace*{-5.05mm}\rotatebox[origin=c]{40}{\textbf{ChatGPT}} & \includegraphics[width=\linewidth]{\detokenize{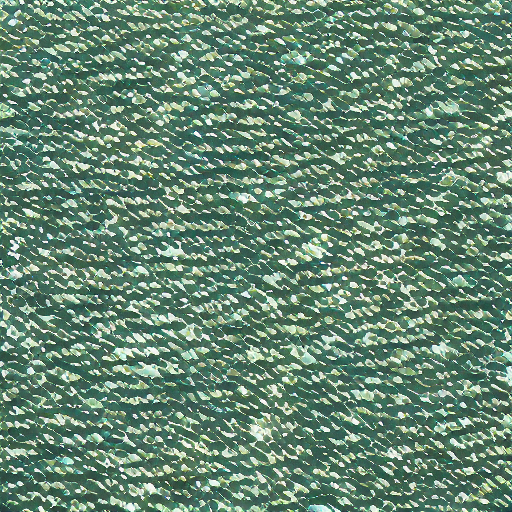}} & \includegraphics[width=\linewidth]{\detokenize{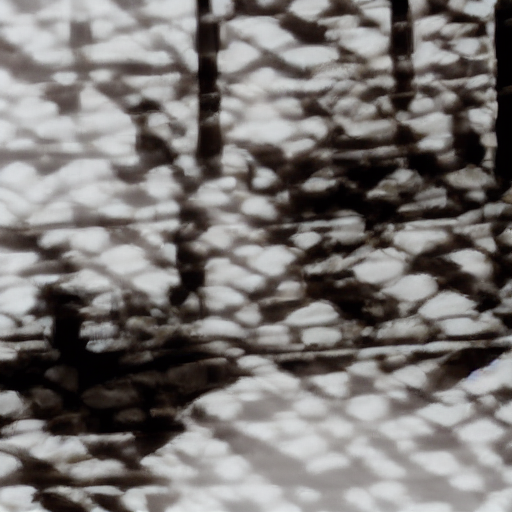}} & \includegraphics[width=\linewidth]{\detokenize{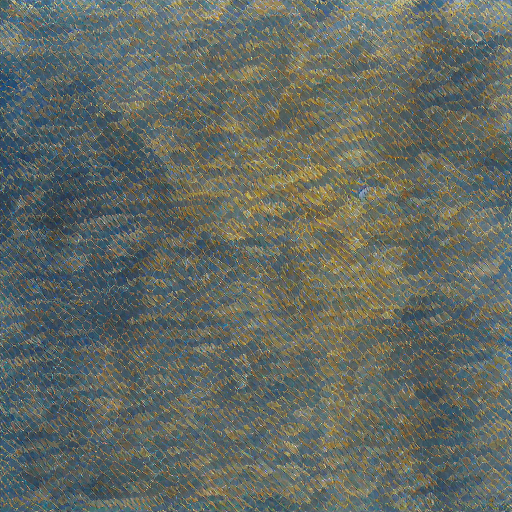}} & \includegraphics[width=\linewidth]{\detokenize{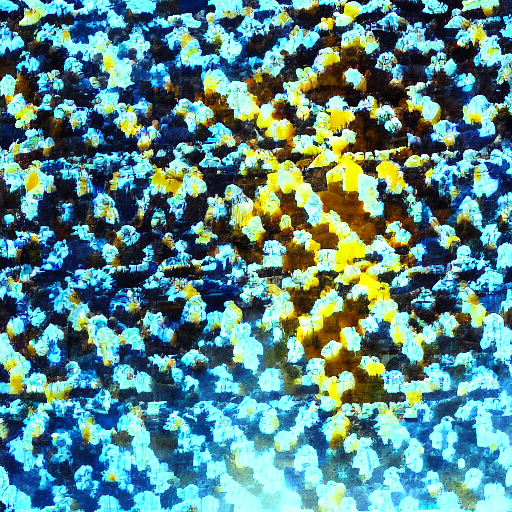}} \\
        \hline
        \vspace*{-5.05mm}\rotatebox[origin=c]{40}{\textbf{DeepSeek}} & \includegraphics[width=\linewidth]{\detokenize{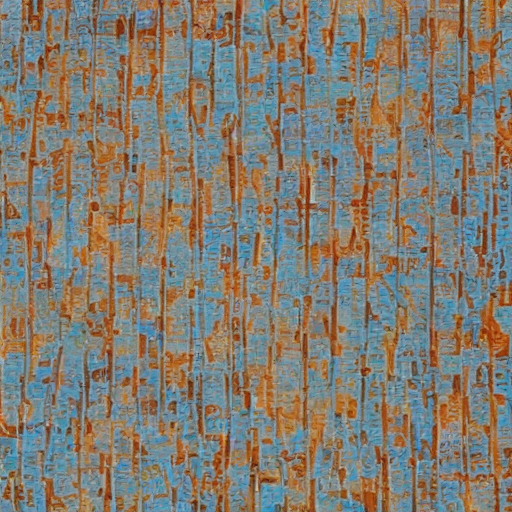}} & \includegraphics[width=\linewidth]{\detokenize{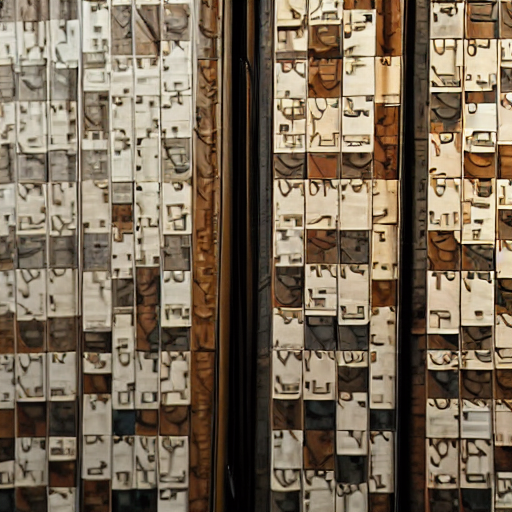}} & \includegraphics[width=\linewidth]{\detokenize{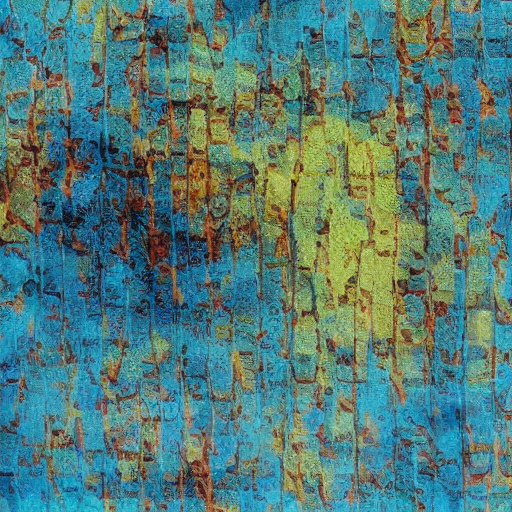}} & \includegraphics[width=\linewidth]{\detokenize{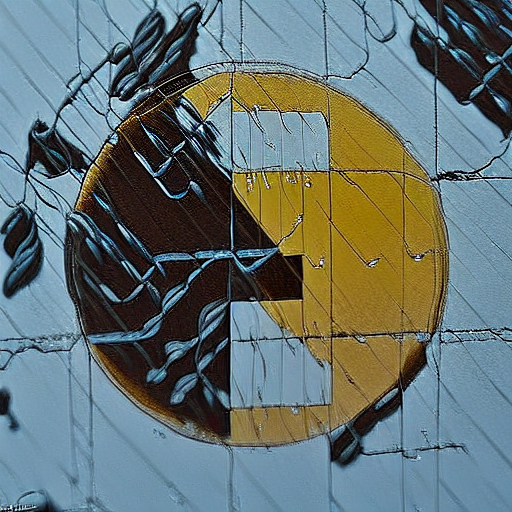}} \\
        \hline
        \vspace*{-3.05mm}\rotatebox[origin=c]{40}{\textbf{Gemini}} & \includegraphics[width=\linewidth]{\detokenize{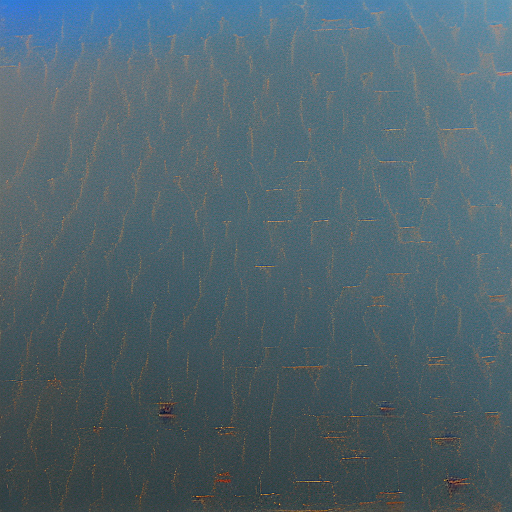}} & \includegraphics[width=\linewidth]{\detokenize{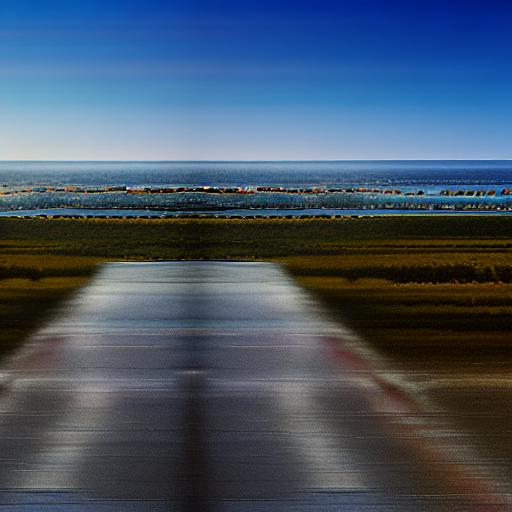}} & \includegraphics[width=\linewidth]{\detokenize{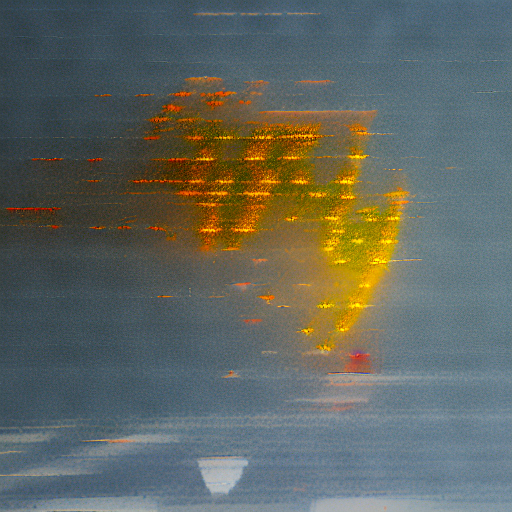}} & \includegraphics[width=\linewidth]{\detokenize{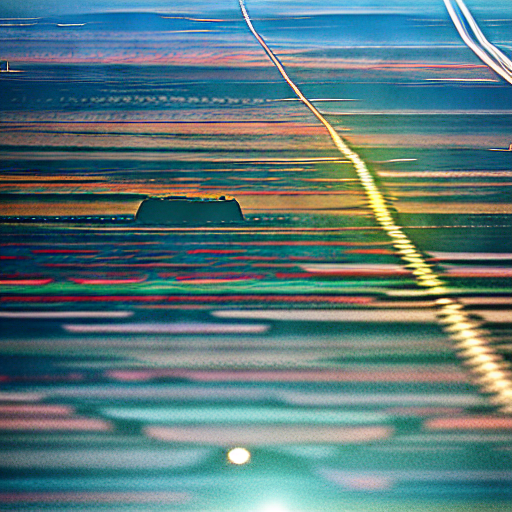}} \\
        \hline
        \vspace*{-2.0mm}\rotatebox[origin=c]{40}{\textbf{Grok}} & \includegraphics[width=\linewidth]{\detokenize{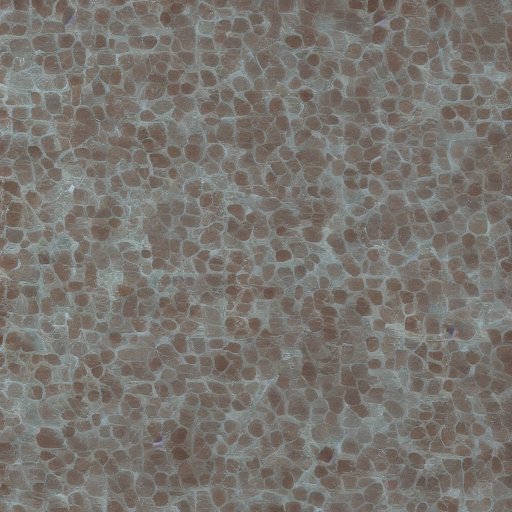}} & \includegraphics[width=\linewidth]{\detokenize{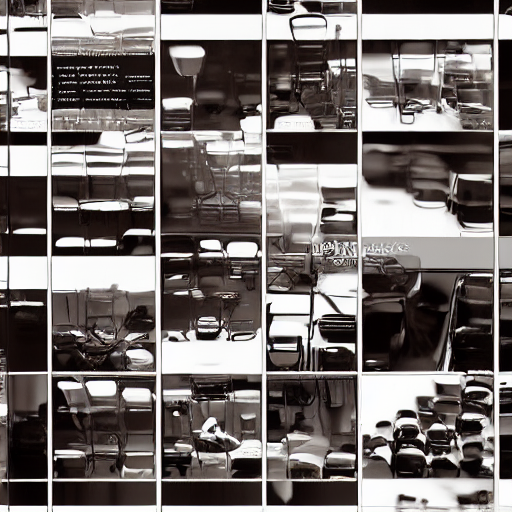}} & \includegraphics[width=\linewidth]{\detokenize{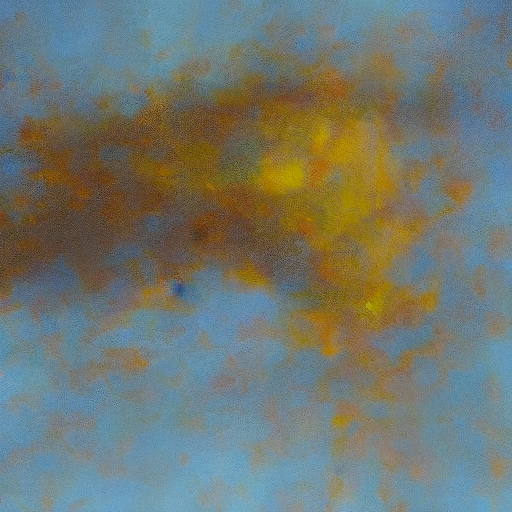}} & \includegraphics[width=\linewidth]{\detokenize{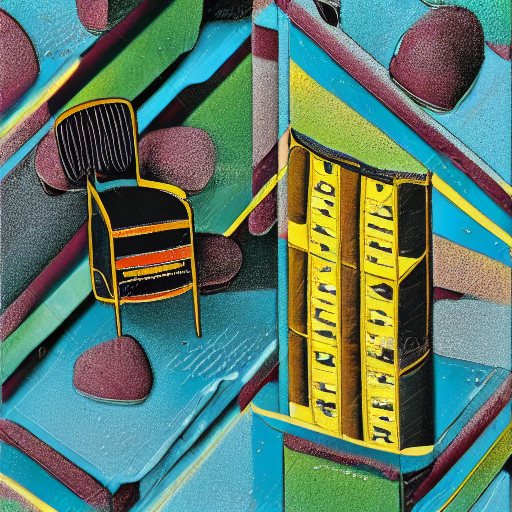}} \\
        \hline
        \vspace*{-6.05mm}\hspace*{-1.5mm}\rotatebox[origin=c]{40}{\textbf{\shortstack{Average\\Chatbots}}} & \includegraphics[width=\linewidth]{\detokenize{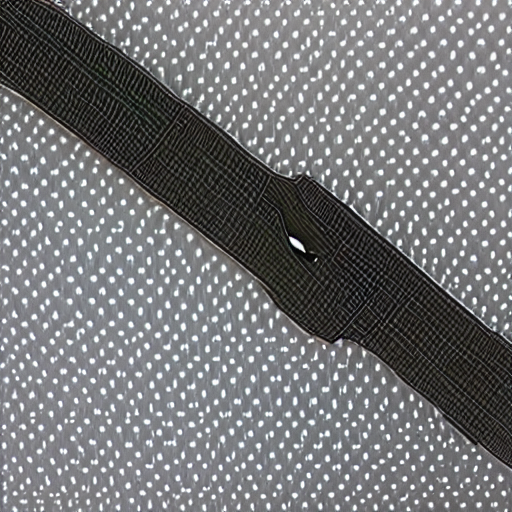}} & \includegraphics[width=\linewidth]{\detokenize{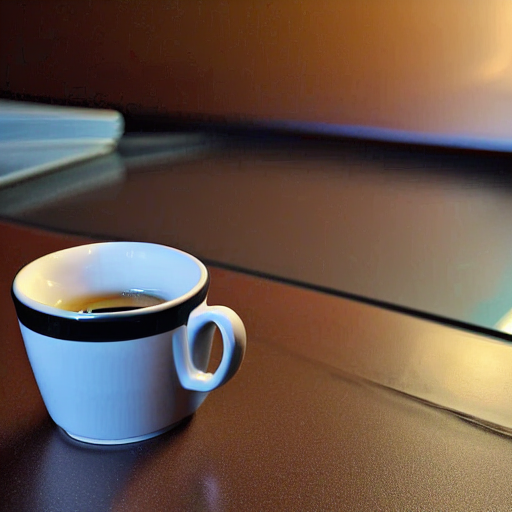}} & \includegraphics[width=\linewidth]{\detokenize{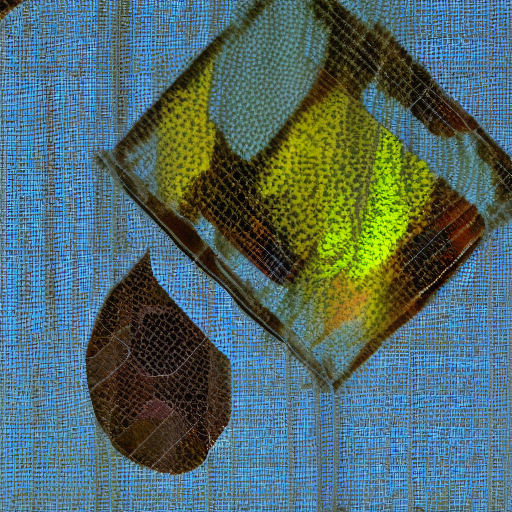}} & \includegraphics[width=\linewidth]{\detokenize{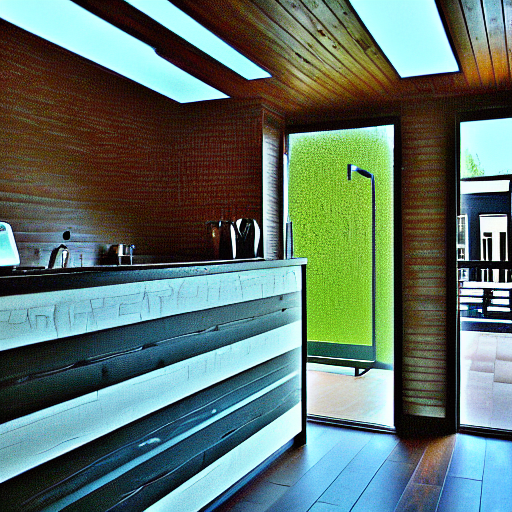}} \\
        \hline
        \vspace*{-6.05mm}\hspace*{-0.5mm}\rotatebox[origin=c]{40}{\textbf{\shortstack{Average\\Everyone}}}& \includegraphics[width=\linewidth]{\detokenize{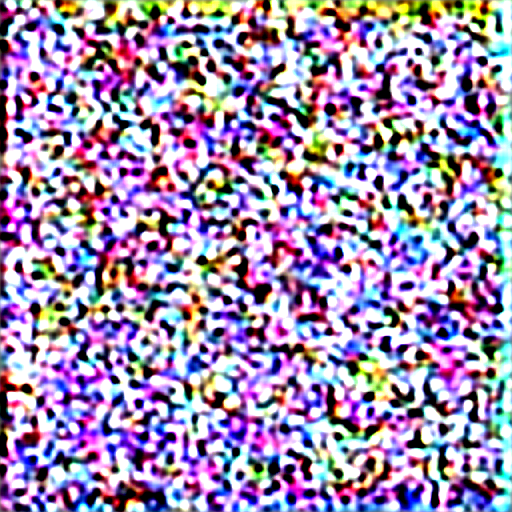}} & \includegraphics[width=\linewidth]{\detokenize{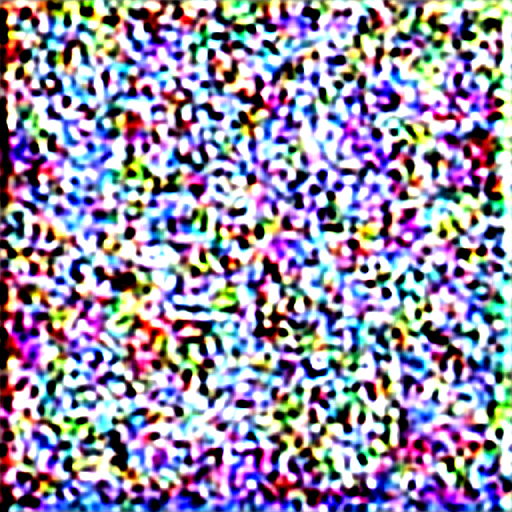}} & \includegraphics[width=\linewidth]{\detokenize{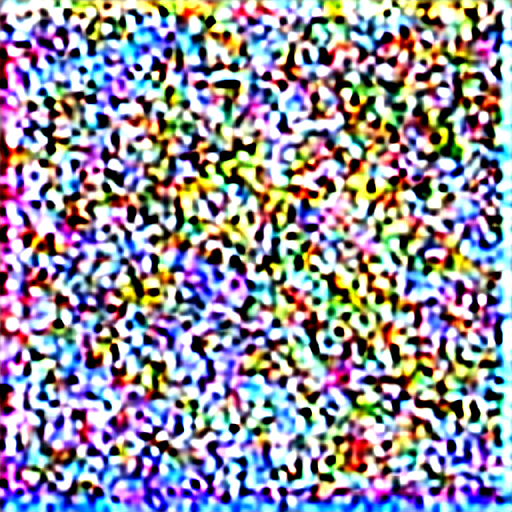}} & \includegraphics[width=\linewidth]{\detokenize{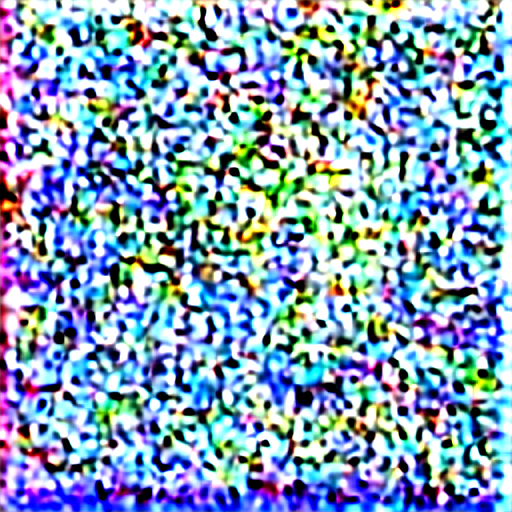}} \\
        \hline
    \end{tabular}
\end{center}
\vspace*{-4mm}
\caption{Sample of generations for the input image shown in Figure \ref{fig:Benchmarks_2}, the text of ``The interior of a coffee shop", and the semantically related audio to the latter (which you can listen \href{https://jorvan758.github.io/A-SD-Alt/Referencias/the\%20interior\%20of\%20a\%20coffee\%20shop.wav}{\blue{\underline{here}}}). \textbf{A} stands for audio, \textbf{I} for image, and \textbf{T} for text. The generations from the \textbf{Average Chatbots} row are based on the average embeddings of all the audio encoders that come from chatbots; while \textbf{Average Everyone} does the same, but also adding ours.}
\label{fig:Generation_2}
\end{figure}

\begin{table}[t!]
\vspace*{-4mm}
\begin{center}
    \renewcommand{\arraystretch}{1.05}
    \begin{tabular}{|wc{4.5em}|wc{5.5em}?wc{0.923em}|wc{0.923em}|wc{0.923em}|wc{0.923em}|wc{0.923em}|wc{0.923em}|wc{0.923em}|wc{0.923em}|wc{0.923em}|wc{0.923em}|wc{0.923em}|wc{0.923em}|wc{0.923em}|}
        \cline{3-15}
        \multicolumn{1}{c}{} & \multicolumn{1}{c|}{} & \rotatebox[origin=c]{80}{ \ A counter \ } & \rotatebox[origin=c]{80}{ \ A cup \ } & \rotatebox[origin=c]{80}{ \ A liquid \ } & \rotatebox[origin=c]{80}{ \ A road \ } & \rotatebox[origin=c]{80}{ \ A wall \ } & \rotatebox[origin=c]{80}{ \ An animal \ } & \rotatebox[origin=c]{80}{ \ Blur \ } & \rotatebox[origin=c]{80}{ \ Chairs \ } & \rotatebox[origin=c]{80}{ \ Dots \ } & \rotatebox[origin=c]{80}{ \ Humans \ } & \rotatebox[origin=c]{80}{ \ Lights \ } & \rotatebox[origin=c]{80}{ \ Stripes \ } & \rotatebox[origin=c]{80}{ \ Tables \ } \\
        \cline{1-2}
        \Cline{2pt}{3-15}
        \multirow{5}{*}{\rotatebox[origin=c]{40}{\textbf{Benchmark}}} & \textbf{Input I} & \noj & \noj & \noj & \noj & \noj & \yesj & \noj & \noj & \yesj & \noj & \noj & \noj & \noj \\
        \cline{2-15}
         & \textbf{R to I} & \noj & \noj & \noj & \noj & \noj & \noj & \yesj & \noj & \noj & \noj & \noj & \yesj & \noj \\
        \cline{2-15}
         & \textbf{R\&I to I} & \noj & \noj & \yesj & \noj & \noj & \yesj & \noj & \noj & \noj & \noj & \noj & \noj & \noj \\
        \cline{2-15}
         & \textbf{T to I} & \yesj & \noj & \noj & \noj & \yesj & \noj & \noj & \yesj & \noj & \yesj & \yesj & \noj & \yesj \\
        \cline{2-15}
         & \textbf{T\&I to I} & \yesj & \yesj & \noj & \noj & \yesj & \noj & \noj & \noj & \noj & \yesj & \yesj & \noj & \noj \\
        \Cline{2pt}{1-15}
        \multirow{4}{*}{\rotatebox[origin=c]{40}{\textbf{Ours}}} & \textbf{A to I} & \noj & \noj & \noj & \noj & \noj & \noj & \noj & \noj & \yesj & \noj & \noj & \noj & \noj \\
        \cline{2-15}
         & \textbf{A\&T to I} & \noj & \noj & \noj & \noj & \noj & \noj & \noj & \noj & \yesj & \noj & \noj & \noj & \noj \\
        \cline{2-15}
         & \textbf{A\&I to I} & \noj & \noj & \noj & \noj & \noj & \noj & \noj & \noj & \yesj & \noj & \noj & \noj & \noj \\
        \cline{2-15}
         & \textbf{A\&T\&I to I} & \noj & \noj & \noj & \noj & \noj & \noj & \noj & \noj & \yesj & \noj & \noj & \noj & \noj \\
        \Cline{2pt}{1-15}
        \multirow{4}{*}{\rotatebox[origin=c]{40}{\textbf{ChatGPT}}} & \textbf{A to I} & \noj & \noj & \noj & \noj & \noj & \noj & \noj & \noj & \yesj & \noj & \noj & \noj & \noj \\
        \cline{2-15}
         & \textbf{A\&T to I} & \noj & \noj & \noj & \noj & \noj & \noj & \noj & \noj & \noj & \noj & \noj & \noj & \noj \\
        \cline{2-15}
         & \textbf{A\&I to I} & \noj & \noj & \noj & \noj & \noj & \noj & \yesj & \noj & \yesj & \noj & \noj & \noj & \noj \\
        \cline{2-15}
         & \textbf{A\&T\&I to I} & \noj & \noj & \noj & \noj & \noj & \noj & \noj & \noj & \yesj & \noj & \noj & \noj & \noj \\
        \Cline{2pt}{1-15}
        \multirow{4}{*}{\rotatebox[origin=c]{40}{\textbf{DeepSeek}}} & \textbf{A to I} & \noj & \noj & \noj & \noj & \yesj & \noj & \noj & \noj & \noj & \noj & \noj & \noj & \noj \\
        \cline{2-15}
         & \textbf{A\&T to I} & \noj & \noj & \noj & \noj & \yesj & \noj & \noj & \noj & \noj & \noj & \noj & \noj & \noj \\
        \cline{2-15}
         & \textbf{A\&I to I} & \noj & \noj & \noj & \noj & \noj & \noj & \noj & \noj & \noj & \noj & \noj & \noj & \noj \\
        \cline{2-15}
         & \textbf{A\&T\&I to I} & \noj & \noj & \noj & \noj & \yesj & \noj & \noj & \noj & \noj & \noj & \noj & \noj & \noj \\
        \Cline{2pt}{1-15}
        \multirow{4}{*}{\rotatebox[origin=c]{40}{\textbf{Gemini}}} & \textbf{A to I} & \noj & \noj & \noj & \noj & \noj & \noj & \noj & \noj & \noj & \noj & \noj & \noj & \noj \\
        \cline{2-15}
         & \textbf{A\&T to I} & \noj & \noj & \yesj & \yesj & \noj & \noj & \noj & \noj & \noj & \noj & \noj & \noj & \noj \\
        \cline{2-15}
         & \textbf{A\&I to I} & \noj & \noj & \noj & \noj & \noj & \noj & \yesj & \noj & \yesj & \noj & \noj & \noj & \noj \\
        \cline{2-15}
         & \textbf{A\&T\&I to I} & \noj & \noj & \noj & \noj & \noj & \noj & \noj & \noj & \noj & \noj & \noj & \yesj & \noj \\
        \Cline{2pt}{1-15}
        \multirow{4}{*}{\rotatebox[origin=c]{40}{\textbf{Grok}}} & \textbf{A to I} & \noj & \noj & \noj & \noj & \noj & \noj & \noj & \noj & \yesj & \noj & \noj & \noj & \noj \\
        \cline{2-15}
         & \textbf{A\&T to I} & \noj & \noj & \noj & \noj & \noj & \noj & \noj & \yesj & \noj & \noj & \noj & \noj & \noj \\
        \cline{2-15}
         & \textbf{A\&I to I} & \noj & \noj & \noj & \noj & \noj & \noj & \yesj & \noj & \noj & \noj & \noj & \noj & \noj \\
        \cline{2-15}
         & \textbf{A\&T\&I to I} & \noj & \noj & \noj & \noj & \noj & \noj & \noj & \yesj & \noj & \noj & \noj & \noj & \noj \\
        \Cline{2pt}{1-15}
        \multirow{4}{*}{\rotatebox[origin=c]{40}{\textbf{\shortstack{Average\\Chatbots}}}} & \textbf{A to I} & \noj & \noj & \noj & \noj & \noj & \noj & \noj & \noj & \yesj & \noj & \noj & \noj & \noj \\
        \cline{2-15}
         & \textbf{A\&T to I} & \yesj & \yesj & \yesj & \noj & \noj & \noj & \noj & \noj & \noj & \noj & \noj & \noj & \noj \\
        \cline{2-15}
         & \textbf{A\&I to I} & \noj & \noj & \noj & \noj & \noj & \noj & \noj & \noj & \noj & \noj & \noj & \yesj & \noj \\
        \cline{2-15}
         & \textbf{A\&T\&I to I} & \yesj & \noj & \noj & \noj & \yesj & \noj & \noj & \noj & \noj & \noj & \yesj & \yesj & \noj \\
        \Cline{2pt}{1-15}
        \multirow{4}{*}{\rotatebox[origin=c]{40}{\textbf{\shortstack{Average\\Everyone}}}} & \textbf{A to I} & \noj & \noj & \noj & \noj & \noj & \noj & \noj & \noj & \yesj & \noj & \noj & \noj & \noj \\
        \cline{2-15}
         & \textbf{A\&T to I} & \noj & \noj & \noj & \noj & \noj & \noj & \noj & \noj & \yesj & \noj & \noj & \noj & \noj \\
        \cline{2-15}
         & \textbf{A\&I to I} & \noj & \noj & \noj & \noj & \noj & \noj & \noj & \noj & \yesj & \noj & \noj & \noj & \noj \\
        \cline{2-15}
         & \textbf{A\&T\&I to I} & \noj & \noj & \noj & \noj & \noj & \noj & \noj & \noj & \yesj & \noj & \noj & \noj & \noj \\
        \hline
    \end{tabular}
\end{center}
\vspace*{-4mm}
\caption{Breakdown of the presence of elements in the images from Figures \ref{fig:Benchmarks_2} and \ref{fig:Generation_2}. \yesj{} means the element is at least somewhat visible in the respective image; otherwise, it is not.}
\label{tab:Generation_elements_2}
\end{table}

Finally, let us compare two representative and distinct cases in our generations.\footnote{We have published all of the images we created (described in Subsubsection \ref{subsubsec:Evaluating_the_Audio_Encoders}), together with all the input material, in the following page: \url{https://jorvan758.github.io/A-SD-Alt/}.}

Let us guide the attention to Figures \ref{fig:Benchmarks_1} and \ref{fig:Generation_1}, and Table \ref{tab:Generation_elements_1} to assess the first case.
Here we have solid evidence that our audio encoder yields values too extreme to actually generate anything coherent (even when averaged with all the other encoders, as we can check in the \textbf{Average Everyone} row).
Although no audio encoder is capable of generating images with quality nor semantic relationship comparable to that of the text encoder on their own, some actually are able to mix better with the embeddings of the text encoder.
The best audio encoder in the latter is the Grok audio encoder, whose generations with the text encoder actually depict kids interacting.
The DeepSeek audio encoder also shows some positive elements, but not as good as Grok.
Regardless, when combining all the audio encoders that come from chatbots (in the \textbf{Average Chatbots} row), we can actually see some more interesting compositions and also a capacity to merge somewhat constructively with the original text encoder.

Now, let us inspect Figures \ref{fig:Benchmarks_2} and \ref{fig:Generation_2}, and Table \ref{tab:Generation_elements_2} to analyze the second and last case.
Once again, proving it was not a fluke, the generations related to our audio encoder are a pure sort of colorful and indistinguishable noise.
In contrast, even if the images adhere less semantically to what one might expect from the input, some audio encoders are still able to collaborate positively with the text encoder.
One more time, the Grok audio encoder seems to do best at the latter, but now even better results are achieved by averaging the embeddings of all the audio encoders.

Similar results to these two cases can be seen in the rest of the generations, so there is no point in reviewing more examples in this document.

In summary, based on the test metrics from Table \ref{tab:Tests}, we would label the Gemini audio encoder as the best one, while in the actual generations the Grok audio encoder seems to have performed better.
Nevertheless, no audio encoder stood out particularly great in any domain and it is slightly worrying that all the architectures of the chatbots are so similar (specially comparing DeepSeek and Grok), while there is no clearly know solution they could have learned it from.
Not having access to the full datasets and architectures of most of the tested chatbots, the latter is a pending question that we must leave to the developers of these models.

Our suspicion is that significantly better results can be achieved with a better architecture, so the task remains open for future chatbots to tackle it.
For new attempts at this, we suggest keeping the 1 \medidas{} length on the audios to truly challenge the respective chatbot(s).
This is because, considering that a larger context window is advisable to get significantly better results, this relatively short length probably will not become massively adopted and thus no universal solution should be defined soon (additionally, the smaller the input, the less parameters the model should require).
However, we suspect that ignoring the noise in our data may have raised the difficulty excessively.
Thus, we recommend a more refined dataset, as well as a longer number of epochs, in order to truly bring out the potential of the designed neural networks.
Possibly, it could be convenient to also include techniques to induce the so called grokking that has been noticed in recent years with certain neural networks \cite{Grokking, Towards_understanding_grokking}.
\section{Conclusions}\label{sec:Conclusions}

In recent years, the field of generative models has seen tremendous advances, yet most works have focused on text-to-image \cite{Text-to-image_Diffusion_Models, High-Resolution_Image_Synthesis, SDXL, Imagen3, Flux}.
Audio-to-image generation remains relatively underexplored, despite evidence that audio signals carry rich semantic information that could guide visual content creation \cite{Audio_deepfakes, Deep_Audio-visual_Learning, A_survey_of_multimodal_deep_generative_models}.

Concurrently, LLMs and chatbots have demonstrated strong coding capabilities, but many benchmarks have become saturated as models rapidly approach perfect scores \cite{Line_Goes_Up, Web-Bench, Inadequacies_of_Large_Language_Model_Benchmarks}.

Motivated by these gaps, in this study, we inspected the coding capabilities of five chatbots (namely, ChatGPT o3-mini \cite{OpenAI_o3-mini}, Claude 3.7 Sonnet \cite{Claude_3.7}, DeepSeek-R1 \cite{DeepSeek-R1}, Gemini 2.5 Pro Preview 03-25 \cite{Gemini_2.5}, and Grok 3 \cite{Grok_3_Beta}) by prompting them to generate audio encoders that replace the text encoder of Stable Diffusion 1.5.
Despite being a novel challenge, most of them were able to accomplish the base task successfully, being Claude 3.7 Sonnet the only one that failed at this.
Regardless, as our tests have shown, the resulting architectures ended up being far from ideal and suspiciously similar (specially compared to one designed by us).
That aside, we found that the audio encoder of Gemini 2.5 Pro Preview 03-25 performed the best overall in the metrics, while the one designed by Grok 3 worked better in the actual image generations (particularly when paired with the original text encoder).

This is the very first iteration of this specific sort of competition and a few questions linger for future editions:

\begin{enumerate}
    \item What happens with the consistency when testing multiple times? Are the changes in the architectures from one attempt to another significant?
    \item How would the performance of the audio encoders we obtained improve with more training epochs, less noisy data and more observations?
    \item How much better would be the architectures proposed by more modern chatbots or even with connection to the internet?
    \item Why did all the chatbots incorporated transformer encoders and why are all their architectures so similar (specially the ones from Grok 3 and DeepSeek-R1)?
    \item What prompt engineering techniques \cite{A_Systematic_Survey_of_Prompt_Engineering} can we leverage to improve the generation prompt for the encoders (either by maximizing the quality of the results, or by giving even fairer conditions to all chatbots)?
\end{enumerate}

Finally, it is a small concern of us that this research will be incorporated in the training pipeline of some chatbots, giving them a some sort of unfair edge.
Due to this, more focused tests like this one should be defined and conducted to keep investigating these chatbots and the new ones to come, with fair and meaningful conditions.

\section*{Acknowledgments}\label{sec:Acknowledgments}

We want to thank the Universidad Adolfo Ibánez (UAI), as the access they provided us to use their high-performance computer (project ANID EQM220152) was indispensable to properly carry out these experiments.

\newpage
%\addcontentsline{toc}{section}{Referencias}
%\printbibliography
\bibliographystyle{plain}
\bibliography{citas.bib}

\begin{thebibliography}{100}

\bibitem{Freesound}
Freesound.
\newblock \url{https://freesound.org/}, 2025.

\bibitem{Pexels}
Pexels.
\newblock \url{https://pexels.com/}, 2025.

\bibitem{Picryl}
Picryl.
\newblock \url{https://picryl.com/}, 2025.

\bibitem{Pixabay}
Pixabay.
\newblock \url{https://pixabay.com/}, 2025.

\bibitem{Rawpixel}
Rawpixel.
\newblock \url{https://rawpixel.com/}, 2025.

\bibitem{MusicLM}
Andrea Agostinelli, Timo~I. Denk, Zalán Borsos, Jesse Engel, Mauro Verzetti,
  Antoine Caillon, Qingqing Huang, Aren Jansen, Adam Roberts, Marco
  Tagliasacchi, Matt Sharifi, Neil Zeghidour, and Christian Frank.
\newblock {MusicLM: Generating Music From Text}.
\newblock {\em {ArXiv}}, 2301.11325, 2023.

\bibitem{Dont_Just_Assume}
Aishwarya Agrawal, Dhruv Batra, Devi Parikh, and Aniruddha Kembhavi.
\newblock {Don't Just Assume; Look and Answer: Overcoming Priors for Visual
  Question Answering}.
\newblock In {\em {Proceedings of the 2018 IEEE Conference on Computer Vision
  and Pattern Recognition}}, pages 4971--4980, 2018.

\bibitem{Mistral_Models}
Mistral AI.
\newblock {Mistral Models}, 2024.

\bibitem{transcripter_generation}
Fatima Ansari, Ramsakal Gupta, Uday Singh, and Fahimur Shaikh.
\newblock {Transcripter-Generation of the transcript from audio to text using
  Deep Learning}.
\newblock {\em {International Journal of Computer Sciences and Engineering}},
  7(1):770--773, 2019.

\bibitem{Claude}
Anthropic.
\newblock {The Claude 3 Model Family: Opus, Sonnet, Haiku}, 2024.

\bibitem{Claude_3.7}
Anthropic.
\newblock {Claude 3.7 Sonnet and Claude Code}, 2025.

\bibitem{AudioSetCaps}
Jisheng Bai, Haohe Liu, Mou Wang, Dongyuan Shi, Mark Plumbley, Woon-Seng Gan,
  and Jianfeng Chen.
\newblock {AudioSetCaps: An Enriched Audio-Caption Dataset using Automated
  Generation Pipeline with Large Audio and Language Models}.
\newblock In {\em {Audio Imagination: NeurIPS 2024 Workshop AI-Driven Speech,
  Music, and Sound Generation}}, 2024.

\bibitem{Are_Models_Biased_on_Text}
Catarina~G Belém, Preethi Seshadri, Yasaman Razeghi, and Sameer Singh.
\newblock {Are Models Biased on Text without Gender-related Language?}
\newblock In {\em {Proceedings of the 12th International Conference on Learning
  Representations}}, 2024.

\bibitem{Image_inpainting}
Marcelo Bertalmío, Guillermo Sapiro, Vicent Caselles, and C.~Ballester.
\newblock {Image inpainting}.
\newblock In {\em {Proceedings of the 27th Internationl Conference on Computer
  Graphics and Interactive Techniques Conference}}, pages 417--424, 2000.

\bibitem{Improving_Image_Generation_with_Better_Captions}
James Betker, Gabriel Goh, Li~Jing, Tim Brooks, Jianfeng Wang, Linjie Li, Long
  Ouyang, Juntang Zhuang, Joyce Lee, Yufei Guo, Wesam Manassra, Prafulla
  Dhariwal, Casey Chu, Yunxin Jiao, and Aditya Ramesh.
\newblock {Improving Image Generation with Better Captions}.
\newblock 2023.

\bibitem{RenAIssance}
Fengxiang Bie, Yibo Yang, Zhongzhu Zhou, Adam Ghanem, Minjia Zhang, Zhewei Yao,
  Xiaoxia Wu, Connor Holmes, Pareesa Golnari, David~A. Clifton, Yuxiong He,
  Dacheng Tao, and Shuaiwen~Leon Song.
\newblock {RenAIssance: A Survey into AI Text-to-Image Generation in the Era of
  Large Model}.
\newblock {\em {ArXiv}}, 2309.00810, 2023.

\bibitem{A_Survey_on_Generative_Diffusion_Models}
Hanqun Cao, Cheng Tan, Zhangyang Gao, Yilun Xu, Guangyong Chen, Pheng-Ann Heng,
  and Stan~Z. Li.
\newblock {A Survey on Generative Diffusion Models}.
\newblock {\em {IEEE Transactions on Knowledge and Data Engineering}},
  36(7):2814--2830, 2024.

\bibitem{A_contemporary_review_on_chatbots}
Avyay Casheekar, Archit Lahiri, Kanishk Rath, Kaushik~Sanjay Prabhakar, and
  Kathiravan Srinivasan.
\newblock {A contemporary review on chatbots, AI-powered virtual conversational
  agents, ChatGPT: Applications, open challenges and future research
  directions}.
\newblock {\em {Computer Science Review}}, 52, 2024.

\bibitem{The_Partial_Least_Squares_Approach_to_Structural}
Wynne Chin and G.~A. Marcoulides.
\newblock {The Partial Least Squares Approach to Structural Equation Modeling}.
\newblock {\em {Modern Methods for Business Research}}, 8:295--358, 1998.

\bibitem{Veo}
Google DeepMind.
\newblock {Veo}, 2024.

\bibitem{DeepSeek-R1}
DeepSeek-AI.
\newblock {DeepSeek-R1: Incentivizing Reasoning Capability in LLMs via
  Reinforcement Learning}.
\newblock {\em {ArXiv}}, 2501.12948, 2025.

\bibitem{A_Survey_of_On-Device_Machine_Learning}
Sauptik Dhar, Junyao Guo, Jiayi~(Jason) Liu, Samarth Tripathi, Unmesh Kurup,
  and Mohak Shah.
\newblock {A Survey of On-Device Machine Learning: An Algorithms and Learning
  Theory Perspective}.
\newblock {\em {ACM Transactions on Internet of Things}}, 2(3), 2021.

\bibitem{Jukebox}
Prafulla Dhariwal, Heewoo Jun, Christine Payne, Jong~Wook Kim, Alec Radford,
  and Ilya Sutskever.
\newblock {Jukebox: A Generative Model for Music}.
\newblock {\em {ArXiv}}, 2005.00341, 2020.

\bibitem{The_Llama_3_Herd_of_Models}
Abhimanyu Dubey, Abhinav Jauhri, Abhinav Pandey, Abhishek Kadian, Ahmad
  Al-Dahle, Aiesha Letman, Akhil Mathur, Alan Schelten, Amy Yang, Angela Fan,
  Anirudh Goyal, Anthony Hartshorn, Aobo Yang, Archi Mitra, Archie Sravankumar,
  Artem Korenev, Arthur Hinsvark, Arun Rao, Aston Zhang, Aurelien Rodriguez,
  Austen Gregerson, Ava Spataru, Baptiste Roziere, Bethany Biron, Binh Tang,
  Bobbie Chern, Charlotte Caucheteux, Chaya Nayak, Chloe Bi, Chris Marra, Chris
  McConnell, Christian Keller, Christophe Touret, Chunyang Wu, Corinne Wong,
  Cristian~Canton Ferrer, Cyrus Nikolaidis, Damien Allonsius, Daniel Song,
  Danielle Pintz, Danny Livshits, David Esiobu, Dhruv Choudhary, Dhruv Mahajan,
  Diego Garcia-Olano, Diego Perino, Dieuwke Hupkes, Egor Lakomkin, Ehab
  AlBadawy, Elina Lobanova, Emily Dinan, Eric~Michael Smith, Filip Radenovic,
  Frank Zhang, Gabriel Synnaeve, Gabrielle Lee, Georgia~Lewis Anderson, Graeme
  Nail, Gregoire Mialon, Guan Pang, Guillem Cucurell, Hailey Nguyen, Hannah
  Korevaar, Hu~Xu, Hugo Touvron, Iliyan Zarov, Imanol~Arrieta Ibarra, Isabel
  Kloumann, Ishan Misra, Ivan Evtimov, Jade Copet, Jaewon Lee, Jan Geffert,
  Jana Vranes, Jason Park, Jay Mahadeokar, Jeet Shah, Jelmer van~der Linde,
  Jennifer Billock, Jenny Hong, Jenya Lee, Jeremy Fu, Jianfeng Chi, Jianyu
  Huang, Jiawen Liu, Jie Wang, Jiecao Yu, Joanna Bitton, Joe Spisak, Jongsoo
  Park, Joseph Rocca, Joshua Johnstun, Joshua Saxe, Junteng Jia, Kalyan~Vasuden
  Alwala, Kartikeya Upasani, Kate Plawiak, Ke~Li, Kenneth Heafield, Kevin
  Stone, Khalid El-Arini, Krithika Iyer, Kshitiz Malik, Kuenley Chiu, Kunal
  Bhalla, Lauren Rantala-Yeary, Laurens van~der Maaten, Lawrence Chen, Liang
  Tan, Liz Jenkins, Louis Martin, Lovish Madaan, Lubo Malo, Lukas Blecher,
  Lukas Landzaat, Luke de~Oliveira, Madeline Muzzi, Mahesh Pasupuleti, Mannat
  Singh, Manohar Paluri, Marcin Kardas, Mathew Oldham, Mathieu Rita, Maya
  Pavlova, Melanie Kambadur, Mike Lewis, Min Si, Mitesh~Kumar Singh, Mona
  Hassan, Naman Goyal, Narjes Torabi, Nikolay Bashlykov, Nikolay Bogoychev,
  Niladri Chatterji, Olivier Duchenne, Onur Çelebi, Patrick Alrassy, Pengchuan
  Zhang, Pengwei Li, Petar Vasic, Peter Weng, Prajjwal Bhargava, Pratik Dubal,
  Praveen Krishnan, Punit~Singh Koura, Puxin Xu, Qing He, Qingxiao Dong,
  Ragavan Srinivasan, Raj Ganapathy, Ramon Calderer, Ricardo~Silveira Cabral,
  Robert Stojnic, Roberta Raileanu, Rohit Girdhar, Rohit Patel, Romain
  Sauvestre, Ronnie Polidoro, Roshan Sumbaly, Ross Taylor, Ruan Silva, Rui Hou,
  Rui Wang, Saghar Hosseini, Sahana Chennabasappa, Sanjay Singh, Sean Bell,
  Seohyun~Sonia Kim, Sergey Edunov, Shaoliang Nie, Sharan Narang, Sharath
  Raparthy, Sheng Shen, Shengye Wan, Shruti Bhosale, Shun Zhang, Simon
  Vandenhende, Soumya Batra, Spencer Whitman, Sten Sootla, Stephane Collot,
  Suchin Gururangan, Sydney Borodinsky, Tamar Herman, Tara Fowler, Tarek
  Sheasha, Thomas Georgiou, Thomas Scialom, Tobias Speckbacher, Todor Mihaylov,
  Tong Xiao, Ujjwal Karn, Vedanuj Goswami, Vibhor Gupta, Vignesh Ramanathan,
  Viktor Kerkez, Vincent Gonguet, Virginie Do, Vish Vogeti, Vladan Petrovic,
  Weiwei Chu, Wenhan Xiong, Wenyin Fu, Whitney Meers, Xavier Martinet, Xiaodong
  Wang, Xiaoqing~Ellen Tan, Xinfeng Xie, Xuchao Jia, Xuewei Wang, Yaelle
  Goldschlag, Yashesh Gaur, Yasmine Babaei, Yi~Wen, Yiwen Song, Yuchen Zhang,
  Yue Li, Yuning Mao, Zacharie~Delpierre Coudert, Zheng Yan, Zhengxing Chen,
  Zoe Papakipos, Aaditya Singh, Aaron Grattafiori, Abha Jain, Adam Kelsey, Adam
  Shajnfeld, Adithya Gangidi, Adolfo Victoria, Ahuva Goldstand, Ajay Menon,
  Ajay Sharma, Alex Boesenberg, Alex Vaughan, Alexei Baevski, Allie Feinstein,
  Amanda Kallet, Amit Sangani, Anam Yunus, Andrei Lupu, Andres Alvarado, Andrew
  Caples, Andrew Gu, Andrew Ho, Andrew Poulton, Andrew Ryan, Ankit Ramchandani,
  Annie Franco, Aparajita Saraf, Arkabandhu Chowdhury, Ashley Gabriel, Ashwin
  Bharambe, Assaf Eisenman, Azadeh Yazdan, Beau James, Ben Maurer, Benjamin
  Leonhardi, Bernie Huang, Beth Loyd, Beto~De Paola, Bhargavi Paranjape, Bing
  Liu, Bo~Wu, Boyu Ni, Braden Hancock, Bram Wasti, Brandon Spence, Brani
  Stojkovic, Brian Gamido, Britt Montalvo, Carl Parker, Carly Burton, Catalina
  Mejia, Changhan Wang, Changkyu Kim, Chao Zhou, Chester Hu, Ching-Hsiang Chu,
  Chris Cai, Chris Tindal, Christoph Feichtenhofer, Damon Civin, Dana Beaty,
  Daniel Kreymer, Daniel Li, Danny Wyatt, David Adkins, David Xu, Davide
  Testuggine, Delia David, Devi Parikh, Diana Liskovich, Didem Foss, Dingkang
  Wang, Duc Le, Dustin Holland, Edward Dowling, Eissa Jamil, Elaine Montgomery,
  Eleonora Presani, Emily Hahn, Emily Wood, Erik Brinkman, Esteban Arcaute,
  Evan Dunbar, Evan Smothers, Fei Sun, Felix Kreuk, Feng Tian, Firat Ozgenel,
  Francesco Caggioni, Francisco Guzmán, Frank Kanayet, Frank Seide,
  Gabriela~Medina Florez, Gabriella Schwarz, Gada Badeer, Georgia Swee, Gil
  Halpern, Govind Thattai, Grant Herman, Grigory Sizov, Guangyi, Zhang, Guna
  Lakshminarayanan, Hamid Shojanazeri, Han Zou, Hannah Wang, Hanwen Zha, Haroun
  Habeeb, Harrison Rudolph, Helen Suk, Henry Aspegren, Hunter Goldman, Ibrahim
  Damlaj, Igor Molybog, Igor Tufanov, Irina-Elena Veliche, Itai Gat, Jake
  Weissman, James Geboski, James Kohli, Japhet Asher, Jean-Baptiste Gaya, Jeff
  Marcus, Jeff Tang, Jennifer Chan, Jenny Zhen, Jeremy Reizenstein, Jeremy
  Teboul, Jessica Zhong, Jian Jin, Jingyi Yang, Joe Cummings, Jon Carvill, Jon
  Shepard, Jonathan McPhie, Jonathan Torres, Josh Ginsburg, Junjie Wang, Kai
  Wu, Kam~Hou U, Karan Saxena, Karthik Prasad, Kartikay Khandelwal, Katayoun
  Zand, Kathy Matosich, Kaushik Veeraraghavan, Kelly Michelena, Keqian Li, Kun
  Huang, Kunal Chawla, Kushal Lakhotia, Kyle Huang, Lailin Chen, Lakshya Garg,
  Lavender A, Leandro Silva, Lee Bell, Lei Zhang, Liangpeng Guo, Licheng Yu,
  Liron Moshkovich, Luca Wehrstedt, Madian Khabsa, Manav Avalani, Manish Bhatt,
  Maria Tsimpoukelli, Martynas Mankus, Matan Hasson, Matthew Lennie, Matthias
  Reso, Maxim Groshev, Maxim Naumov, Maya Lathi, Meghan Keneally, Michael~L.
  Seltzer, Michal Valko, Michelle Restrepo, Mihir Patel, Mik Vyatskov, Mikayel
  Samvelyan, Mike Clark, Mike Macey, Mike Wang, Miquel~Jubert Hermoso,
  Mo~Metanat, Mohammad Rastegari, Munish Bansal, Nandhini Santhanam, Natascha
  Parks, Natasha White, Navyata Bawa, Nayan Singhal, Nick Egebo, Nicolas
  Usunier, Nikolay~Pavlovich Laptev, Ning Dong, Ning Zhang, Norman Cheng, Oleg
  Chernoguz, Olivia Hart, Omkar Salpekar, Ozlem Kalinli, Parkin Kent, Parth
  Parekh, Paul Saab, Pavan Balaji, Pedro Rittner, Philip Bontrager, Pierre
  Roux, Piotr Dollar, Polina Zvyagina, Prashant Ratanchandani, Pritish Yuvraj,
  Qian Liang, Rachad Alao, Rachel Rodriguez, Rafi Ayub, Raghotham Murthy, Raghu
  Nayani, Rahul Mitra, Raymond Li, Rebekkah Hogan, Robin Battey, Rocky Wang,
  Rohan Maheswari, Russ Howes, Ruty Rinott, Sai~Jayesh Bondu, Samyak Datta,
  Sara Chugh, Sara Hunt, Sargun Dhillon, Sasha Sidorov, Satadru Pan, Saurabh
  Verma, Seiji Yamamoto, Sharadh Ramaswamy, Shaun Lindsay, Shaun Lindsay, Sheng
  Feng, Shenghao Lin, Shengxin~Cindy Zha, Shiva Shankar, Shuqiang Zhang,
  Shuqiang Zhang, Sinong Wang, Sneha Agarwal, Soji Sajuyigbe, Soumith Chintala,
  Stephanie Max, Stephen Chen, Steve Kehoe, Steve Satterfield, Sudarshan
  Govindaprasad, Sumit Gupta, Sungmin Cho, Sunny Virk, Suraj Subramanian,
  Sy~Choudhury, Sydney Goldman, Tal Remez, Tamar Glaser, Tamara Best, Thilo
  Kohler, Thomas Robinson, Tianhe Li, Tianjun Zhang, Tim Matthews, Timothy
  Chou, Tzook Shaked, Varun Vontimitta, Victoria Ajayi, Victoria Montanez,
  Vijai Mohan, Vinay~Satish Kumar, Vishal Mangla, Vítor Albiero, Vlad Ionescu,
  Vlad Poenaru, Vlad~Tiberiu Mihailescu, Vladimir Ivanov, Wei Li, Wenchen Wang,
  Wenwen Jiang, Wes Bouaziz, Will Constable, Xiaocheng Tang, Xiaofang Wang,
  Xiaojian Wu, Xiaolan Wang, Xide Xia, Xilun Wu, Xinbo Gao, Yanjun Chen, Ye~Hu,
  Ye~Jia, Ye~Qi, Yenda Li, Yilin Zhang, Ying Zhang, Yossi Adi, Youngjin Nam,
  Yu, Wang, Yuchen Hao, Yundi Qian, Yuzi He, Zach Rait, Zachary DeVito, Zef
  Rosnbrick, Zhaoduo Wen, Zhenyu Yang, and Zhiwei Zhao.
\newblock {The Llama 3 Herd of Models}.
\newblock {\em {ArXiv}}, 2407.21783, 2024.

\bibitem{Grok_Gemini_ChatGPT_and_DeepSeek}
Murillo Edson~de Carvalho~Souza and Li~Weigang.
\newblock {Grok, Gemini, ChatGPT and DeepSeek: Comparison and Applications in
  Conversational Artificial Intelligence}.
\newblock 2025.

\bibitem{Image_Generation}
Mohamed Elasri, Omar Elharrouss, Somaya Al-Maadeed, and Hamid Tairi.
\newblock {Image Generation: A Review}.
\newblock {\em {Neural Processing Letters}}, 54(5):4609--4646, 2022.

\bibitem{Stable_diffusion_3}
Patrick Esser, Sumith Kulal, Andreas Blattmann, Rahim Entezari, Jonas Müller,
  Harry Saini, Yam Levi, Dominik Lorenz, Axel Sauer, Frederic Boesel, Dustin
  Podell, Tim Dockhorn, Zion English, Kyle Lacey, Alex Goodwin, Yannik Marek,
  and Robin Rombach.
\newblock {Scaling Rectified Flow Transformers for High-Resolution Image
  Synthesis}.
\newblock {\em {ArXiv}}, 2403.03206, 2024.

\bibitem{Learning_From_Noisy_Correspondence}
Zerun Feng, Zhimin Zeng, Caili Guo, Zheng Li, and Lin Hu.
\newblock {Learning From Noisy Correspondence With Tri-Partition for
  Cross-Modal Matching}.
\newblock {\em {IEEE Transactions on Multimedia}}, 26:3884--3896, 2024.

\bibitem{Line_Goes_Up}
James Fodor.
\newblock {Line Goes Up? Inherent Limitations of Benchmarks for Evaluating
  Large Language Models}.
\newblock {\em {ArXiv}}, 2502.14318, 2025.

\bibitem{Creativity_and_Machine_Learning}
Giorgio Franceschelli and Mirco Musolesi.
\newblock {Creativity and Machine Learning: A Survey}.
\newblock {\em {ArXiv}}, 2104.02726, 2022.

\bibitem{The_Pile}
Leo Gao, Stella Biderman, Sid Black, Laurence Golding, Travis Hoppe, Charles
  Foster, Jason Phang, Horace He, Anish Thite, Noa Nabeshima, Shawn Presser,
  and Connor Leahy.
\newblock {The Pile: An 800GB Dataset of Diverse Text for Language Modeling}.
\newblock {\em {ArXiv}}, 2101.00027, 2020.

\bibitem{Imagebind}
Rohit Girdhar, Alaaeldin El-Nouby, Zhuang Liu, Mannat Singh, Kalyan~Vasudev
  Alwala, Armand Joulin, and Ishan Misra.
\newblock {ImageBind: One Embedding Space To Bind Them All}.
\newblock {\em {ArXiv}}, 2305.05665, 2023.

\bibitem{Mamba}
Albert Gu and Tri Dao.
\newblock {Mamba: Linear-Time Sequence Modeling with Selective State Spaces}.
\newblock {\em {ArXiv}}, 2312.00752, 2024.

\bibitem{AudioCLIP}
Andrey Guzhov, Federico Raue, Jörn Hees, and Andreas Dengel.
\newblock {AudioCLIP: Extending CLIP to Image, Text and Audio}.
\newblock {\em {ArXiv}}, 2106.13043, 2021.

\bibitem{Deep_Residual_Learning_for_Image}
Kaiming He, Xiangyu Zhang, Shaoqing Ren, and Jian Sun.
\newblock {Deep Residual Learning for Image Recognition}.
\newblock In {\em {Proceedings of the 2016 IEEE Conference on Computer Vision
  and Pattern Recognition}}, pages 770--778, 2016.

\bibitem{The_Use_of_Partial_Least_Squares_Path_Modeling}
Jörg Henseler, Christian~M. Ringle, and Rudolf~R. Sinkovics.
\newblock {The Use of Partial Least Squares Path Modeling in International
  Marketing}.
\newblock {\em {Advances in International Marketing}}, 20:277--319, 2009.

\bibitem{Intuitive_Multilingual}
Joanna Hong, Se~Park, and Yong Ro.
\newblock {Intuitive Multilingual Audio-Visual Speech Recognition with a
  Single-Trained Model}.
\newblock In {\em {Findings of the 2023 Conference on Empirical Methods in
  Natural Language Processing}}, pages 4886--4890, 2023.

\bibitem{Make-an-audio}
Rongjie Huang, Jiawei Huang, Dongchao Yang, Yi~Ren, Luping liu, Mingze Li,
  Zhenhui Ye, Jinglin Liu, Xiang Yin, and Zhou Zhao.
\newblock {Make-an-audio: text-to-audio generation with prompt-enhanced
  diffusion models}.
\newblock In {\em {Proceedings of the 40th International Conference on Machine
  Learning}}, pages 13916 -- 13932, 2023.

\bibitem{Nlip}
Runhui Huang, Yanxin Long, Jianhua Han, Hang Xu, Xiwen Liang, Chunjing Xu, and
  Xiaodan Liang.
\newblock {NLIP: Noise-Robust Language-Image Pre-training}.
\newblock In {\em {Proceedings of the 37th AAAI Conference on Artificial
  Intelligence}}, pages 926--934, 2023.

\bibitem{Large_Language_Models_for_Code_Generation}
Nam Huynh and Beiyu Lin.
\newblock {Large Language Models for Code Generation: A Comprehensive Survey of
  Challenges, Techniques, Evaluation, and Applications}.
\newblock {\em {ArXiv}}, 2503.01245, 2025.

\bibitem{Imagen3}
Imagen-Team-Google, :, Jason Baldridge, Jakob Bauer, Mukul Bhutani, Nicole
  Brichtova, Andrew Bunner, Kelvin Chan, Yichang Chen, Sander Dieleman, Yuqing
  Du, Zach Eaton-Rosen, Hongliang Fei, Nando de~Freitas, Yilin Gao, Evgeny
  Gladchenko, Sergio~Gómez Colmenarejo, Mandy Guo, Alex Haig, Will Hawkins,
  Hexiang Hu, Huilian Huang, Tobenna~Peter Igwe, Christos Kaplanis, Siavash
  Khodadadeh, Yelin Kim, Ksenia Konyushkova, Karol Langner, Eric Lau, Shixin
  Luo, Soňa Mokrá, Henna Nandwani, Yasumasa Onoe, Aäron van~den Oord, Zarana
  Parekh, Jordi Pont-Tuset, Hang Qi, Rui Qian, Deepak Ramachandran, Poorva
  Rane, Abdullah Rashwan, Ali Razavi, Robert Riachi, Hansa Srinivasan,
  Srivatsan Srinivasan, Robin Strudel, Benigno Uria, Oliver Wang, Su~Wang,
  Austin Waters, Chris Wolff, Auriel Wright, Zhisheng Xiao, Hao Xiong, Keyang
  Xu, Marc van Zee, Junlin Zhang, Katie Zhang, Wenlei Zhou, Konrad Zolna, Ola
  Aboubakar, Canfer Akbulut, Oscar Akerlund, Isabela Albuquerque, Nina
  Anderson, Marco Andreetto, Lora Aroyo, Ben Bariach, David Barker, Sherry Ben,
  Dana Berman, Courtney Biles, Irina Blok, Pankil Botadra, Jenny Brennan, Karla
  Brown, John Buckley, Rudy Bunel, Elie Bursztein, Christina Butterfield, Ben
  Caine, Viral Carpenter, Norman Casagrande, Ming-Wei Chang, Solomon Chang,
  Shamik Chaudhuri, Tony Chen, John Choi, Dmitry Churbanau, Nathan Clement,
  Matan Cohen, Forrester Cole, Mikhail Dektiarev, Vincent Du, Praneet Dutta,
  Tom Eccles, Ndidi Elue, Ashley Feden, Shlomi Fruchter, Frankie Garcia, Roopal
  Garg, Weina Ge, Ahmed Ghazy, Bryant Gipson, Andrew Goodman, Dawid Górny,
  Sven Gowal, Khyatti Gupta, Yoni Halpern, Yena Han, Susan Hao, Jamie Hayes,
  Amir Hertz, Ed~Hirst, Tingbo Hou, Heidi Howard, Mohamed Ibrahim, Dirichi
  Ike-Njoku, Joana Iljazi, Vlad Ionescu, William Isaac, Reena Jana, Gemma
  Jennings, Donovon Jenson, Xuhui Jia, Kerry Jones, Xiaoen Ju, Ivana Kajic,
  Christos Kaplanis, Burcu~Karagol Ayan, Jacob Kelly, Suraj Kothawade,
  Christina Kouridi, Ira Ktena, Jolanda Kumakaw, Dana Kurniawan, Dmitry Lagun,
  Lily Lavitas, Jason Lee, Tao Li, Marco Liang, Maggie Li-Calis, Yuchi Liu,
  Javier~Lopez Alberca, Peggy Lu, Kristian Lum, Yukun Ma, Chase Malik, John
  Mellor, Inbar Mosseri, Tom Murray, Aida Nematzadeh, Paul Nicholas,
  João~Gabriel Oliveira, Guillermo Ortiz-Jimenez, Michela Paganini, Tom~Le
  Paine, Roni Paiss, Alicia Parrish, Anne Peckham, Vikas Peswani, Igor
  Petrovski, Tobias Pfaff, Alex Pirozhenko, Ryan Poplin, Utsav Prabhu, Yuan Qi,
  Matthew Rahtz, Cyrus Rashtchian, Charvi Rastogi, Amit Raul, Ali Razavi,
  Sylvestre-Alvise Rebuffi, Susanna Ricco, Felix Riedel, Dirk Robinson, Pankaj
  Rohatgi, Bill Rosgen, Sarah Rumbley, Moonkyung Ryu, Anthony Salgado, Sahil
  Singla, Florian Schroff, Candice Schumann, Tanmay Shah, Brendan Shillingford,
  Kaushik Shivakumar, Dennis Shtatnov, Zach Singer, Evgeny Sluzhaev, Valerii
  Sokolov, Thibault Sottiaux, Florian Stimberg, Brad Stone, David Stutz,
  Yu-Chuan Su, Eric Tabellion, Shuai Tang, David Tao, Kurt Thomas, Gregory
  Thornton, Andeep Toor, Cristian Udrescu, Aayush Upadhyay, Cristina
  Vasconcelos, Alex Vasiloff, Andrey Voynov, Amanda Walker, Luyu Wang, Miaosen
  Wang, Simon Wang, Stanley Wang, Qifei Wang, Yuxiao Wang, Ágoston Weisz,
  Olivia Wiles, Chenxia Wu, Xingyu~Federico Xu, Andrew Xue, Jianbo Yang, Luo
  Yu, Mete Yurtoglu, Ali Zand, Han Zhang, Jiageng Zhang, Catherine Zhao, Adilet
  Zhaxybay, Miao Zhou, Shengqi Zhu, Zhenkai Zhu, Dawn Bloxwich, Mahyar Bordbar,
  Luis~C. Cobo, Eli Collins, Shengyang Dai, Tulsee Doshi, Anca Dragan, Douglas
  Eck, Demis Hassabis, Sissie Hsiao, Tom Hume, Koray Kavukcuoglu, Helen King,
  Jack Krawczyk, Yeqing Li, Kathy Meier-Hellstern, Andras Orban, Yury Pinsky,
  Amar Subramanya, Oriol Vinyals, Ting Yu, and Yori Zwols.
\newblock {Imagen 3}.
\newblock {\em {ArXiv}}, 2408.07009, 2024.

\bibitem{A_Survey_on_Large_Language_Models_for_Code_Generation}
Juyong Jiang, Fan Wang, Jiasi Shen, Sungju Kim, and Sunghun Kim.
\newblock {A Survey on Large Language Models for Code Generation}.
\newblock {\em {ArXiv}}, 2406.00515, 2024.

\bibitem{TimbreCLIP}
Nicolas Jonason and Bob L.~T. Sturm.
\newblock {TimbreCLIP: Connecting Timbre to Text and Images}.
\newblock {\em {ArXiv}}, 2211.11225, 2022.

\bibitem{Noise_Aware_Learning}
Wooyoung Kang, Jonghwan Mun, Sungjun Lee, and Byungseok Roh.
\newblock {Noise-Aware Learning from Web-Crawled Image-Text Data for Image
  Captioning}.
\newblock In {\em {Proceedings of the 2023 IEEE International Conference on
  Computer Vision}}, pages 2942--2952, 2023.

\bibitem{Gemini_2.5}
Koray Kavukcuoglu.
\newblock {Gemini 2.5: Our most intelligent AI model}, 2025.

\bibitem{Audio_deepfakes}
Zahra Khanjani, Gabrielle Watson, and Vandana~P. Janeja.
\newblock {Audio deepfakes: A survey}.
\newblock {\em {Frontiers in Big Data}}, 5, 2023.

\bibitem{Auto-Encoding_Variational_Bayes}
Diederik~P. Kingma and Max Welling.
\newblock {Auto-Encoding Variational Bayes}.
\newblock In {\em {Proceedings of the 2nd International Conference on Learning
  Representations}}, 2014.

\bibitem{Benchmarking_Cognitive_Biases}
Ryan Koo, Minhwa Lee, Vipul Raheja, Jong~Inn Park, Zae~Myung Kim, and Dongyeop
  Kang.
\newblock {Benchmarking Cognitive Biases in Large Language Models as
  Evaluators}.
\newblock {\em {ArXiv}}, 2309.17012, 2023.

\bibitem{Do_Large_Language_Models_Pay_Similar_Attention_Like_Human_Programmers}
Bonan Kou, Shengmai Chen, Zhijie Wang, Lei Ma, and Tianyi Zhang.
\newblock {Do Large Language Models Pay Similar Attention Like Human
  Programmers When Generating Code?}
\newblock {\em {Proceedings of the ACM on Software Engineering}},
  1(FSE):2261--2284, 2024.

\bibitem{AudioGen}
Felix Kreuk, Gabriel Synnaeve, Adam Polyak, Uriel Singer, Alexandre Défossez,
  Jade Copet, Devi Parikh, Yaniv Taigman, and Yossi Adi.
\newblock {AudioGen: Textually Guided Audio Generation}.
\newblock {\em {ArXiv}}, 2209.15352, 2023.

\bibitem{BindDiffusion}
Sea~AI Lab.
\newblock {BindDiffusion: One Diffusion Model to Bind Them All}, 2024.

\bibitem{Flux}
Black~Forest Labs.
\newblock {FLUX}, 2024.

\bibitem{Obtaining_acoustic_visual_and_textual_data}
Jorge~E. León and Miguel Carrasco.
\newblock {Effectively obtaining acoustic, visual and textual data from
  videos}.
\newblock {\em {ArXiv}}, 2509.05786, 2025.

\bibitem{BLIP-2}
Junnan Li, Dongxu Li, Silvio Savarese, and Steven Hoi.
\newblock {BLIP-2: Bootstrapping Language-Image Pre-training with Frozen Image
  Encoders and Large Language Models}.
\newblock {\em {ArXiv}}, 2301.12597, 2023.

\bibitem{BLIP}
Junnan Li, Dongxu Li, Caiming Xiong, and Steven Hoi.
\newblock {BLIP: Bootstrapping Language-Image Pre-training for Unified
  Vision-Language Understanding and Generation}.
\newblock {\em {ArXiv}}, 2201.12086, 2022.

\bibitem{Word_Level_Explanations}
Alexander Lin, Lucas~Monteiro Paes, Sree~Harsha Tanneru, Suraj Srinivas, and
  Himabindu Lakkaraju.
\newblock {Word-Level Explanations for Analyzing Bias in Text-to-Image Models}.
\newblock {\em {ArXiv}}, 2306.05500, 2023.

\bibitem{AudioLDM}
Haohe Liu, Zehua Chen, Yi~Yuan, Xinhao Mei, Xubo Liu, Danilo Mandic, Wenwu
  Wang, and Mark~D. Plumbley.
\newblock {AudioLDM: Text-to-Audio Generation with Latent Diffusion Models}.
\newblock In {\em {Proceedings of the 40th International Conference on Machine
  Learning}}, pages 21450--21474, 2023.

\bibitem{sora_analisis}
Yixin Liu, Kai Zhang, Yuan Li, Zhiling Yan, Chujie Gao, Ruoxi Chen, Zhengqing
  Yuan, Yue Huang, Hanchi Sun, Jianfeng Gao, Lifang He, and Lichao Sun.
\newblock {Sora: A Review on Background, Technology, Limitations, and
  Opportunities of Large Vision Models}.
\newblock {\em {ArXiv}}, 2402.17177, 2024.

\bibitem{Towards_understanding_grokking}
Ziming Liu, Ouail Kitouni, Niklas Nolte, Eric~J. Michaud, Max Tegmark, and Mike
  Williams.
\newblock {Towards understanding grokking: an effective theory of
  representation learning}.
\newblock In {\em {Proceedings of the 36th International Conference on Neural
  Information Processing Systems}}, pages 34651--34663, 2024.

\bibitem{BLAP}
Nathanaël~Perraudin Luca A~Lanzendörfer, Constantin~Pinkl and Roger
  Wattenhofer.
\newblock {BLAP: Bootstrapping Language-Audio Pre-training for Music
  Captioning}.
\newblock In {\em {Audio Imagination: NeurIPS 2024 Workshop AI-Driven Speech,
  Music, and Sound Generation}}, 2024.

\bibitem{Look_Listen_and_Answer}
Jie Ma, Min Hu, Pinghui Wang, Wangchun Sun, Lingyun Song, Hongbin Pei, Jun Liu,
  and Youtian Du.
\newblock {Look, Listen, and Answer: Overcoming Biases for Audio-Visual
  Question Answering}.
\newblock {\em {ArXiv}}, 2404.12020, 2024.

\bibitem{SD_Akashic_Records}
Maks-s.
\newblock {Stable Diffusion Akashic Records}, 2023.

\bibitem{Inadequacies_of_Large_Language_Model_Benchmarks}
Timothy~R McIntosh, Teo Susnjak, Nalin Arachchilage, Tong Liu, Dan Xu, Paul
  Watters, and Malka~N Halgamuge.
\newblock {Inadequacies of Large Language Model Benchmarks in the Era of
  Generative Artificial Intelligence}.
\newblock {\em {IEEE Transactions on Artificial Intelligence}}, pages 1--18,
  2025.

\bibitem{Mustango}
Jan Melechovsky, Zixun Guo, Deepanway Ghosal, Navonil Majumder, Dorien
  Herremans, and Soujanya Poria.
\newblock {Mustango: Toward Controllable Text-to-Music Generation}.
\newblock In {\em {Proceedings of the 2024 North American Chapter of the
  Association for Computational Linguistics}}, page 8293–8316, 2024.

\bibitem{From_Classical_Machine_Learning_to_Deep_Neural_Networks}
Ravil~I. Mukhamediev, Adilkhan Symagulov, Yan Kuchin, Kirill Yakunin, and
  Marina Yelis.
\newblock {From Classical Machine Learning to Deep Neural Networks: A
  Simplified Scientometric Review}.
\newblock {\em {Applied Sciences}}, 11(12), 2021.

\bibitem{DALLE_3}
OpenAI.
\newblock {DALL·E 3 System Card}, 2023.

\bibitem{sora_reporte}
OpenAI.
\newblock {Video generation models as world simulators}, 2024.

\bibitem{OpenAI_o3-mini}
OpenAI.
\newblock {OpenAI o3-mini}, 2025.

\bibitem{Image-to-Image_Translation}
Yingxue Pang, Jianxin Lin, Tao Qin, and Zhibo Chen.
\newblock {Image-to-Image Translation: Methods and Applications}.
\newblock {\em {IEEE Transactions on Multimedia}}, 24:3859--3881, 2022.

\bibitem{SDXL}
Dustin Podell, Zion English, Kyle Lacey, Andreas Blattmann, Tim Dockhorn, Jonas
  Müller, Joe Penna, and Robin Rombach.
\newblock {SDXL: Improving Latent Diffusion Models for High-Resolution Image
  Synthesis}.
\newblock {\em {ArXiv}}, 2307.01952, 2023.

\bibitem{Grokking}
Alethea Power, Yuri Burda, Harrison Edwards, Igor Babuschkin, and Vedant Misra.
\newblock {Grokking: Generalization Beyond Overfitting on Small Algorithmic
  Datasets}.
\newblock In {\em {Proceedings of the 1st Mathematical Reasoning in General
  Artificial Intelligence Workshop}}, 2021.

\bibitem{Mirrorgan}
Tingting Qiao, Jing Zhang, Duanqing Xu, and Dacheng Tao.
\newblock {MirrorGAN: Learning Text-To-Image Generation by Redescription}.
\newblock In {\em {Proceedings of the 2019 IEEE Conference on Computer Vision
  and Pattern Recognition}}, pages 1505--1514, 2019.

\bibitem{CLIP}
Alec Radford, Jong~Wook Kim, Chris Hallacy, Aditya Ramesh, Gabriel Goh,
  Sandhini Agarwal, Girish Sastry, Amanda Askell, Pamela Mishkin, Jack Clark,
  Gretchen Krueger, and Ilya Sutskever.
\newblock {Learning Transferable Visual Models From Natural Language
  Supervision}.
\newblock {\em {ArXiv}}, 2103.00020, 2021.

\bibitem{Learning_Transferable_Visual_Models_From_NL}
Alec Radford, Jong~Wook Kim, Chris Hallacy, Aditya Ramesh, Gabriel Goh,
  Sandhini Agarwal, Girish Sastry, Amanda Askell, Pamela Mishkin, Jack Clark,
  Gretchen Krueger, and Ilya Sutskever.
\newblock {Learning Transferable Visual Models From Natural Language
  Supervision}.
\newblock {\em {ArXiv}}, 2103.00020, 2024.

\bibitem{whisper}
Alec Radford, Jong~Wook Kim, Tao Xu, Greg Brockman, Christine McLeavey, and
  Ilya Sutskever.
\newblock {Robust Speech Recognition via Large-Scale Weak Supervision}.
\newblock In {\em {Proceedings of the 40th International Conference on Machine
  Learning}}, pages 28492--28518, 2023.

\bibitem{Zero-Shot_Text-to-Image}
Aditya Ramesh, Mikhail Pavlov, Gabriel Goh, Scott Gray, Chelsea Voss, Alec
  Radford, Mark Chen, and Ilya Sutskever.
\newblock {Zero-Shot Text-to-Image Generation}.
\newblock {\em {ArXiv}}, 2102.12092, 2021.

\bibitem{gemini_1_5}
Machel Reid, Nikolay Savinov, Denis Teplyashin, Dmitry Lepikhin, Timothy
  Lillicrap, Jean baptiste Alayrac, Radu Soricut, Angeliki Lazaridou, Orhan
  Firat, Julian Schrittwieser, Ioannis Antonoglou, Rohan Anil, Sebastian
  Borgeaud, Andrew Dai, Katie Millican, Ethan Dyer, Mia Glaese, Thibault
  Sottiaux, Benjamin Lee, Fabio Viola, Malcolm Reynolds, Yuanzhong Xu, James
  Molloy, Jilin Chen, Michael Isard, Paul Barham, Tom Hennigan, Ross McIlroy,
  Melvin Johnson, Johan Schalkwyk, Eli Collins, Eliza Rutherford, Erica
  Moreira, Kareem Ayoub, Megha Goel, Clemens Meyer, Gregory Thornton, Zhen
  Yang, Henryk Michalewski, Zaheer Abbas, Nathan Schucher, Ankesh Anand,
  Richard Ives, James Keeling, Karel Lenc, Salem Haykal, Siamak Shakeri, Pranav
  Shyam, Aakanksha Chowdhery, Roman Ring, Stephen Spencer, Eren Sezener, Luke
  Vilnis, Oscar Chang, Nobuyuki Morioka, George Tucker, Ce~Zheng, Oliver
  Woodman, Nithya Attaluri, Tomas Kocisky, Evgenii Eltyshev, Xi~Chen, Timothy
  Chung, Vittorio Selo, Siddhartha Brahma, Petko Georgiev, Ambrose Slone,
  Zhenkai Zhu, James Lottes, Siyuan Qiao, Ben Caine, Sebastian Riedel, Alex
  Tomala, Martin Chadwick, Juliette Love, Peter Choy, Sid Mittal, Neil Houlsby,
  Yunhao Tang, Matthew Lamm, Libin Bai, Qiao Zhang, Luheng He, Yong Cheng,
  Peter Humphreys, Yujia Li, Sergey Brin, Albin Cassirer, Yingjie Miao, Lukas
  Zilka, Taylor Tobin, Kelvin Xu, Lev Proleev, Daniel Sohn, Alberto Magni,
  Lisa~Anne Hendricks, Isabel Gao, Santiago Ontañón, Oskar Bunyan, Nathan
  Byrd, Abhanshu Sharma, Biao Zhang, Mario Pinto, Rishika Sinha, Harsh Mehta,
  Dawei Jia, Sergi Caelles, Albert Webson, Alex Morris, Becca Roelofs, Yifan
  Ding, Robin Strudel, Xuehan Xiong, Marvin Ritter, Mostafa Dehghani, Rahma
  Chaabouni, Abhijit Karmarkar, Guangda Lai, Fabian Mentzer, Bibo Xu, YaGuang
  Li, Yujing Zhang, Tom~Le Paine, Alex Goldin, Behnam Neyshabur, Kate Baumli,
  Anselm Levskaya, Michael Laskin, Wenhao Jia, Jack~W. Rae, Kefan Xiao, Antoine
  He, Skye Giordano, Lakshman Yagati, Jean-Baptiste Lespiau, Paul Natsev,
  Sanjay Ganapathy, Fangyu Liu, Danilo Martins, Nanxin Chen, Yunhan Xu, Megan
  Barnes, Rhys May, Arpi Vezer, Junhyuk Oh, Ken Franko, Sophie Bridgers, Ruizhe
  Zhao, Boxi Wu, Basil Mustafa, Sean Sechrist, Emilio Parisotto,
  Thanumalayan~Sankaranarayana Pillai, Chris Larkin, Chenjie Gu, Christina
  Sorokin, Maxim Krikun, Alexey Guseynov, Jessica Landon, Romina Datta,
  Alexander Pritzel, Phoebe Thacker, Fan Yang, Kevin Hui, Anja Hauth, Chih-Kuan
  Yeh, David Barker, Justin Mao-Jones, Sophia Austin, Hannah Sheahan, Parker
  Schuh, James Svensson, Rohan Jain, Vinay Ramasesh, Anton Briukhov, Da-Woon
  Chung, Tamara von Glehn, Christina Butterfield, Priya Jhakra, Matthew
  Wiethoff, Justin Frye, Jordan Grimstad, Beer Changpinyo, Charline~Le Lan,
  Anna Bortsova, Yonghui Wu, Paul Voigtlaender, Tara Sainath, Charlotte Smith,
  Will Hawkins, Kris Cao, James Besley, Srivatsan Srinivasan, Mark Omernick,
  Colin Gaffney, Gabriela Surita, Ryan Burnell, Bogdan Damoc, Junwhan Ahn,
  Andrew Brock, Mantas Pajarskas, Anastasia Petrushkina, Seb Noury, Lorenzo
  Blanco, Kevin Swersky, Arun Ahuja, Thi Avrahami, Vedant Misra, Raoul
  de~Liedekerke, Mariko Iinuma, Alex Polozov, Sarah York, George van~den
  Driessche, Paul Michel, Justin Chiu, Rory Blevins, Zach Gleicher, Adrià
  Recasens, Alban Rrustemi, Elena Gribovskaya, Aurko Roy, Wiktor Gworek, Séb
  Arnold, Lisa Lee, James Lee-Thorp, Marcello Maggioni, Enrique Piqueras,
  Kartikeya Badola, Sharad Vikram, Lucas Gonzalez, Anirudh Baddepudi, Evan
  Senter, Jacob Devlin, James Qin, Michael Azzam, Maja Trebacz, Martin Polacek,
  Kashyap Krishnakumar, Shuo yiin Chang, Matthew Tung, Ivo Penchev, Rishabh
  Joshi, Kate Olszewska, Carrie Muir, Mateo Wirth, Ale~Jakse Hartman, Josh
  Newlan, Sheleem Kashem, Vijay Bolina, Elahe Dabir, Joost van Amersfoort,
  Zafarali Ahmed, James Cobon-Kerr, Aishwarya Kamath, Arnar~Mar Hrafnkelsson,
  Le~Hou, Ian Mackinnon, Alexandre Frechette, Eric Noland, Xiance Si, Emanuel
  Taropa, Dong Li, Phil Crone, Anmol Gulati, Sébastien Cevey, Jonas Adler, Ada
  Ma, David Silver, Simon Tokumine, Richard Powell, Stephan Lee, Michael Chang,
  Samer Hassan, Diana Mincu, Antoine Yang, Nir Levine, Jenny Brennan, Mingqiu
  Wang, Sarah Hodkinson, Jeffrey Zhao, Josh Lipschultz, Aedan Pope, Michael~B.
  Chang, Cheng Li, Laurent~El Shafey, Michela Paganini, Sholto Douglas, Bernd
  Bohnet, Fabio Pardo, Seth Odoom, Mihaela Rosca, Cicero~Nogueira dos Santos,
  Kedar Soparkar, Arthur Guez, Tom Hudson, Steven Hansen, Chulayuth
  Asawaroengchai, Ravi Addanki, Tianhe Yu, Wojciech Stokowiec, Mina Khan,
  Justin Gilmer, Jaehoon Lee, Carrie~Grimes Bostock, Keran Rong, Jonathan
  Caton, Pedram Pejman, Filip Pavetic, Geoff Brown, Vivek Sharma, Mario
  Lučić, Rajkumar Samuel, Josip Djolonga, Amol Mandhane, Lars~Lowe Sjösund,
  Elena Buchatskaya, Elspeth White, Natalie Clay, Jiepu Jiang, Hyeontaek Lim,
  Ross Hemsley, Jane Labanowski, Nicola~De Cao, David Steiner, Sayed~Hadi
  Hashemi, Jacob Austin, Anita Gergely, Tim Blyth, Joe Stanton, Kaushik
  Shivakumar, Aditya Siddhant, Anders Andreassen, Carlos Araya, Nikhil Sethi,
  Rakesh Shivanna, Steven Hand, Ankur Bapna, Ali Khodaei, Antoine Miech,
  Garrett Tanzer, Andy Swing, Shantanu Thakoor, Zhufeng Pan, Zachary Nado,
  Stephanie Winkler, Dian Yu, Mohammad Saleh, Loren Maggiore, Iain Barr, Minh
  Giang, Thais Kagohara, Ivo Danihelka, Amit Marathe, Vladimir Feinberg,
  Mohamed Elhawaty, Nimesh Ghelani, Dan Horgan, Helen Miller, Lexi Walker,
  Richard Tanburn, Mukarram Tariq, Disha Shrivastava, Fei Xia, Chung-Cheng
  Chiu, Zoe Ashwood, Khuslen Baatarsukh, Sina Samangooei, Fred Alcober, Axel
  Stjerngren, Paul Komarek, Katerina Tsihlas, Anudhyan Boral, Ramona Comanescu,
  Jeremy Chen, Ruibo Liu, Dawn Bloxwich, Charlie Chen, Yanhua Sun, Fangxiaoyu
  Feng, Matthew Mauger, Xerxes Dotiwalla, Vincent Hellendoorn, Michael Sharman,
  Ivy Zheng, Krishna Haridasan, Gabe Barth-Maron, Craig Swanson, Dominika
  Rogozińska, Alek Andreev, Paul~Kishan Rubenstein, Ruoxin Sang, Dan Hurt,
  Gamaleldin Elsayed, Renshen Wang, Dave Lacey, Anastasija Ilić, Yao Zhao,
  Lora Aroyo, Chimezie Iwuanyanwu, Vitaly Nikolaev, Balaji Lakshminarayanan,
  Sadegh Jazayeri, Raphaël~Lopez Kaufman, Mani Varadarajan, Chetan Tekur, Doug
  Fritz, Misha Khalman, David Reitter, Kingshuk Dasgupta, Shourya Sarcar, Tina
  Ornduff, Javier Snaider, Fantine Huot, Johnson Jia, Rupert Kemp, Nejc Trdin,
  Anitha Vijayakumar, Lucy Kim, Christof Angermueller, Li~Lao, Tianqi Liu,
  Haibin Zhang, David Engel, Somer Greene, Anaïs White, Jessica Austin, Lilly
  Taylor, Shereen Ashraf, Dangyi Liu, Maria Georgaki, Irene Cai, Yana
  Kulizhskaya, Sonam Goenka, Brennan Saeta, Kiran Vodrahalli, Christian Frank,
  Dario de~Cesare, Brona Robenek, Harry Richardson, Mahmoud Alnahlawi,
  Christopher Yew, Priya Ponnapalli, Marco Tagliasacchi, Alex Korchemniy, Yelin
  Kim, Dinghua Li, Bill Rosgen, Zoe Ashwood, Kyle Levin, Jeremy Wiesner,
  Praseem Banzal, Praveen Srinivasan, Hongkun Yu, Çağlar Ünlü, David Reid,
  Zora Tung, Daniel Finchelstein, Ravin Kumar, Andre Elisseeff, Jin Huang, Ming
  Zhang, Rui Zhu, Ricardo Aguilar, Mai Giménez, Jiawei Xia, Olivier Dousse,
  Willi Gierke, Soheil~Hassas Yeganeh, Damion Yates, Komal Jalan, Lu~Li, Eri
  Latorre-Chimoto, Duc~Dung Nguyen, Ken Durden, Praveen Kallakuri, Yaxin Liu,
  Matthew Johnson, Tomy Tsai, Alice Talbert, Jasmine Liu, Alexander Neitz, Chen
  Elkind, Marco Selvi, Mimi Jasarevic, Livio~Baldini Soares, Albert Cui, Pidong
  Wang, Alek~Wenjiao Wang, Xinyu Ye, Krystal Kallarackal, Lucia Loher, Hoi Lam,
  Josef Broder, Dan Holtmann-Rice, Nina Martin, Bramandia Ramadhana, Daniel
  Toyama, Mrinal Shukla, Sujoy Basu, Abhi Mohan, Nick Fernando, Noah Fiedel,
  Kim Paterson, Hui Li, Ankush Garg, Jane Park, DongHyun Choi, Diane Wu,
  Sankalp Singh, Zhishuai Zhang, Amir Globerson, Lily Yu, John Carpenter,
  Félix de~Chaumont~Quitry, Carey Radebaugh, Chu-Cheng Lin, Alex Tudor,
  Prakash Shroff, Drew Garmon, Dayou Du, Neera Vats, Han Lu, Shariq Iqbal, Alex
  Yakubovich, Nilesh Tripuraneni, James Manyika, Haroon Qureshi, Nan Hua,
  Christel Ngani, Maria~Abi Raad, Hannah Forbes, Anna Bulanova, Jeff Stanway,
  Mukund Sundararajan, Victor Ungureanu, Colton Bishop, Yunjie Li, Balaji
  Venkatraman, Bo~Li, Chloe Thornton, Salvatore Scellato, Nishesh Gupta,
  Yicheng Wang, Ian Tenney, Xihui Wu, Ashish Shenoy, Gabriel Carvajal,
  Diana~Gage Wright, Ben Bariach, Zhuyun Xiao, Peter Hawkins, Sid Dalmia,
  Clement Farabet, Pedro Valenzuela, Quan Yuan, Chris Welty, Ananth Agarwal,
  Mia Chen, Wooyeol Kim, Brice Hulse, Nandita Dukkipati, Adam Paszke, Andrew
  Bolt, Elnaz Davoodi, Kiam Choo, Jennifer Beattie, Jennifer Prendki, Harsha
  Vashisht, Rebeca Santamaria-Fernandez, Luis~C. Cobo, Jarek Wilkiewicz, David
  Madras, Ali Elqursh, Grant Uy, Kevin Ramirez, Matt Harvey, Tyler Liechty,
  Heiga Zen, Jeff Seibert, Clara~Huiyi Hu, Mohamed Elhawaty, Andrey Khorlin,
  Maigo Le, Asaf Aharoni, Megan Li, Lily Wang, Sandeep Kumar, Alejandro Lince,
  Norman Casagrande, Jay Hoover, Dalia~El Badawy, David Soergel, Denis Vnukov,
  Matt Miecnikowski, Jiri Simsa, Anna Koop, Praveen Kumar, Thibault Sellam,
  Daniel Vlasic, Samira Daruki, Nir Shabat, John Zhang, Guolong Su, Jiageng
  Zhang, Jeremiah Liu, Yi~Sun, Evan Palmer, Alireza Ghaffarkhah, Xi~Xiong,
  Victor Cotruta, Michael Fink, Lucas Dixon, Ashwin Sreevatsa, Adrian
  Goedeckemeyer, Alek Dimitriev, Mohsen Jafari, Remi Crocker, Nicholas
  FitzGerald, Aviral Kumar, Sanjay Ghemawat, Ivan Philips, Frederick Liu,
  Yannie Liang, Rachel Sterneck, Alena Repina, Marcus Wu, Laura Knight, Marin
  Georgiev, Hyo Lee, Harry Askham, Abhishek Chakladar, Annie Louis, Carl Crous,
  Hardie Cate, Dessie Petrova, Michael Quinn, Denese Owusu-Afriyie, Achintya
  Singhal, Nan Wei, Solomon Kim, Damien Vincent, Milad Nasr, Christopher~A.
  Choquette-Choo, Reiko Tojo, Shawn Lu, Diego de~Las~Casas, Yuchung Cheng,
  Tolga Bolukbasi, Katherine Lee, Saaber Fatehi, Rajagopal Ananthanarayanan,
  Miteyan Patel, Charbel Kaed, Jing Li, Jakub Sygnowski, Shreyas~Rammohan
  Belle, Zhe Chen, Jaclyn Konzelmann, Siim Põder, Roopal Garg, Vinod
  Koverkathu, Adam Brown, Chris Dyer, Rosanne Liu, Azade Nova, Jun Xu, Slav
  Petrov, Demis Hassabis, Koray Kavukcuoglu, Jeffrey Dean, and Oriol Vinyals.
\newblock {Gemini 1.5: Unlocking multimodal understanding across millions of
  tokens of context}.
\newblock {\em {ArXiv}}, 2403.05530, 2024.

\bibitem{Stable_Diffusion}
Robin Rombach, Andreas Blattmann, Dominik Lorenz, Patrick Esser, and Björn
  Ommer.
\newblock {Stable Diffusion}, 2021.

\bibitem{High-Resolution_Image_Synthesis}
Robin Rombach, Andreas Blattmann, Dominik Lorenz, Patrick Esser, and Björn
  Ommer.
\newblock {High-Resolution Image Synthesis with Latent Diffusion Models}.
\newblock {\em {ArXiv}}, 2112.10752, 2022.

\bibitem{Stable_Diffusion_v1-5_Model_Card}
Robin Rombach and Patrick Esser.
\newblock {Stable Diffusion v1-5 Model Card}, 2024.

\bibitem{U-Net}
Olaf Ronneberger, Philipp Fischer, and Thomas Brox.
\newblock {U-Net: Convolutional Networks for Biomedical Image Segmentation}.
\newblock In {\em {Proceedings of the 18th International Conference on Medical
  Image Computing and Computer-Assisted Intervention}}, pages 234--241, 2015.

\bibitem{Gen3}
Runway.
\newblock {Introducing Gen-3 Alpha: A New Frontier for Video Generation}, 2024.

\bibitem{Photorealistic_Text-to-Image_Diffusion_Models}
Chitwan Saharia, William Chan, Saurabh Saxena, Lala Li, Jay Whang, Emily
  Denton, Seyed Kamyar~Seyed Ghasemipour, Burcu~Karagol Ayan, S~Sara Mahdavi,
  Raphael Gontijo-Lopes, Tim Salimans, Jonathan Ho, David~J Fleet, and Mohammad
  Norouzi.
\newblock {Photorealistic Text-to-Image Diffusion Models with Deep Language
  Understanding}.
\newblock pages 36479--36494, 2024.

\bibitem{A_Systematic_Survey_of_Prompt_Engineering}
Pranab Sahoo, Ayush~Kumar Singh, Sriparna Saha, Vinija Jain, Samrat Mondal, and
  Aman Chadha.
\newblock {A Systematic Survey of Prompt Engineering in Large Language Models:
  Techniques and Applications}.
\newblock {\em {ArXiv}}, 2402.07927, 2025.

\bibitem{Comparison_and_Analysis_of_Image-to-Image}
Sagar Saxena and Mohammad~Nayeem Teli.
\newblock {Comparison and Analysis of Image-to-Image Generative Adversarial
  Networks: A Survey}.
\newblock {\em {ArXiv}}, 2112.12625, 2022.

\bibitem{What_is_noise}
John Scales and Roel Snieder.
\newblock {What is noise?}
\newblock {\em {Geophysics}}, 63(4):1122--1124, 1998.

\bibitem{Large_pre_trained_language_models}
Patrick Schramowski, Cigdem Turan-Schwiewager, Nico Andersen, Constantin
  Rothkopf, and Kristian Kersting.
\newblock {Large pre-trained language models contain human-like biases of what
  is right and wrong to do}.
\newblock {\em {Nature Machine Intelligence}}, 4:258--268, 2022.

\bibitem{A_comprehensive_review_of_large_language_models}
Tariq Shahzad, Tehseen Mazhar, Muhammad~Usman Tariq, Wasim Ahmad, Khmaies
  Ouahada, and Habib Hamam.
\newblock {A comprehensive review of large language models: issues and
  solutions in learning environments}.
\newblock {\em {Discover Sustainability}}, 6, 2025.

\bibitem{I_Hear_Your_True_Colors}
Roy Sheffer and Yossi Adi.
\newblock {I Hear Your True Colors: Image Guided Audio Generation}.
\newblock {\em {ArXiv}}, 2211.03089, 2023.

\bibitem{A_Survey_on_Audio_Synthesis}
Zhaofeng Shi.
\newblock {A Survey on Audio Synthesis and Audio-Visual Multimodal Processing}.
\newblock {\em {ArXiv}}, 2108.00443, 2021.

\bibitem{Audio-to-Visual_Cross-Modal}
Joo~Yong Shim, Joongheon Kim, and Jong-Kook Kim.
\newblock {Audio-to-Visual Cross-Modal Generation of Birds}.
\newblock {\em {IEEE Access}}, 11:27719--27729, 2023.

\bibitem{Outpainting_Images_and_Videos}
Shailendra Singh, Nainish Aggarwal, Udit Jain, and Hrithik Jaiswal.
\newblock {Outpainting Images and Videos using GANs}.
\newblock {\em {International Journal of Computer Trends and Technology}},
  68(5):24--29, 2020.

\bibitem{Pre_trained_Speech_Processing_Models}
Isaac Slaughter, Craig Greenberg, Reva Schwartz, and Aylin Caliskan.
\newblock {Pre-trained Speech Processing Models Contain Human-Like Biases that
  Propagate to Speech Emotion Recognition}.
\newblock In {\em {Findings of the 2023 Conference on Empirical Methods in
  Natural Language Processing}}, pages 8967--8989, 2023.

\bibitem{A_survey_of_multimodal_deep_generative_models}
Masahiro Suzuki and Yutaka Matsuo.
\newblock {A survey of multimodal deep generative models}.
\newblock {\em {Advanced Robotics}}, 36(5-6):261--278, 2022.

\bibitem{Codi2}
Zineng Tang, Ziyi Yang, Mahmoud Khademi, Yang Liu, Chenguang Zhu, and Mohit
  Bansal.
\newblock {CoDi-2: In-Context, Interleaved, and Interactive Any-to-Any
  Generation}.
\newblock {\em {ArXiv}}, 2311.18775, 2023.

\bibitem{Any-to-Any_Generation}
Zineng Tang, Ziyi Yang, Chenguang Zhu, Michael Zeng, and Mohit Bansal.
\newblock {Any-to-any generation via composable diffusion}.
\newblock In {\em {Proceedings of the 37th International Conference on Neural
  Information Processing Systems}}, pages 16083--16099, 2024.

\bibitem{Movie_Gen}
{The Movie Gen team}.
\newblock {Movie Gen: A Cast of Media Foundation Models}, 2024.

\bibitem{Llama}
Hugo Touvron, Thibaut Lavril, Gautier Izacard, Xavier Martinet, Marie-Anne
  Lachaux, Timothée Lacroix, Baptiste Rozière, Naman Goyal, Eric Hambro,
  Faisal Azhar, Aurelien Rodriguez, Armand Joulin, Edouard Grave, and Guillaume
  Lample.
\newblock {LLaMA: Open and Efficient Foundation Language Models}.
\newblock {\em {ArXiv}}, 2302.13971, 2023.

\bibitem{Structural_Equation_Modeling_in_Information_Systems}
Nils Urbach and Frederik Ahlemann.
\newblock {Structural Equation Modeling in Information Systems Research Using
  Partial Least Squares}.
\newblock {\em {Journal of Information Technology Theory and Application}},
  11(2):5--40, 2010.

\bibitem{Fugatto}
Rafael Valle, Rohan Badlani, Zhifeng Kong, Sang gil Lee, Arushi Goel, Sungwon
  Kim, Joao~Felipe Santos, Shuqi Dai, Siddharth Gururani, Aya AIJa'fari, Alex
  Liu, Kevin Shih, Wei Ping, Huck Yang, and Bryan Catanzaro.
\newblock {Fugatto 1 - Foundational Generative Audio Transformer Opus 1}, 2024.

\bibitem{Attention_is_All_you_Need}
Ashish Vaswani, Noam~M. Shazeer, Niki Parmar, Jakob Uszkoreit, Llion Jones,
  Aidan~N. Gomez, Lukasz Kaiser, and Illia Polosukhin.
\newblock {Attention is All you Need}.
\newblock In {\em {Proceedings of the 31st International Conference on Neural
  Information Processing Systems}}, page 6000–6010, 2017.

\bibitem{Audio_Describing_Sound}
Gert Vercauteren and Nina Reviers.
\newblock {Audio Describing Sound – What Sounds are Described and How?:
  Results from a Flemish case study}.
\newblock {\em {Journal of Audiovisual Translation}}, 5(2):114–133, 2022.

\bibitem{Audiobox}
Apoorv Vyas, Bowen Shi, Matthew Le, Andros Tjandra, Yi-Chiao Wu, Baishan Guo,
  Jiemin Zhang, Xinyue Zhang, Robert Adkins, William Ngan, Jeff Wang, Ivan
  Cruz, Bapi Akula, Akinniyi Akinyemi, Brian Ellis, Rashel Moritz, Yael
  Yungster, Alice Rakotoarison, Liang Tan, Chris Summers, Carleigh Wood, Joshua
  Lane, Mary Williamson, and Wei-Ning Hsu.
\newblock {Audiobox: Unified Audio Generation with Natural Language Prompts}.
\newblock {\em {ArXiv}}, 2312.15821, 2023.

\bibitem{Vall_e}
Chengyi Wang, Sanyuan Chen, Yu~Wu, Ziqiang Zhang, Long Zhou, Shujie Liu, Zhuo
  Chen, Yanqing Liu, Huaming Wang, Jinyu Li, Lei He, Sheng Zhao, and Furu Wei.
\newblock {Neural Codec Language Models are Zero-Shot Text to Speech
  Synthesizers}.
\newblock {\em {ArXiv}}, 2301.02111, 2023.

\bibitem{From_Association_to_Generation}
Junyang Wang, Ming Yan, Yi~Zhang, and Jitao Sang.
\newblock {From Association to Generation: Text-only Captioning by Unsupervised
  Cross-modal Mapping}.
\newblock In {\em {Proceedings of the 32nd International Joint Conference on
  Artificial Intelligence}}, pages 4326--4334, 2023.

\bibitem{A_Systematic_Review_and_Comprehensive_Analysis_of_Pioneering_AI_Chatbot_Models}
Ketmanto Wangsa, Shakir Karim, Ergun Gide, and Mahmoud Elkhodr.
\newblock {A Systematic Review and Comprehensive Analysis of Pioneering AI
  Chatbot Models from Education to Healthcare: ChatGPT, Bard, Llama, Ernie and
  Grok}.
\newblock {\em {Future Internet}}, 16(7), 2024.

\bibitem{Towards_audio_language_modeling}
Haibin Wu, Xuanjun Chen, Yi-Cheng Lin, Kai wei Chang, Ho-Lam Chung,
  Alexander~H. Liu, and Hung yi~Lee.
\newblock {Towards audio language modeling -- an overview}.
\newblock {\em {ArXiv}}, 2402.13236, 2024.

\bibitem{Audio_Text_Models_Do_Not_Yet}
Ho-Hsiang Wu, Oriol Nieto, Juan~Pablo Bello, and Justin Salamon.
\newblock {Audio-Text Models Do Not Yet Leverage Natural Language}.
\newblock In {\em {Proceedings of the 2023 IEEE International Conference on
  Acoustics, Speech and Signal Processing}}, pages 1--5, 2023.

\bibitem{Wav2CLIP}
Ho-Hsiang Wu, Prem Seetharaman, Kundan Kumar, and Juan~Pablo Bello.
\newblock {Wav2CLIP: Learning Robust Audio Representations from Clip}.
\newblock In {\em {Proceedings of the 2022 IEEE International Conference on
  Acoustics, Speech and Signal Processing}}, pages 4563--4567, 2022.

\bibitem{NExT_GPT}
Shengqiong Wu, Hao Fei, Leigang Qu, Wei Ji, and Tat-Seng Chua.
\newblock {NExT-GPT: Any-to-Any Multimodal LLM}.
\newblock {\em {ArXiv}}, 2309.05519, 2024.

\bibitem{Grok_3_Beta}
xAI.
\newblock {Grok 3 Beta — The Age of Reasoning Agents}, 2025.

\bibitem{Web-Bench}
Kai Xu, YiWei Mao, XinYi Guan, and ZiLong Feng.
\newblock {Web-Bench: A LLM Code Benchmark Based on Web Standards and
  Frameworks}.
\newblock {\em {ArXiv}}, 2505.07473, 2025.

\bibitem{Multimodal_Learning_With_Transformers}
Peng Xu, Xiatian Zhu, and David~A. Clifton.
\newblock {Multimodal Learning With Transformers: A Survey}.
\newblock {\em {IEEE Transactions on Pattern Analysis and Machine
  Intelligence}}, pages 1--20, 2023.

\bibitem{BLAT}
Xuenan Xu, Zhiling Zhang, Zelin Zhou, Pingyue Zhang, Zeyu Xie, Mengyue Wu, and
  Kenny~Q. Zhu.
\newblock {BLAT: Bootstrapping Language-Audio Pre-training based on AudioSet
  Tag-guided Synthetic Data}.
\newblock In {\em {Proceedings of the 31st ACM International Conference on
  Multimedia}}, page 2756–2764, 2023.

\bibitem{Bicro}
Shuo Yang, Zhaopan Xu, Kai Wang, Yang You, Hongxun Yao, Tongliang Liu, and Min
  Xu.
\newblock {BiCro: Noisy Correspondence Rectification for Multi-modality Data
  via Bi-directional Cross-modal Similarity Consistency}.
\newblock In {\em {Proceedings of the 2023 IEEE Conference on Computer Vision
  and Pattern Recognition}}, pages 19883--19892, 2023.

\bibitem{The_Dawn_of_LMMs}
Zhengyuan Yang, Linjie Li, Kevin Lin, Jianfeng Wang, Chung-Ching Lin, Zicheng
  Liu, and Lijuan Wang.
\newblock {The Dawn of LMMs: Preliminary Explorations with GPT-4V(ision)}.
\newblock {\em {ArXiv}}, 2309.17421, 2023.

\bibitem{AudioToken}
Guy Yariv, Itai Gat, Lior Wolf, Yossi Adi, and Idan Schwartz.
\newblock {AudioToken: Adaptation of Text-Conditioned Diffusion Models for
  Audio-to-Image Generation}.
\newblock {\em {ArXiv}}, 2305.13050, 2023.

\bibitem{Multimodal_Image_Synthesis_and_Editing}
Fangneng Zhan, Yingchen Yu, Rongliang Wu, Jiahui Zhang, Shijian Lu, Lingjie
  Liu, Adam Kortylewski, Christian Theobalt, and Eric Xing.
\newblock {Multimodal Image Synthesis and Editing: The Generative AI Era}.
\newblock {\em {ArXiv}}, 2112.13592, 2023.

\bibitem{Text-to-image_Diffusion_Models}
Chenshuang Zhang, Chaoning Zhang, Mengchun Zhang, and In~So Kweon.
\newblock {Text-to-image Diffusion Models in Generative AI: A Survey}.
\newblock {\em {ArXiv}}, 2303.07909, 2023.

\bibitem{Remote_Sensing_Image_Generation_From_Audio}
Zhiyuan Zheng, Jun Chen, Xiangtao Zheng, and Xiaoqiang Lu.
\newblock {Remote Sensing Image Generation From Audio}.
\newblock {\em {IEEE Geoscience and Remote Sensing Letters}}, 18(6):994--998,
  2021.

\bibitem{Cacophony}
Ge~Zhu and Zhiyao Duan.
\newblock {Cacophony: An Improved Contrastive Audio-Text Model}.
\newblock {\em {ArXiv}}, 2402.06986, 2024.

\bibitem{Deep_Audio-visual_Learning}
Hao Zhu, Man-Di Luo, Rui Wang, Ai-Hua Zheng, and Ran He.
\newblock {Deep Audio-visual Learning: A Survey}.
\newblock {\em {International Journal of Automation and Computing}},
  18:351–376, 2021.

\bibitem{On_Some_Biases_Encountered}
Sławomir Zieliński, Francis Rumsey, and Søren Bech.
\newblock {On Some Biases Encountered in Modern Audio Quality Listening Tests -
  A Review}.
\newblock {\em {Journal of the Audio Engineering Society}}, 56(6):427--451,
  2008.

\bibitem{Audio-to-Image_Cross-Modal}
Maciej Żelaszczyk and Jacek Mańdziuk.
\newblock {Audio-to-Image Cross-Modal Generation}.
\newblock {\em {ArXiv}}, 2109.13354, 2021.

\end{thebibliography}

%\newpage
%\appendix

\end{document}